\documentclass[longauth]{aa} 
\usepackage{graphicx}
\usepackage{natbib}
\bibpunct{(}{)}{;}{a}{}{,} 
\usepackage[toc,page]{appendix}
\usepackage{txfonts}
\usepackage{multicol}
\usepackage{amsmath}
\usepackage{hyperref}
\usepackage{lscape}
\usepackage{longtable}
\begin{document}
\title{Gliese 49: Activity evolution and detection of a super-Earth\thanks{Based on observations made with the Italian TNG, operated on the island of La Palma, Spain; the CARMENES instrument installed at the 3.5\,m telescope of the Calar Alto Observatory, Spain; the robotic APT2 located at Serra La Nave on Mt. Etna, Italy; and the T90 telescope at Sierra Nevada Observatory, Spain; Table\,\ref{TT1} is only available in electronic form at the CDS via anonymous ftp to cdsarc.u-strasbg.fr (130.79.128.5) or via http://cdsweb.u-strasbg.fr/cgi-bin/qcat?J/A+A/.}}
\subtitle{A HADES and CARMENES collaboration}
\author{M.~Perger\inst{1,2}
        \and G.~Scandariato\inst{3}  	
        \and I.~Ribas\inst{1,2}
        \and J.~C.~Morales\inst{1,2}
		\and L.~Affer\inst{4}
		\and M.~Azzaro\inst{5}
        \and P.~J.~Amado\inst{6}    
        \and G.~Anglada-Escud{\'e}\inst{6,7}    
        \and D.~Baroch\inst{1,2}
        \and D.~Barrado\inst{8}
        \and F.~F.~Bauer\inst{6}
	    \and V.~J.~S.~B{\'e}jar\inst{9,10} 
        \and J.~A.~Caballero\inst{8}
        \and M.~Cort{\'e}s-Contreras\inst{8} 
        \and M.~Damasso\inst{11} 
		\and S.~Dreizler\inst{12} 
		\and L.~Gonz{\'a}lez-Cuesta\inst{9,10}  
		\and J.~I.~Gonz{\'a}lez Hern{\'a}ndez\inst{9,10}   
        \and E.~W.~Guenther\inst{13}
		\and T.~Henning\inst{14}  
        \and E.~Herrero\inst{1,2} 
		\and S.V.~Jeffers\inst{12}   
		\and A.~Kaminski\inst{15} 
		\and M.~K{\"u}rster\inst{14} 
        \and M.~Lafarga\inst{1,2}   
        \and G.~Leto\inst{3} 
        \and M.~J.~L{\'o}pez-Gonz{\'a}lez\inst{6} 
        \and J.~Maldonado\inst{4}
        \and G.~Micela\inst{4}
		\and D.~Montes\inst{16} 
		\and M.~Pinamonti\inst{11}          
		\and A.~Quirrenbach\inst{15} 
		\and R.~Rebolo\inst{9,10,17} 
		\and A.~Reiners\inst{12}  
        \and E.~Rodr{\'i}guez\inst{6}
        \and C.~Rodr{\'i}guez-L{\'o}pez\inst{6}
		\and J.~H.~M.~M.~Schmitt\inst{18}
		\and A.~Sozzetti\inst{11}
		\and A.~Su{\'a}rez Mascare{\~n}o\inst{9,19}
		\and B.~Toledo-Padr{\'o}n\inst{9,10}
        \and R.~Zanmar S{\'a}nchez\inst{3}
        \and M.~R.~Zapatero Osorio\inst{20}
		\and M.~Zechmeister\inst{12}
 }
\offprints{Manuel Perger, \email{perger@ice.cat}}
\institute{
		\inst{1}Institut de Ci$\grave{\rm e}$ncies de l'Espai (ICE, CSIC), Campus UAB, Carrer de Can Magrans s/n, 08193 Bellaterra, Spain\\
        \inst{2}Institut d'Estudis Espacials de Catalunya (IEEC), 08034 Barcelona, Spain\\
        \inst{3}INAF - Osservatorio Astrofisico di Catania, via S. Sofia 78, 95123 Catania, Italy\\
		\inst{4}INAF - Osservatorio Astronomico di Palermo, Piazza del Parlamento 1, 90134 Palermo, Italy\\		
		\inst{5}Centro Astron\'omico Hispano-Alem\'an (CAHA). Calar Alto Observatory, c/ Jes\'us Durb\'an Rem\'on 2-2, 04004 Almer\'ia, Spain\\
		\inst{6}Instituto de Astrofísica de Andaluc\'ia (IAA-CSIC), Glorieta de la Astronom\'ia s/n, 18008 Granada, Spain\\
		\inst{7}School of Physics and Astronomy, Queen Mary University of London, 327 Mile End Rd, E1 4NS London, United Kingdom\\
		\inst{8}Centro  de  Astrobiolog\'ia  (CSIC-INTA),  Camino  Bajo  del  Castillo  s/n,  ESAC  Campus,  28692, Villanueva, Madrid, Spain\\
		\inst{9}Instituto de Astrof\'isica de Canarias (IAC), 38205 La Laguna, Tenerife, Spain\\
		\inst{10}Universidad de La Laguna (ULL), Departamento de Astrof\'isica, 38206 La Laguna, Tenerife, Spain\\
		\inst{11}INAF - Osservatorio Astrofisico di Torino, via Osservatorio 20, 10025 Pino Torinese, Italy\\ 
		\inst{12}Institut für Astrophysik - Georg-August-Universit\"at G\"ottingen, Friedrich-Hund-Platz 1, 37077 G\"ottingen, Germany\\	
		\inst{13}Th\"uringer Landessternwarte Tautenburg, Sternwarte 5, 07778 Tautenburg, Germany\\
		\inst{14}Max-Planck-Institut für Astronomie, K\"onigstuhl 17, 69117 Heidelberg, Germany\\
		\inst{15}Landessternwarte, Zentrum f\"ur Astronomie der Universt\"at Heidelberg, K\"onigstuhl 12, 69117 Heidelberg, Germany\\
		\inst{16}Departamento de F\'isica de la Tierra y Astrof\'isica \& UPARCOS-UCM, Facultad de Ciencias F\'isicas, Universidad Complutense de Madrid, 28040 Madrid, Spain\\ 
		\inst{17}Consejo Superior de Investigaciones Cient\'ificas (CSIC), 28006 Madrid, Spain\\ 		
		\inst{18}Hamburger Sternwarte, Universit\"at Hamburg, Gojenbergsweg 112, 21029 Hamburg, Germany\\
		\inst{19}Observatoire Astronomique de l'Universit\'e de Gen$\grave{\rm e}$ve, 1290 Versoix, Switzerland\\
		\inst{20}Centro de Astrobiolog\'ia (CSIC-INTA), Carretera de Ajalvir km 4, 28850 Torrej\'on de Ardoz, Madrid, Spain\\
}
\date{Accepted: 11 March 2019}
\abstract
{Small planets around low-mass stars often show orbital periods in a range that corresponds to the temperate zones of their host stars which are therefore of prime interest for planet searches. Surface phenomena such as spots and faculae create periodic signals in radial velocities and in observational activity tracers in the same range, so they can mimic or hide true planetary signals.}
{We aim to detect Doppler signals corresponding to planetary companions, determine their most probable orbital configurations, and understand the stellar activity and its impact on different datasets.}
{We analyzed 22\,years of data of the M1.5\,V-type star Gl\,49 (BD+61\,195) including HARPS-N and CARMENES spectrographs, complemented by APT2 and SNO photometry. Activity indices are calculated from the observed spectra, and all datasets are analyzed with periodograms and noise models. We investigated how the variation of stellar activity imprints on our datasets. We further tested the origin of the signals and investigate phase shifts between the different sets. To search for the best-fit model we maximize the likelihood function in a Markov chain Monte Carlo approach.}
{As a result of this study, we are able to detect the super-Earth Gl\,49b with a minimum mass of 5.6\,M$_{\oplus}$. It orbits its host star with a period of 13.85\,d at a semi-major axis of 0.090\,au and we calculate an equilibrium temperature of 350\,K and a transit probability of 2.0\,\%. The contribution from the spot-dominated host star to the different datasets is complex, and includes signals from the stellar rotation at 18.86\,d, evolutionary timescales of activity phenomena at 40--80\,d, and a long-term variation of at least four years.}
{}
\maketitle
%
\section{Introduction} \index{int} \label{int}

Time-series observations with high-resolution spectrographs such as the High Accuracy Radial velocity Planet Searcher of the southern \cite[HARPS;][]{2003Msngr.114...20M} and northern \cite[HARPS-N;][]{2012SPIE.8446E..1VC} hemispheres, the Calar Alto high-Resolution search for M dwarfs with Exoearths with Near-infrared and optical echelle Spectrographs \citep[CARMENES;][]{2018SPIE10702E..0WQ}, or the iodine-cell HIgh Resolution echelle Spectrograph \cite[HIRES;][]{1994SPIE.2198..362V} are used to detect and confirm planetary companions of stars by the Doppler shifts of their spectra. The measured radial velocities (RVs) show variations over time that are induced by the Keplerian orbit of the planet, or planets, and a contribution of the stellar surface. The latter can include actual motions on the surface such as oscillations or surface granulation with different short-term timescales \citep{2011A&A...525A.140D}, but also phenomena such as dark spots and bright plages, which introduce RV variations by reducing the number of photons from either the blue- or the red-shifted side of the rotating host star. Those phenomena are caused by the magnetic field of the star and its detailed structure and variability. The typical lifetime of those surface phenomena follows a parabolic decay law depending on size \citep{1997SoPh..176..249P} and can last for up to several rotation periods \citep{2014ApJ...795...79B, 2017A&A...598A..28S} or even longer \citep[see, e.g.,][]{2015ApJ...806..212D}. In the Sun, we also observe a long-term magnetic cycle of approximately 11\,years\footnote{22\,years if considering polarity} that is induced by the migration of the spot patterns across different latitudes and by the variation of the spot number. Such cycles are expected for all stellar types \citep{2011arXiv1107.5325L, 2016A&A...595A..12S}. The stellar contribution to the RVs, therefore, can be described partly as uncorrelated noise, but also possibly correlated with the rotational period of the star, the lifetime of the activity phenomena, or any long-term magnetic cycle. 

The disentangling of activity-induced variations and planetary signals is of great interest, especially due to the numerous planet searches around M dwarfs that are currently being conducted. The rotational periods of these cool stars range from some days to months and coincide with the orbital periods of planets in hot or temperate orbits which are, therefore, prime targets for detection and characterization \citep{2007AcA....57..149K, 2016ApJ...821L..19N, 2019A&A...621A.126D}. Planets of a given mass and period induce larger RV amplitudes in low-mass stars but these stars are also magnetically more active \citep[e.g.,][]{2015RSPTA.37340259T}, with longer-living surface phenomena \citep{2017MNRAS.472.1618G} and less influence of surface granulation than stars of earlier type \citep{2016MNRAS.459.3551B}. They are the most numerous stars in the Galaxy and in the immediate solar vicinity (75\% of all stars closer than 10\,pc are M dwarfs according to the RECONS survey\footnote{\url{www.recons.org}}), giving us the possibility to probe our neighboring planetary population. Most importantly, the surveys around M dwarfs are the best possibility to date to access the domain of rocky planets in order to study their atmospheres and to apply first statistical calculations \citep[e.g.,][]{2011arXiv1109.2497M, 2013ApJ...767...95D, 2013A&A...549A.109B, 2017A&A...598A..26P}. 

Such surveys include the HARPS-N red Dwarf Exoplanet Survey \citep[HADES;][]{2016A&A...593A.117A}, which monitored regularly around 80 M0- to M3-type stars and detected and confirmed seven Neptune- to Earth-like planets \citep{2017A&A...605A..92S, 2017A&A...608A..63P, 2018A&A...617A.104P, 2019A&A...622A.193A}. The HADES program is a collaboration between Italian and Spanish institutes, and has also extensively explored the rotational and magnetic behavior of those low-mass stars \citep{2017A&A...598A..27M, 2017A&A...598A..28S, 2018A&A...612A..89S, 2019arXiv190204868G}. Another survey is conducted with the CARMENES instrument, which is monitoring regularly more than 300 M-type dwarfs selected from the input catalog Carmencita \citep{2016csss.confE.148C}. The program has already confirmed a number of planets \citep{2018A&A...609A.117T, 2018AJ....155..257S} and detected up to seven previously unknown planets \citep{2018A&A...609L...5R, 2018A&A...618A.115K, 2018A&A...620A.171L,2019A&A...622A.153N}, including a cold companion of the nearby Barnard's Star \citep{2018Natur.563..365R}. The instrument is an effort of German and Spanish institutes to fully explore for the first time a large and complete sample of nearby M dwarfs of all spectral subtypes. With its wide range of wavelengths, CARMENES is especially capable of disentangling different sources of RV variability. Whereas a planetary orbit shifts the stellar light of all wavelengths equally, the contribution from surface phenomena is connected to physical processes that usually depend on wavelength. 

In this work we have combined observations from both the HADES and CARMENES programs in order to search for planetary companions around Gl\,49. This is a low-mass star with spectral type M1.5\,V located in the Cassiopeia constellation. Its kinematics and large proper motions suggest a membership to the young disk population of the Milky Way \citep{PhDCortes-Contreras, 2017A&A...598A..27M}. The star shows a moderate activity level and Sun-like metallicity, and was found to rotate with a period of 18.4$\pm$0.7\,d by HARPS-N activity indices \citep{2018A&A...612A..89S} and 19.9$\pm$0.4\,d by MEarth photometry \citep{2019A&A...621A.126D}. We provide an overview of its basic properties in Table~\ref{T1}. Gl~49 forms the wide binary system WNO\,51 \citep[Washington Double Star catalog;][]{2001AJ....122.3466M} with a fainter proper-motion companion at 293.1\,arcsec to the east, namely Gl~51 (V388~Cas, Karm J01033+623). The companion is a flaring M5.0\,V star with a relatively large amplitude of RV variations due to activity \citep{2015A&A...577A.128A, 2018A&A...614A.122T, 2018A&A...614A..76J}, but for which reliable membership in young moving groups, and thus an age estimate, has not been reported \citep[e.g.,][]{2015ApJS..219...33G}. 

\begin{table} 
        \caption{Basic parameters of Gliese\,49 (BD+61\,195, Karm J01026+623).}  \label{T1}
        \small
        \centering
        \begin{tabular}{lll}
                \hline \hline
                \noalign{\smallskip}
                Parameter & Value & Reference \\
                \hline
                \noalign{\smallskip}
                $\alpha$ (J2000) &  01h 02m 40.5s & (1) \\ 
                $\delta$ (J2000) & +62$^{\circ}$ 20' 44'' & (1)  \\  
                Sp. type & M1.5\,V & (2,3) \\	
                $d$ & 9.856$\pm$0.003\,pc  &  (1) \\   
                $M$ & 0.515$\pm$0.019~M$_{\odot}$  & (4)\\	
		        $R$ & 0.511$\pm$0.018~R$_{\odot}$  & (4)\\   
                $T_{\rm eff}$  & 3805$\pm$51~K  & (4)\\		      
                $L_{\rm bol}$  & 0.04938$\pm$0.00090 L$_{\odot}$ & (4) \\ 
                $\log{g}$ & 4.69$\pm$0.07\,dex  & (4)\\			           
                $\rm [Fe/H]$ & 0.13$\pm$0.16  & (4)\\          
                $\log {R}'_{\rm HK}$ & $-$4.83$\pm$0.03\,dex & (5) \\
                pEW(H$\alpha$) & $-$0.044$\pm$0.087\,$\AA{}$ & (6) \\
                $\log{L_{\rm X}/L_{bol}}$  & $-$4.70$\pm$0.09\,dex  & (7) \\    
                $V$ & 9.56$\pm$0.02~mag & (8) \\
                $G$ & 8.66\,mag & (1)\\   
                $J$ & 6.230$\pm$0.021~mag & (9)\\  
                $\mu_{\alpha} \cos{\delta}$ & $+$731.134$\pm$0.041~mas~a$^{-1}$ & (1)\\ 
                $\mu_{\delta}$  & $+$90.690$\pm$0.048~mas~a$^{-1}$ & (1) \\  
                $V_{\rm r}$ & -5.777$\pm$0.066\,km\,s$^{-1}$ & (10) \\
                dv/dt  & 12.304$\pm$0.004~cm~s$^{-1}$a$^{-1}$ & (1) \\ 
                $P_{\rm rot}$ & 18.4$\pm$0.7\,d, 19.9$\pm$0.4\,d & (5,11) \\             
                $v \sin{i}$ & $<$2~km~s$^{-1}$ & (12)\\ 
                HZ & 0.18-0.49~au, 39-172~d & (13) \\ 
                \noalign{\smallskip}
                \hline 
        \end{tabular}
        \tablefoot{
        	\tablefoottext{1}{\cite{2018A&A...616A...1G}},
        	\tablefoottext{2}{\cite{2015A&A...577A.128A}},
			\tablefoottext{3}{\cite{2017A&A...598A..27M}},
			\tablefoottext{4}{\cite{Schweitzer2019}},			
			\tablefoottext{5}{\cite{2018A&A...612A..89S}},			
			\tablefoottext{6}{\cite{2019A&A...623A..44S}},
			\tablefoottext{7}{\cite{2019arXiv190204868G}},	
			\tablefoottext{8}{\cite{1998A&A...335L..65H}},
			\tablefoottext{9}{\cite{2006AJ....131.1163S}},
			\tablefoottext{10}{this work},	
			\tablefoottext{11}{\cite{2019A&A...621A.126D}},			
			\tablefoottext{12}{\cite{2018A&A...612A..49R}},
			\tablefoottext{13}{recent Venus/early Mars habitable zones (HZ) following \cite{2013ApJ...765..131K}.}
			}		 
\end{table}

In Sect.\,\ref{dat:ob} we present the spectroscopic observations of Gl\,49 and their treatment. In Sect.\,\ref{ain} we analyze the variation and evolution of the activity of the star with additional photometric observations and various time-series data derived from the observed spectra. We study their properties by investigating their periodicities, calculating phase shifts and model the data as correlated noise in a Gaussian process (GP) framework. We describe further in Sect.\,\ref{sig} how we apply different state-of-the-art models to our system in order to ascertain the existence of a planet, and to find the most likely parameters for the proposed system. A detailed discussion about our findings regarding the discovered system and the evolution of the stellar activity over the last six years is given in Sect.\,\ref{dis}. We conclude the work in Sect.\,\ref{con}.

\section{Spectroscopic observations} \index{dat:ob} \label{dat:ob}

\begin{figure*}
 	\resizebox{\hsize}{!}{\includegraphics[width=\textwidth]{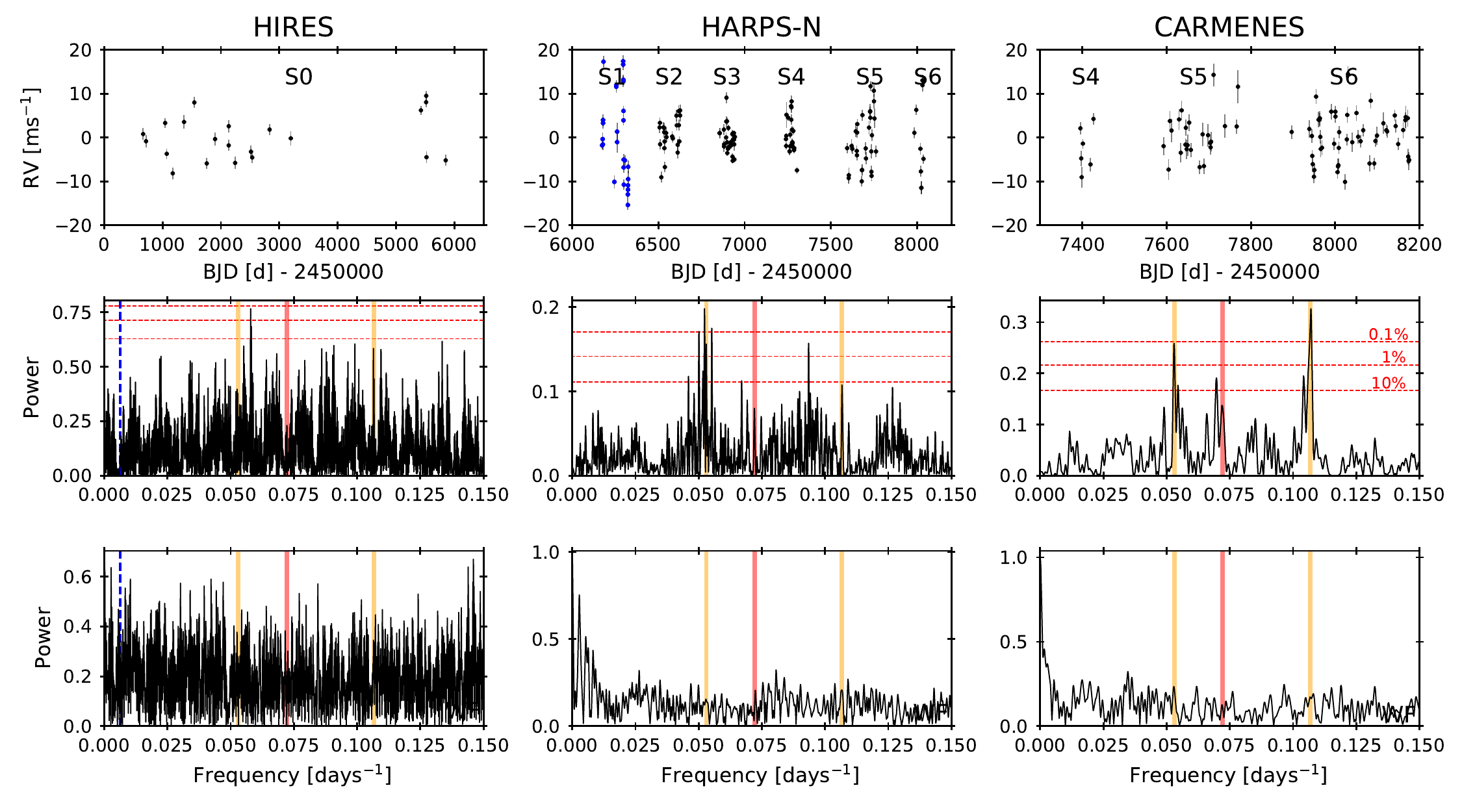}}
 	\caption{RV time series ({\it top panels}), GLS periodograms ({\it middle panels}), and window functions ({\it bottom panels}) of the HIRES ({\it left column}), HARPS-N ({\it middle column}) and CARMENES ({\it right column}) RV data. The different observational seasons are labeled from S0 to S6, and the unusual season S1 is colored in blue. In the periodograms, we mark the periods at 9.37 and 18.86\,d with orange vertical lines, the period at 13.85\,d with a red vertical line, and the 0.1, 1, and 10\,\% analytical FAP levels by red horizontal dashed lines. The blue vertical dashed line for the HIRES data shows the Nyquist frequency at (159\,d)$^{-1}$=0.00629\,d$^{-1}$.}
 	\centering
 	\label{F1}
 \end{figure*}

We obtained 137 RVs from optical spectra of the HADES program. They were observed over six seasons (S1 to S6) between 3 Sep 2012 and 11 Oct 2017 with HARPS-N. The instrument is installed since 2012 at the 3.58\,m Telescopio Nazionale Galileo (TNG) located at the Roque de Los Muchachos Observatory in La Palma, Spain, and is connected to its Nasmyth B focus through a front-end unit. It is a fiber-fed, cross-dispersed echelle spectrograph with a spectral resolution of 115\,000, covering a wavelength range from 3\,830 to 6\,900~$\AA{}$. In the HADES framework, we observed all targets with fixed integration times of 900\,s to obtain data of sufficient signal-to-noise ratio and to average over short-term variations. The data were reduced with the data reduction pipeline DRS \citep{2007A&A...468.1115L}. We extracted RVs using the Java-based Template-Enhanced Radial velocity Re-analysis Application code \cite[TERRA;][]{2012ApJS..200...15A}, which has been shown to deliver RV time-series of lower dispersion than the binary mask technique used by the DRS, at least in the case of early M-type stars \citep{2017A&A...598A..26P}. As outlined by those authors, we added quadratically a value of 1.0\,m\,s$^{-1}$ to the RV uncertainty if no correction for RV drifts was applied by observing simultaneously a Th-Ar lamp with the second fiber. The DRS pipeline delivers a value for the absolute RV shift for Gl\,49 of -5.777$\pm$0.066\,km\,s$^{-1}$. The statistics of the datasets are shown in Table\,\ref{T2}, and the RV values are given in Table\,\ref{TT1} and visualized in the top panels of Fig.\,\ref{F1}. The whole HARPS-N set shows small RV uncertainties but a strong variation in RV scatter over the seasons. Assuming a semi-periodic behavior, we would set a lower limit of recurring apparent RV amplitudes at approximately 1500\,d, with a minimum at S3 to S4. Additionally, S1 shows an exceptionally large RV scatter, pointing toward a rather variable stellar contribution. 

Furthermore, we obtained spectroscopic observations with the CARMENES instrument, installed since 2015 at the 3.51\,m telescope of the Calar Alto Observatory in Spain. Its wavelength coverage ranges from 5\,200 to 9\,600~$\AA$ in the optical and up to 17\,100~$\AA$ in the near-infrared, with resolutions of 94\,600 and 80\,400, respectively \citep{2016SPIE.9908E..12Q, 2018SPIE10702E..0WQ}. The visual channel is capable of reaching RV precisions of 1 to 2~\,m\,s$^{-1}$ \citep[e.g.,][]{2018A&A...609A.117T}. As reduction pipeline, we used the CARMENES Reduction And CALibration software \citep[CARACAL;][]{2016SPIE.9910E..0EC}, and to calculate the RVs, the template-based SpEctrum Radial Velocity AnaLyser \citep[SERVAL;][]{2018A&A...609A..12Z}. In the framework of the CARMENES Guaranteed Time Observation program, the target was observed 80 times from 7 Jan 2016 to 25 Feb 2018, covering the three seasons S4, S5, and S6 of the HARPS-N observations. The optical echelle spectra were obtained with exposure times from 100 to 800\,s in order to reach a signal-to-noise ratio of 150. The optical data set shows a low RV scatter. The data of the near-infrared channel of CARMENES, on the other hand, are not used in this study since the RV amplitudes that we are studying for our early M-dwarf target are expected to be smaller than the RV precision.

Gl\,49 was also observed with the HIRES instrument, installed since the late 1990s at the Keck I telescope located in Hawaii, USA \citep{1994SPIE.2198..362V}. The instrument observes from 3\,000 to 10\,000~$\AA$ with a resolution of up to 85\,000 and a precision of down to 1\,m\,s$^{-1}$. HIRES uses a iodine cell to monitor and correct for the RV drifts introduced by temperature changes. In total, data were taken at 21 epochs between 6 Aug 1996 and 17 Oct 2011 with, on average, one observation every 259\,d. The data were released by \cite{2017AJ....153..208B}, but we used the corrections applied by \cite{2019MNRAS.484L...8T}. We refer to this set of data in the following as season S0, since it does not overlap with our HARPS-N and CARMENES observations and since it is not visibly separated into the seasons of observability. The data show small RV scatter and uncertainty.
We considered for the RVs of all three instruments instrumental RV drifts, a barycentric correction following \citet{2014PASP..126..838W}, and the small secular acceleration of about 0.12~m\,s$^{-1}$yr$^{-1}$ of Gl\,49 (see Table\,\ref{T1}).

\begin{table*}[htb]
	\caption{Basic statistics and main periodicities of the different RV datasets.}  \label{T2}
	\small
	\centering
	\begin{tabular}{lcccccl}
		\hline \hline
		\noalign{\smallskip}
		Data set  & $N_{\rm obs}$ & rms RV [m\,s$^{-1}$] & dRV [m\,s$^{-1}$] & $\Delta T$ [d] & $\delta t$  [d] & Main periodicities [d] \\ 
		\noalign{\smallskip}
	    \hline	 
		\noalign{\smallskip}
		HIRES S0 & 21 & 4.99 & 1.31 & 5185 & 259.3 & 17.3 \\
		HARPS-N S1-S6 & 137 & 6.27 & 1.18 & 1864 & 13.1 & 19.1, 18.9, 9.6, 13.4, (18.3) \\ 
		\hspace{0.2cm} HARPS-N S1 & 27 & 9.99 & 1.35 & 151 & 5.8 & 18.6, 7.6, 14.5 \\
		\hspace{0.2cm} HARPS-N S2-S6 & 110 & 4.94 & 1.14 & 1531 & 14.1 & 9.4, 19.0, 14.4, (18.9)\\ 
		CARMENES S4-S6 & 80 & 4.97 & 1.78 & 780 & 10.0 & 9.3, 18.9, (14.4) \\
		\noalign{\smallskip}
		\hline 
	\end{tabular}
	  \tablefoot{dRV refers to the RV uncertainty, $\Delta T$ is the total observational timespan, and $\delta t$ is the mean separation between different epochs. Periodicities are sorted by their occurence in the prewhitening process. After the significant periods (FAP$<$0.1\%), we show the tentative signals (0.1\%$<$FAP$<$1\%) in parentheses.}	
\end{table*}
 
\section{Stellar activity} \index{ain} \label{ain}

The stellar contribution to RV variations can be very complex and mimic or mask a planetary companion. In order to understand those effects more profoundly and to be able to include this knowledge in our RV data modeling, we analyzed the activity level of Gl\,49 and collected and derived various additional time series, which are generally understood as tracers for stellar activity effects of the star and on its surface.

\subsection{Activity level} \index{ain:al} \label{ain:al}

Gl\,49 shows a moderate activity level, which is expressed by its calcium and H$\alpha$ indices as shown in Table\,\ref{T1} with $\log {R}'_{\rm HK}$=$-$4.8\,dex and pEW(H$\alpha$)=$-$0.04\,$\AA{}$. In the recent {\it RASS} (ROSAT All-Sky Survey) source catalog by \citet{2016A&A...588A.103B}, the source 2RXS\,J010318.3+622140 is close to the position of the companion Gl\,51. Its count rate of 0.24\,ct\,s$^{-1}$ leads to an X-ray luminosity of about 1.7$\times$10$^{28}$\,erg\,s$^{-1}$ for the whole system. This value refers predominantly to Gl\,51, however, a close inspection of the images shows an extended source with an elongation toward Gl\,49. The observation of Gl\,51 (Obs-Id 0742230501) in the XMM-Newton archive contains Gl\,49 in its field-of-view. The two sources are clearly separated and the X-ray flux and luminosity can be determined separately. For Gl\,49, \cite{2019arXiv190204868G} estimate an X-ray luminosity of about 3.9$\times$10$^{27}$\,erg\,s$^{-1}$ (see Table\,\ref{T1}), which accounts for about 20\% of the total flux, essentially consistent with the RASS measurements. We then find Gl\,49 located in the upper quartile of the early M-dwarf X-ray luminosity distribution function \citep{1995ApJ...450..392S}.

\subsection{Photometric data} \index{ain:ph} \label{ain:ph}

We obtained photometric observations in the framework of the HADES and CARMENES consortia. The basic properties and the temporal distributions of the data are shown in Table\,\ref{T3} and in Fig.\,\ref{A3}.

Simultaneously to the spectroscopic observations during seasons S5 and S6, we collected data within the EXORAP (EXOplanetary systems Robotic APT2 Photometry) program. It is carried out at INAF-Catania Astrophysical Observatory with an 80-cm f/8 Ritchey-Chr{\'e}tien robotic telescope (Automated Photoelectric Telescope, APT2) located at Serra la Nave on Mt. Etna. $BVRI$ photometry of the star was collected over 157 nights between 7 Jun 2016 and 18 Oct 2017. It covers 499\,d with an average of one observation every 2.0 to 2.3\,d for $BVRI$ filters. To obtain differential photometry, we started with an ensemble of about ten stars, the brightest that were close to Gl\,49. We checked the variability of each of them by building their differential light curves using the rest of the sample as a reference. In that way, we selected the four least variable stars of the sample for filters $B$, $V$, and $I$ (five stars in $R$). The average rms of the ensemble stars was 15, 15, 24, 21 ~mmag in the $B$, $V$, $R$, and $I$ filters, respectively. We obtained 250, 239, 230, and 223 data points, respectively, and rejected very few of them by a 5$\sigma$ clip. We calculated average photometric uncertainties as sky plus Poisson noise. As visible in Fig.\,\ref{A3}, Gl\,49 is slightly fainter in S5 than in S6 for EXORAP filter $B$, but the relationship seems to shift going to longer wavelengths until the star is brighter in S5 for filter $I$.

We also used the 90\,cm Ritchey-Chr{\'e}tien T90 telescope at Sierra Nevada Observatory (SNO), Spain. It is equipped with a 2k$\times$2k CCD camera and a field of view of 13.2$\times$13.2\,arcmin$^{2}$ \citep{2010MNRAS.408.2149R}. The observations were collected in Johnson $V$ and $R$ filters on 44 nights in 2018 during the period from 11 Jun to 21 Sep. The observations were carried out after the RV campaigns in order to search for a possible planetary transit. The measurements were obtained by the method of synthetic aperture photometry using no binning and in average 25 observations of 10 to 25\,s per night. Each CCD frame was corrected in a standard way for bias and flat-fielding. Different aperture sizes were also tested in order to choose the best one for our observations (16 pixels). A number of carefully selected nearby and relatively bright stars within the frames were used as reference stars. Outlier points due to bad weather conditions or high airmass were removed and a nightly average applied, since no transit event could be detected and no short-periodic signal was expected. The data of the two filters show a correlation with similar variations (see Fig\,\ref{A3}), but slightly larger errors and a larger apparent amplitude in $R$ band. As visible in Table\,\ref{T3}, the average uncertainties exceed the rms of the data of both filters.

Besides the photometry provided by the HADES and CARMENES consortia, we analyzed 11 years of observations from Las Campanas Observatory, Chile, with the ASAS-3N system \citep[All Sky Automated Survey;][]{1997AcA....47..467P}. The 440 data points cover 434 nights from 3 Jun 2006 to 22 Jun 2017. This is a timespan of 4037\,d with on average one observation every 9.2\,d. The data also cover season S1, S2, S3, and S5 of our RV observations, but show average uncertainties similar to the overall data scatter. The survey delivers a standard magnitude of $V_{\rm ASAS}$=9.631$\pm$0.025\,mag using a five-pixel aperture.

We also used MEarth data \citep{2012AJ....144..145B} obtained at the Fred Lawrence Whipple Observatory on Mount Hopkins, Arizona, USA, and provided to us in three sets \citep[see also][]{2019A&A...621A.126D}. Set A includes observations on 7 nights from 13 Oct 2008 to 3 Mar 2010. Set B includes observations on 53 nights from 19 Nov 2010 to 23 Jun 2011, and is split into the two seasons of visibility. Set C covers 79 nights from 22 Oct 2011 to 12 Nov 2015 and overlaps with the S1 to S4 seasons of our spectroscopic campains (see Fig\,\ref{A3}). We did not find significant periodicities of less than 1\,d in the data and applied nightly binning after a 3$\sigma$ clipping of the values and the uncertainties. The uncertainties are similar to the scatter.

\begin{table*}[htb]
	\caption{Basic properties and results of the analysis of different photometric observations ({\it top}), and activity indices of HARPS-N ({\it middle}) and CARMENES observations ({\it bottom}).}  \label{T3}
	\centering
	\tiny
	\begin{tabular}{l|ccl|ccc|ccc|cccc}
		\hline \hline
		\noalign{\smallskip}
		& \multicolumn{3}{c|}{Basic properties} & \multicolumn{3}{|c|}{Periodogram analysis} & \multicolumn{3}{|c|}{GP with {\tt celerite}} & \multicolumn{4}{|c}{GP with {\tt george}} \\
		\noalign{\smallskip}
		Index  & rms  & error & unit & $P_{\rm rot}$ [d] & $P_{\rm add}$ [d] & Phase [deg] & $P_{\rm SHO}$ [d] & $P_{\rm life}$ [d] & $\Delta \ln{L}$ & $P_{\rm QP}$ [d] & $\lambda$ [d] & w & $\Delta \ln{L}$ \\ 
		\noalign{\smallskip}
		\hline
		\noalign{\smallskip}
		EXORAP $\Delta B$ & 17.1 & 1.3 & mmag & 18.87 & 221.6 & 50.2 & 17.07 & 43.1 & 62.9 & 18.86 & 73.7 & 1.1 & 102.7 \\
		EXORAP $\Delta V$ & 14.6 & 1.5 & mmag & 18.91 & 259.1 & 41.4 & 18.24 & 39.0 & 21.3 & 19.16 & 133.7 & 3.2 & 307.0 \\
		EXORAP $\Delta R$ & 13.4 & 1.4 & mmag & 18.85 & 240.3 & 78.0 & 18.54 & 53.2 & 37.3 & 18.81 & 66.1 & 1.3 & 56.0 \\
		EXORAP $\Delta I$ &  14.3 & 1.0 & mmag & 18.94 & 623.3 & 61.7 & 20.50 & 31.6 & 47.6 & 18.98 & 48.8 & 1.2 & 90.1 \\
		SNO $\Delta V$ & 10.3 & 13.0 & mmag & 18.86  & 173.8 & 343.2 & 19.03 & 302.4 & 215.1 & 19.25  &  429.5 & 2.3 & 213.3 \\ 
		SNO $\Delta R$ & 10.3 & 16.8 & mmag & 19.04 & 133.3 & 336.1 & 19.21 & 278.6 &  199.1 & 19.09 &  239.5 & 2.3 & 197.3  \\
		ASAS $\Delta V$ & 25.3 & 27.4 & mmag & * & ... & 77.0 & ... & ... & ... & ... & ... & ... & ... \\
		MEarth $\Delta V$ & 7.8 & 7.3 & mmag & 20.02 & ... & 51.7 & ... & ... & ... & ... & ... & ... & ... \\
		\noalign{\smallskip}
		\hline
		\noalign{\smallskip}
		CaHK & 0.350 & 0.014 & ... & ... & 419.3 & 79.3 & 19.98 & 60.5 & 20.0 & 19.75 & 157.2 & 2.6 & 39.0 \\
		H$\alpha$ & 0.0284 & 0.0021 & ... & ... & 409.0 & 77.2 & 21.93 & 35.6 & 46.2 & 18.72 & 263.8 & 0.7 & 57.4 \\
		NaI & 0.0052 & 0.0034 & ... & * & 416.6 & 80.7 & ... & ... & ... & ... & ... & ... & ... \\
		FWHM & 0.040 & 0.013 & km\,s$^{-1}$ & ... & >3060 & 117.5 & ... & ... & ... & ... & ... & ... & ... \\
		CON  & 0.453 & 0.081 & \% & * & 425.5 & 229.6 & ... & ... & ... & ... & ... & ... & ... \\
		BIS & 2.04 & 0.69 & m\,s$^{-1}$ & 19.1 & ... & 356.0 & ... & ... & ... & 18.62 & 110.5 & 6.1 & 13.5 \\
		CRX  & 10.5 & 7.9 & m\,s$^{-1}$\,N$_{\rm p}^{-1}$ & * & >3062 & 227.1 & ... & ... & ... & 18.91 & 762.3 & 1.3 & 12.2 \\
		\noalign{\smallskip}
		\hline
		\noalign{\smallskip}
		CaIRT & 0.0417 & 0.0025 & ... & 18.8 & 538.6 & 79.8 & 18.99 & 83.2 & 16.8 & 18.98 & 166.0 & 2.3 & 25.8 \\
		H$\alpha$ & 0.0300 & 0.0015 & ... & 18.8 & 569.1 & 76.7 & 18.63 & 234.9 & 15.4 & 18.90 & 225.2 & 2.7 & 27.2 \\
		NaI & 0.0390 & 0.0040 & ... & * & 437.1 & 66.0 & ... & ... & ... & 18.17 & 111.1 & 3.4 & 12.7 \\
		FWHM & 0.019 & 0.021 & km\,s$^{-1}$ & 7.6 & >920 & 50.6 & ... & ... & ... & ... & ... & ... & ... \\
		CON & 0.124 & 0.060 & \% & * & 428.9 & 18.5 & ... & ... & ... & 19.10 & 140.8 & 1.0 & 13.6 \\
		BIS & 6.33 & 0.94 & m\,s$^{-1}$ & 18.3 & ... & 121.7 & 18.59 & 60.5 & 11.1 & 18.91 & 55.5 & 1.8 & 14.3 \\
		CRX & 20.8 & 14.5 & m\,s$^{-1}$\,N$_{\rm p}^{-1}$ & * & 90.3 & 243.9 & ... & ... & ... & ... & ... & ... & ... \\
		\noalign{\smallskip}
		\hline 
	\end{tabular}
	\tablefoot{We show the tentative and significant signals of the periodogram analysis, and the parameters of the best solution of an MCMC approach using Gaussian Processes (GP) with the simple harmonic oscillator (SHO) as kernel and the {\tt celerite} code, and with the quasi-periodic harmonic oscillator (QP) as kernel of the {\tt george} package (see Sect.\,\ref{ain:cn} for the definition of the hyper-parameters). The $\Delta \ln{L}$ value refers to the difference in $\ln{L}$ compared to a {\tt null} model.\\
		*: the peak in the periodogram connected to the rotational period of Gl\,49 stands out from the noise, but does not reach the 1\% FAP level.}
\end{table*}

\subsection{Spectroscopic indices} \index{ain:sa} \label{ain:sa}

We derive a number of additional properties from the HARPS-N and CARMENES spectra. We do not attempt to mix values from different instruments and reduction pipelines. An overview and a visualization of all the time-series data after a 3$\sigma$-clip on measurement uncertainties are given in Table\,\ref{T3} and in the left panels of Figs.\,\ref{A1} (HARPS-N) and \ref{A2} (CARMENES). The data are given in Table\,\ref{TT1}.

We calculate the cross-correlation function (CCF) between each observed spectrum and a binary mask constructed from the average spectrum of all observations of Gl\,49. The mask is a selection of lines of the combined spectra of certain depth and width, excluding regions of strong atmospheric influences. The CCF is then a representation of the average line of the spectrum in the velocity space and can be approximated by a Gaussian function. The variations over time of the velocity span of the half-maximum power (FWHM) and the maximum depth or contrast (CON) of this fit should therefore correlate directly with velocity variations across the stellar surface. \citet{2014ApJ...789..154D} associate such variations with the bright faculae on the stellar surface. We also calculate the bisector inverse slope \citep[BIS,][]{2001A&A...379..279Q}, which measures the asymmetry of the CCF. This value gives an insight into the structure of the photosphere and should correlate with differences in temperature and pressure, both connected strongly to the activity level of the star. The HARPS-N time-series data show in its temporal distribution already that {\it (i)} a long-term variation of $>$1500\,d is present in FWHM and CON, {\it (ii)} the BIS value is constant, and {\it (iii)} S1 does not stand out in those indices as in the RVs. The CARMENES time-series, on the other hand, are less variable over the different seasons, but match the long-term trend described by the HARPS-N data .

Many magnetically-sensitive line features are present in the optical spectra of both instruments. They are formed in the hot plasma of the chromosphere and hence vary with the strength of the stellar magnetic field. Those features are connected to elements such as calcium, hydrogen, sodium, helium, and iron and are some of the most important tracers for stellar activity in the literature. The \ion{Ca}{II} ion shows the Fraunhofer H \& K emission lines at 3\,934 and 3\,969~$\AA$ in a domain of low signal-to-noise ratio in the HARPS-N spectra of our target. We quantify those lines, which measure the lower chromosphere \citep{2011A&A...534A..30G}, with the S index \citep[CaHK;][]{1991ApJS...76..383D}. The infrared triplet of the same ion is located in the CARMENES spectral range at 8\,498, 8\,542, and 8\,662~$\AA$ (CaIRT). We measure and sum up their fluxes with respect to the continuum. We additionally measure H$\alpha$ at 6\,563~$\AA$, which is absorbed rather in the upper part of the chromosphere \citep{2011A&A...534A..30G}. We include the measurements of the \ion{Na}{I} D1 and D2 absorption doublet at 5\,890 and 5\,896~$\AA$. \citet{2007MNRAS.378.1007D} proposed that these lines could be used to follow the chromospheric activity level of very active late-type stars and that they provide good complementary information about the conditions in the middle-to-lower chromosphere \citep{2000ApJ...539..858M}. The HARPS-N indices show the long-period variation $>$1500\,d already described for the RV and CCF index time-series data. In contrast to the CCF indices, the data points show the different behavior of S1 (marked in green in Fig.\,\ref{A1}), but with a strong consistency between the different indices. The CARMENES datasets follow the long-term trend of the HARPS-N sets nicely with an increasing and decreasing activity level for S5 and S6, respectively.

With the publication of the data reduction pipeline SERVAL, the chromatic index (CRX) was introduced as a new tracer for activity phenomena \citep{2018A&A...609A..12Z}. For every spectrum, it measures the slope of the RVs calculated for each different echelle order when they are plotted against wavelength. A non-zero and time-variable CRX is therefore indicative of RV variations due to stellar surface phenomena. As shown in Figs.\,\ref{A1}, and \ref{A2} the index follows the long-term trend described by, for example the FWHM or CON indices.

\subsection{Periodogram analysis} \index{ain:aa} \label{ain:aa}

In order to get a first overview of the periodicities in our data sets, we use the Generalized Lomb-Scargle (GLS) periodogram \citep{2009A&A...496..577Z}. As thresholds for a signal to be tentative or significant, we use at this stage the 1 and 0.1\,\% False Alarm Probability (FAP), respectively, calculated analytically with the formula by \citet{1986ApJ...302..757H}. If the signal shows FAPs below those thresholds, we subtract it by a sinusoidal best-fit and analyze the residuals in the same manner; this procedure is customarily dubbed prewhitening. We show the periodograms of the activity tracers in the right panels of Figs.\,\ref{A3} (photometry), \ref{A1} (HARPS-N), and \ref{A2} (CARMENES). The red horizontal lines indicate the 0.1, 1, and 10\,\% FAP level, whereas the two orange vertical lines stand for the periodicities at 9.37 and 18.86\,d, and the red vertical line marks the period at 13.85\,d. All significant signals of the different datasets are given in Table\,\ref{T3} under $P_{\rm rot}$, if the signal is close to the 19-d period, and under $P_{\rm add}$, if it is not. We mark the first with *, if the signal is distinguishable from the noise but above the 1\,\% FAP level.

The various photometric datasets reveal the period at 18.9\,d very well. The first harmonic, on the other hand, is mostly not seen due to the low amplitude of the signal. For the EXORAP filters, the data show an additional strong periodicity at $>$220\,d. We connect it in part to the time-sampling, but also identify the variation in the data point distribution. The SNO photometry shows also a strong periodicity of $>$150\,d. In the noisy ASAS-3 data, we are able to see two statistically insignificant periodicities at 18.94\,d and 177\,d. Also from MEarth data we are able to clearly detect the rotational periodicity.

It seems that the HARPS-N S1 data are dominated by a strongly varying activity level of Gl\,49. This is seen clearly in the large RV scatter of that season (Fig.\,\ref{F1}) and in the varying values for the chromospheric indices (Fig.\,\ref{A1}). Since in all cases this data subset (20\% of all data points) only adds noise to the periodograms, we exclude it from this analysis and the periodograms in Fig.\,\ref{A1}. Even then, we identify in only four indices low-significance signals at periodicities of 19 or 9.4\,d. All sets show at 400 to 425\,d a signal which is probably a combination of the long-term periodicity of $>$1500\,d and the strong yearly alias created by the seasons of visibility as seen in the window function, that is, the Fourier transformation of the time-sampling (see Fig.\,\ref{F1}). Given the distribution of all the HARPS-N datasets, we assume the star to be magnetically rather active at S2 and by the end of S5. The minimum activity level at approximately Mar 2015 (BJD=2\,457\,100\,d) coincides with the minimum scatter in the RV measurements. But such signals of longer periods are not found by the periodogram approach.

The data distribution of the CARMENES indices is, on average, very similar between the S4 to S6 seasons, but indicating a slightly less active state of Gl\,49 by the end of S6 than in the previous seasons and showing an upward/downward trend for S5/S6. More clearly than for the HARPS-N indices, all their periodograms show features at around 19, and/or 9.4\,d, which might in part be due to the higher observing cadence. The alias at 365.25\,d shows at longer periods since it is probably as well mixed with the unresolved long-term trend already described.

\subsection{Secondary periodicity} \index{times} \label{times}

\begin{figure}
 	\resizebox{\hsize}{!}{\includegraphics[width=\textwidth]{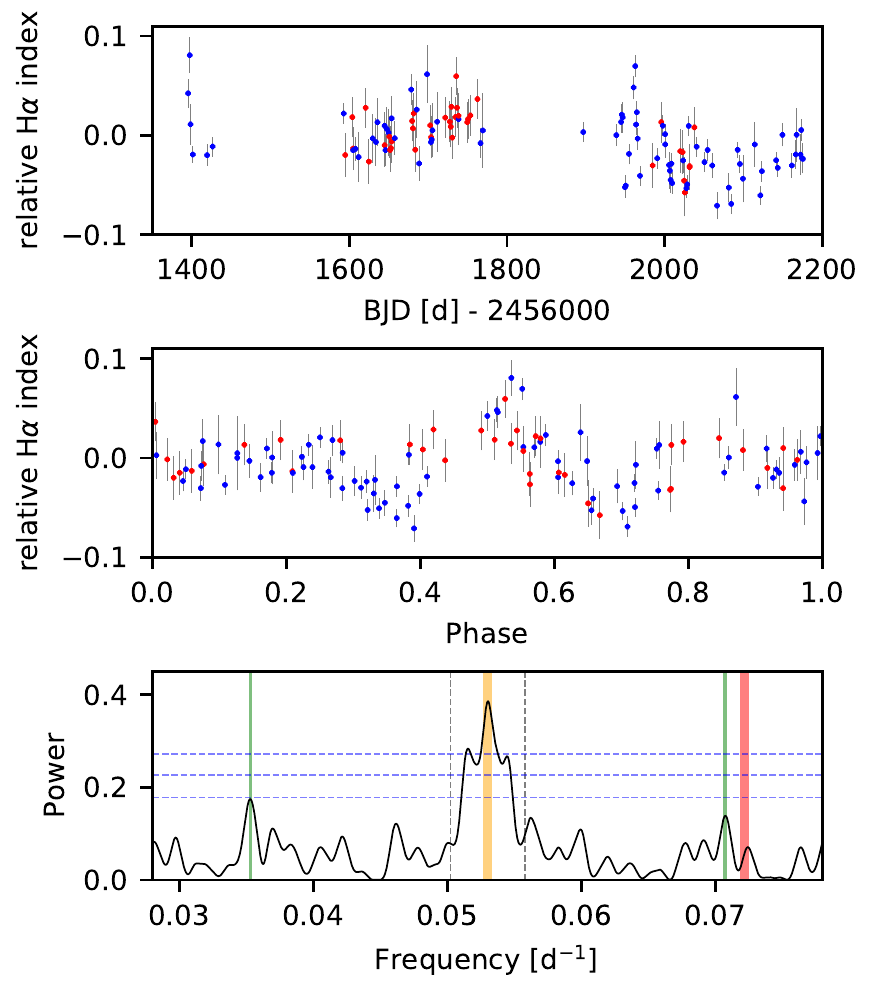}}
 	\caption{H$\alpha$ time-series data of CARMENES (blue) and HARPS-N (red) spectra of seasons S4, S5, and S6 (top panel). In the middle panel, we show the data points phase folded to 56.44\,d. The bottom panel shows the periodogram of the CARMENES index as in Fig.\,\ref{A2}. We mark the periods at 18.86 (orange vertical line) and 13.85\,d (red vertical line), their yearly aliases at frequencies $\nu = P^{-1} \pm$ 365.25$^{-1}$\,d$^{-1}$ (black vertical dashed lines), and the frequencies at 18.86$^{-1} \pm$56.44$^{-1}$\,d$^{-1}$ (green vertical lines), where the symmetry of the index data strongly imprints. The 0.1, 1, and 10\,\% FAP levels are shown by the blue horizontal dashed lines (from top to bottom).}
 	\centering
 	\label{FigB}
 \end{figure}

The presence and stellar nature of a 19-d periodicity and its first harmonic in Gl\,49 is shown by the periodogram analysis of the different activity indices. It is also reflected in the strong symmetry that the periodograms of especially the CARMENES CaIRT and H$\alpha$ indices show around this period (see Fig.\,\ref{A2} and, for the latter, bottom panel of Fig.\,\ref{FigB}). Besides the important periodicities at P=18.86\,d (orange) and 13.85\,d (red), we show the yearly aliases of the first at frequencies P$^{-1} \pm$365.25\,d$^{-1}$ (black dashed lines) and the analytical 0.1, 1, and 10\,\% FAP levels (from top to bottom) with the blue horizontal dashed lines. The symmetry results most prominently in two periodogram peaks above the noise level at 14.1 and 28.3\,d (green lines), corresponding to frequencies P$^{-1}\pm$56.44\,d$^{-1}$. This is indicative of a 56\,d-variation of the amplitude of the 19\,d-signal. We show the whole CARMENES dataset in the top panel of Fig.\,\ref{FigB} in blue. Since this secondary periodicity is not seen in the periodogram but the value is close to 3P, we fold the data to 56.44\,days and show it in the middle panel. There, we observe one very strong variation between phase 0.3 and 0.8, proving this secondary 56.4\,d-periodicity of the stellar contribution at S5 and S6. We also include the HARPS-N values, after the fit of an adequate offset, which then fall nicely in our picture.

\subsection{Time-series data correlations} \index{ain:lc} \label{ain:lc}

We identify the rotational periodicity in all photometric and most of the activity index datasets, as well as in the RVs (see Sect.\,\ref{sig}). To be able to study their possible correlations, we calculate the phase shifts with respect to the RV time-series data \citep{2017MNRAS.468.4772S}. To achieve this, we apply to all sets the best-fit period of the interval of 18\,d$<P<$20\,d of the RV data of HARPS-N (19.11\,d), and CARMENES (18.92\,d), respectively (see Table\,\ref{T2}). The resulting shift for each set is given in Table\,\ref{T3}. We expect only the FWHM and CON values to be in phase with the RVs. For all other indices, including the photometry, a phase shift of 45\,deg is theoretically expected for a simple one-starspot model. A closer look at the results reveals a great consistency only for the chromospheric indices with phase shifts between 75 and 80\,deg and for the CRX index with 230 to 240\,deg. We additionally calculate the phase shifts of the photometric data using the HARPS-N periodicity. The results are pointing to a value of 40 to 80\,deg. The differences for the SNO photometry could be related to an unknown activity event after S6 season.

We then calculate Spearman and Pearson factors and find highly significant linear correlations only for the chromospheric indices. For CaHK/CaIRT and H$\alpha$ indices we obtain Spearman factors of 0.88 and 0.95 for HARPS-N and CARMENES data, respectively. For the calcium index with the \ion{Na}{I} index it is 0.73 and 0.68, and for H$\alpha$ with the sodium index, we calculate 0.70 and 0.72. We do not find correlations for any other combination of time-series data which is, in most cases, due to the large phase shifts between them. Linear correlations are also not found for RV/CON of CARMENES or CRX/CON of HARPS-N observations, where small phase differences are measured.

\begin{figure}
	\resizebox{\hsize}{!}{\includegraphics[width=\textwidth]{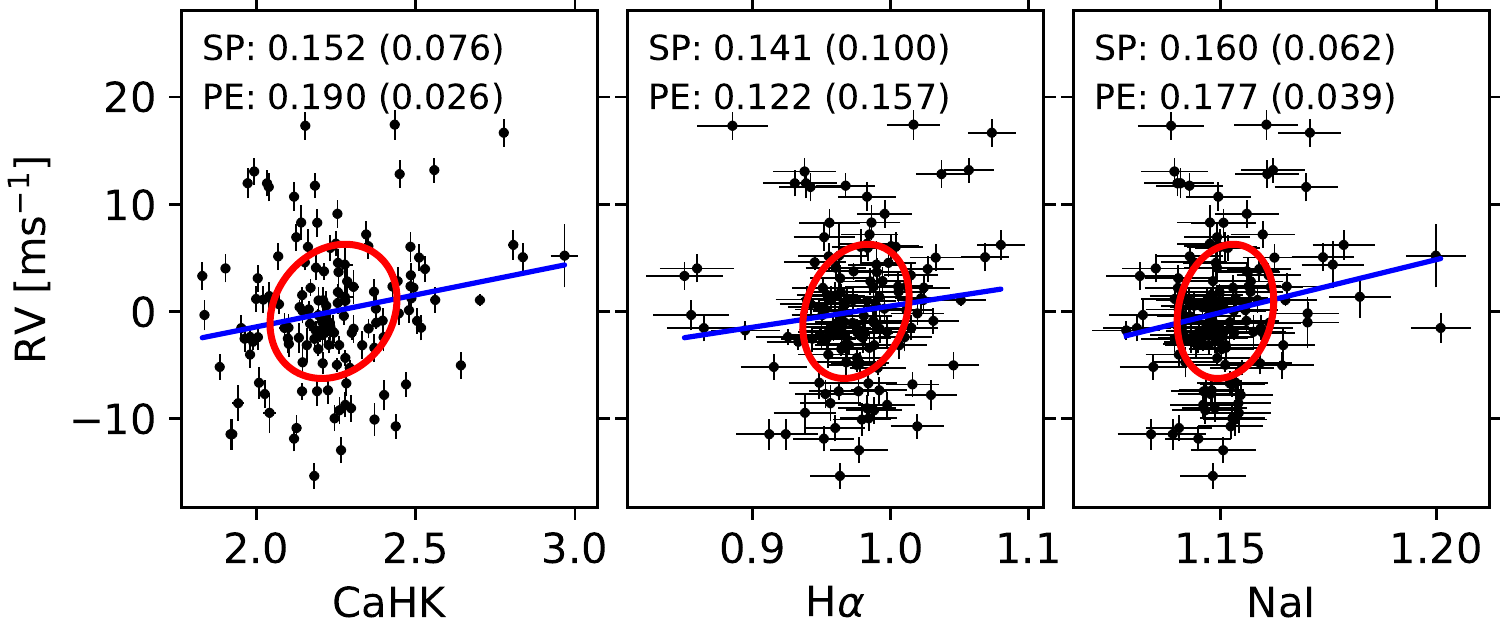}} 
	\resizebox{\hsize}{!}{\includegraphics[width=\textwidth]{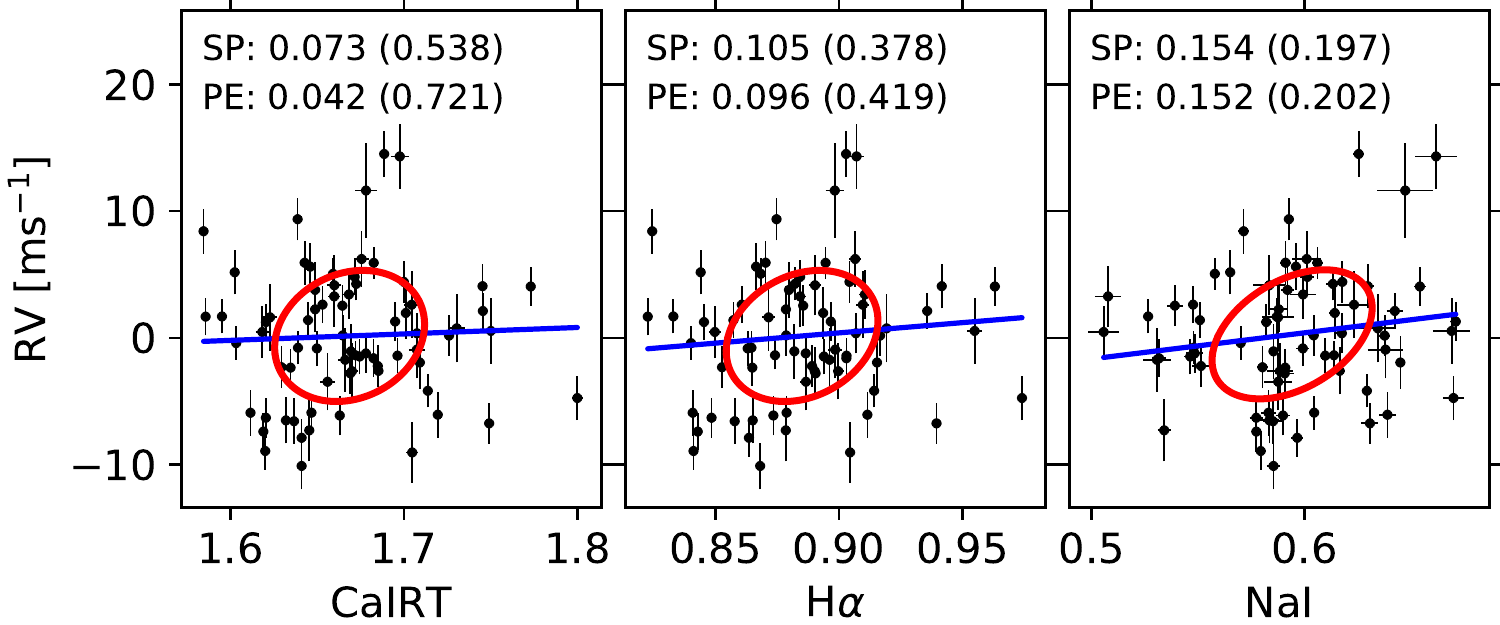}}
	\caption{Correlations of the HARPS-N ({\it top}) and CARMENES ({\it bottom}) chromospheric line indices with the RV data. The red ellipses show the theoretical phase-shifted correlation using the phase shifts of periodicities of 19.11\,d (HARPS-N) and 18.92\,d (CARMENES), respectively, and the rms of the data as the amplitude of the signal. The blue lines represent the best linear fit, i.e., the Spearman factor. At the top of each image, we show the Spearman (SP) and Pearson (PE) factors together with their respective p-values in parenthesis.}
	\centering
	\label{F5}
\end{figure}

To illustrate this procedure, we show the correlations of the RV data with the chromospheric activity indices in Fig.\,\ref{F5}. We show, from left to right, the calcium, H$\alpha$, and NaI indices for HARPS-N (top panels), and CARMENES (bottom panels) instruments, respectively.

\subsection{Modeling correlated noise} \index{ain:cn} \label{ain:cn}

The phase shifts of the activity indices imply that the information about the stellar contribution monitored by each index cannot be implemented easily in the modeling of our RV data. We therefore model the data as correlated noise using GP regression \citep{2015MNRAS.452.2269R, Roberts20110550}. The covariance matrix of the GP contains all the information we want to apply regarding the correlation of the data points, which is consistent with the picture of active regions rotating with the star and evolving in time. In our case this is a quasi-periodic kernel, similar to a damped oscillator, consisting of a periodic component (described by hyperparameters period $P_{\rm QP}$ and amplitude $K_{\rm QP}$) connected to the rotational period of the star, and a damping factor $\lambda$ (connected to the evolutionary timescale of surface phenomena). An additional scaling factor $w$ (generally $<$1) is used to measure the relation of both kernel parts. A small $w$ shows large influence of $P_{\rm QP}$ and small influence of the exponential decay. Details on this method using the {\tt george} code by D.\,Foreman-Mackey are described in \citet{2015ITPAM..38..252A}. In another approach, we use the {\tt celerite} code by the same authors calculating the covariance matrix by a stochastically-driven simple harmonic oscillator \citep{2017AJ....154..220F}. In this code, we model the rotational periodicity with the two hyperparameters $P_{\rm SHO}$ and $K_{\rm SHO}$, and a hyperparameter $P_{\rm life}$, which is supposed to be connected to the lifetime of the active regions. 

To compare different models consistently, we use the log-likelihood ($\ln{L}$) statistics \citep[e.g.,][]{2013A&A...556A.126A}. The likelihood function shows the probability of the data matching a certain model, and generalizes the $\chi^{2}$ statistics introducing an additional jitter term $\sigma$. Details on the definition for a given dataset and noise models are given in, for example \citet{2018Natur.563..365R}. At this point, we consider a model compared to a second one (including a {\tt null} model) significant if the difference in $\ln{L}$ exceeds a value of 15, which roughly corresponds to a FAP of 0.1\% in our measurements. A difference of $\Delta \ln{L}<$5 is considered noise. For a detailed discussion on the usage of this statistics for model comparison see \citet{2013MNRAS.429.2052B}, and for an example implementation of this procedure see \citet{2018Natur.563..365R}. We provide later (see Sect.\,\ref{sig:me}) an empirical determination of the $\Delta \ln{L}$ threshold required to claim a confident detection for a certain model.

We further use the {\tt emcee} code by D.\,Foreman-Mackey to explore the parameter space with a Markov Chain Monte Carlo procedure \citep[MCMC,][]{2013PASP..125..306F}. For the index time-series data, we apply 10\,000 steps on each of 100 walkers. As a comparison, we maximize the $\ln{L}$ using only offset and additional jitter ({\tt null}-model). In a second step, we then apply the two GP algorithms and their respective covariance matrices. Since we already know the rotational period of Gl\,49 and have analyzed both the periodograms and the temporal distribution of the different time-series data, we set boundaries to the uniform priors of the hyperparameters $P_{\rm QP}$ and $P_{\rm SHO}$ from 5 to 30\,d. To all other priors we apply reasonable boundaries (see Tab.\,\ref{T10}). We show the results of this procedure in Table\,\ref{T3} only if $\Delta \ln{L} >$10 compared to the {\tt null} model, to include also tentative solutions.

The CaHK/CaIRT and H$\alpha$ indices and the EXORAP photometry show the most significant and consistent results and retrieve the rotational period very well. But the interpretation of the hyperparameters $P_{\rm life}$, $\lambda$, and $w$ is more complicated. As already mentioned, $w \gg$1 (e.g., HARPS-N BIS) shows the dominance of the exponential decay over the rotation. On the other hand, $\lambda$ and $P_{\rm life}$ are generally not constrained consistently. The latter shows best solutions in the interval 2P$_{\rm rot}$--4P$_{\rm rot}$ for nearly all the indices, confirming the value of 56.4\,d, which we found as secondary period in some activity indices of seasons S5 and S6. $\lambda$ could instead be summarized as an overall larger value working similar for the EXORAP photometry (2P$_{\rm rot}$--4P$_{\rm rot}$), but showing instead 150 to 250\,d for the other activity indices. The large values for the SNO photometry are due to the stability of the 19\,d-variation over the short time-period of the observations.

\section{Radial velocities} \index{sig} \label{sig}

\subsection{First analysis of individual datasets} \index{sig:fa} \label{sig:fa}

To get an overview of the periodicities present in the different RV datasets, we analyze them individually with the GLS method and show the results in Fig.\,\ref{F1} for HIRES, HARPS-N, and CARMENES.
The 21 HIRES S0 RV measurements are not able to sample the periodicities that we are interested in, since a lower limit is given by the Nyquist frequency with approximately 160\,d. If we search for periodicities down to 1\,d, we find, however, a tentative signal at 17.3\,d (see column 7 of Table\,\ref{T2}), which we assume to be connected to the rotational period. 
In the prewhitening process using the 137 epochs of HARPS-N S1 to S6 observations, we find three strong signals in the range of the rotational period at 18.3, 18.9, and 19.1\,d. Furthermore, we find a periodicity close to the first harmonic of the rotation at approximately 9.6\,d. The facts that those periodicities are not easily cleaned with simple sinusoidal fits, that they influence each other, and that the supposed harmonic is so prominent point to a rather complex RV contribution from the stellar rotation. The additional significant signal at 13.4\,d will be shown to be connected to the orbital period of a proposed planet. If we exclude the S1 dataset in our prewhitening periodogram analysis, the main periodicity of the remaining dataset is now the 9.4\,d-signal, as shown in Table\,\ref{T2}, and a signal at 14.4\,d rather than 13.4\,d. The S1 season with the large data scatter shows significant peaks in all three interesting time intervals of the periodogram.
The analysis of the 80 CARMENES data points reveals three periodicities at 9.3, 18.9, and 14.4\,d. The periodogram is more easily cleaned by sinusoidal fits, indicating a more stable stellar contribution in S4 to S6 than in S1 to S3. Like in the case of the S2-S6 HARPS-N set, the most prominent periodicity in the CARMENES set is connected to the first harmonic of the rotational period, that is 9.3\,d, and a remaining signal at 14.4\,d is visible.

\subsection{Methodology} \index{sig:me} \label{sig:me}

Using the knowledge of the periodicities in our datasets, we search in the following for the best model to fit to the RV data, in order to estimate the best parameters for a possible planetary companion. We exclude the first season of the HARPS-N observations, since we assume it to be affected by strong variations in stellar activity. We exclude further the HIRES data, since the time sampling was very poor. Although the rotational period in the RVs is very stable, these datasets add more noise than signal to our data and might distort any fitted parameter. In the following search for the most probable fit parameters, we then use the HARPS-N S2 to S6 and CARMENES datasets, that is 80\,\% of the available RV measurements. 

We fit different models to be able to compare their probabilities through their $\ln{L}$ values. We use again the {\tt emcee} code for the MCMC analysis as outlined in Sect.\,\ref{ain:cn} with 100 walkers and 10\,000 steps. If not mentioned differently, uniform priors with reasonable boundaries (see Table\,\ref{T10}) are used, in order to fully exploit the parameter space and to have a different approach to evaluate the probability and significance of the fitted parameters. 

We begin by fitting a constant {\tt null} model, including offsets and jitters for both datasets, and obtaining as a best-fit $\ln{L}$=$-$574.0 (see Table\,\ref{T5}). Based on this value, we built a periodogram by considering for every period the Keplerian fit ({\it f}($P$,$K$,$e\sin{\omega}$, $e\cos{\omega}$, $T_{\rm per}$), with $P$ as period, $K$ as amplitude, $e$ as eccentricity, $\omega$ as true anomaly, and $T_{\rm per}$ as time of periastron of the respective orbit) with the largest $\ln{L}$. As a significance test, we apply bootstrap randomization \citep{1993ApJ...413..349M} using 10\,000 permutations of the data points and reach a 0.1 and 1\% FAP with $\Delta \ln{L}$=16.11 and 14.24, respectively. Those numbers confirm the $\Delta \ln{L}$=15 mentioned above, but are connected to the one-Keplerian model. Nevertheless, in the following, a more detailed model, that is including a correlated noise term (GP) or an additional Keplerian (Kep), is considered tentative or significant if it shows $\Delta \ln{L}>$14.24 and 16.11, respectively, compared to the former model. We also confirm the noise level of $\Delta \ln{L}$=5.

\begin{table}[]
	\caption{Overview of the statistics on the different models applied to the RV data of Gl\,49 from HARPS-N S2-S6 and CARMENES datasets.}  \label{T5}
	\small
	\centering
	\begin{tabular}{clcc}
		\hline \hline
		\noalign{\smallskip} 
		$N_{\rm signal}$ & model & $N_{\rm param}$ & $\ln{L}$ \\ \hline 
		\noalign{\smallskip} 
		0 & {\tt null} & 4 & $-$574.0 \\ 
		 & {\tt null} + GP$_{\rm QP}$ & 8 & $-$519.9 \\ 
		 & {\tt null} + GP$_{\rm SHO}$ & 7 & $-$527.8 \\
		\noalign{\smallskip}
		\noalign{\smallskip}
		1 & Keplerian & 9 & $-$550.9 \\
		 & Keplerian + GP$_{\rm QP}$& 13 & $-$491.8 \\ 
		 & \hspace{0.25cm} seasonal & 20 & $-$491.3 \\ 
		 & \hspace{0.25cm} circular & 11 & $-$496.9 \\ 
		 & \hspace{0.25cm} with HARPS-N S1& 15 & ($-$499.3) \\
		 & \hspace{0.25cm} with HIRES/HARPS-N S1& 17 & ($-$502.3) \\ 
		 & Keplerian + GP$_{\rm SHO}$ & 12 & $-$492.2 \\ 
		\noalign{\smallskip}
		\noalign{\smallskip}				
		 2 & Keplerians & 14 & $-$531.7 \\ 
		  & Keplerians + GP$_{\rm QP}$ & 18 & $-$482.8 \\ 
		  & Keplerians + GP$_{\rm SHO}$ & 17 & $-$484.7 \\ 		
		\noalign{\smallskip}
		\noalign{\smallskip}
		3 & Keplerians & 19 & $-$514.4 \\
		\noalign{\smallskip}
		\noalign{\smallskip}
		4 & Keplerians & 24 & $-$496.7 \\
		\noalign{\smallskip}  
		\noalign{\smallskip} 
		5 & Keplerians & 29 & $-$495.3 \\ 
		\noalign{\smallskip}
		\hline
	\end{tabular}
	\tablefoot{$N_{\rm signal}$ indicates the number of fitted Keplerians; $N_{\rm param}$ shows the number of parameters fitted in the respective model; GP$_{\rm QP}$ and GP$_{\rm SHO}$ are the GP noise terms with the quasi-periodic kernel of {\tt george} and the harmonic oscillator kernel of {\tt celerite}, respectively. The parenthesis indicate the $\ln{L}$ values interpolated from fits with different number of data points.}
\end{table}
		
\subsection{Beat frequency} \index{dat:rv} \label{dat:rv}

\begin{figure}
	\resizebox{\hsize}{!}{\includegraphics[width=\textwidth]{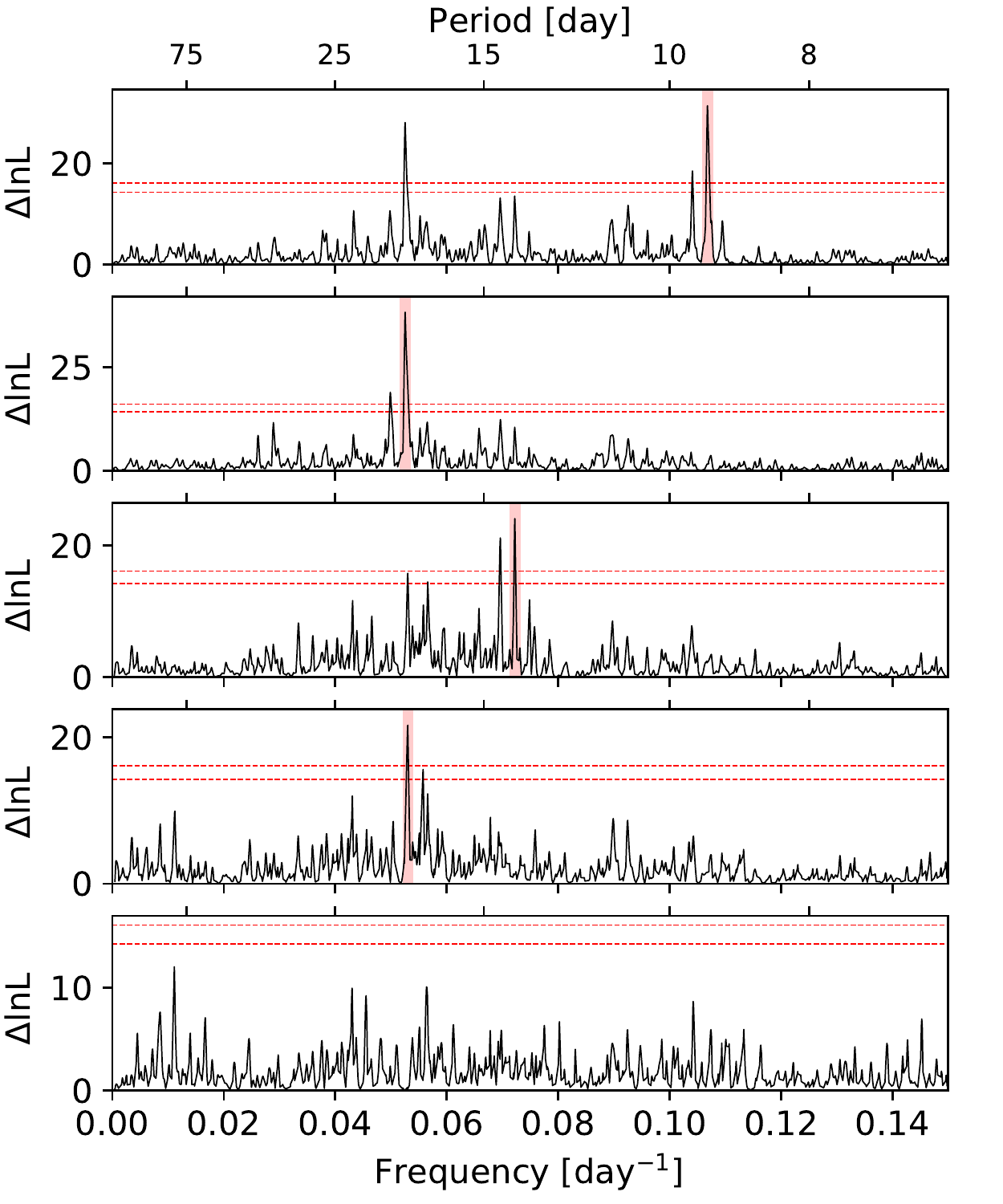}}
	\caption{Region of interest of the $\Delta \ln{L}$ periodograms of the combined RV datasets of Gl\,49 (HARPS-N S2-S6 and CARMENES). The top panel shows the difference in $\ln{L}$ from a linear {\tt null} model including additional jitters and offsets, to a model including one additional Keplerian (top panel). The following panel then shows the periodogram of the data with this Keplerian subtracted. From the second panel from the top to the bottom panel we have subtracted a best-fit Keplerian curve with periods of 9.37, 19.09, 13.85 and 18.91\,d, as shown by the broad red vertical lines. The dashed horizontal red lines indicate the 0.1 and1\% FAP values calculated by bootstrap randomization using a single Keplerian model on the datasets (see Sect.\,\ref{sig:me}).}
	\centering
	\label{F2}
\end{figure}

Disregarding the knowledge of the nature of the RV signals, we fit up to five Keplerians to the datasets. Each of those curves implements five fit parameters, which add to the four parameters already in place for RV offsets and additional jitters (see $N_{\rm param}$ in Table\,\ref{T5}). The increase in $\ln{L}$ is significant until the fourth Keplerian ($\ln{L}_{\rm 4Kep} - \ln{L}_{\rm 3Kep} >$16.11), where 24 parameters are fitted. The periodograms of this procedure are shown in Fig.\,\ref{F2} for the whole dataset, and the residuals after subtracting the best-fit of one, two and three Keplerians, respectively.

\begin{figure}
	\resizebox{\hsize}{!}{\includegraphics[width=\textwidth]{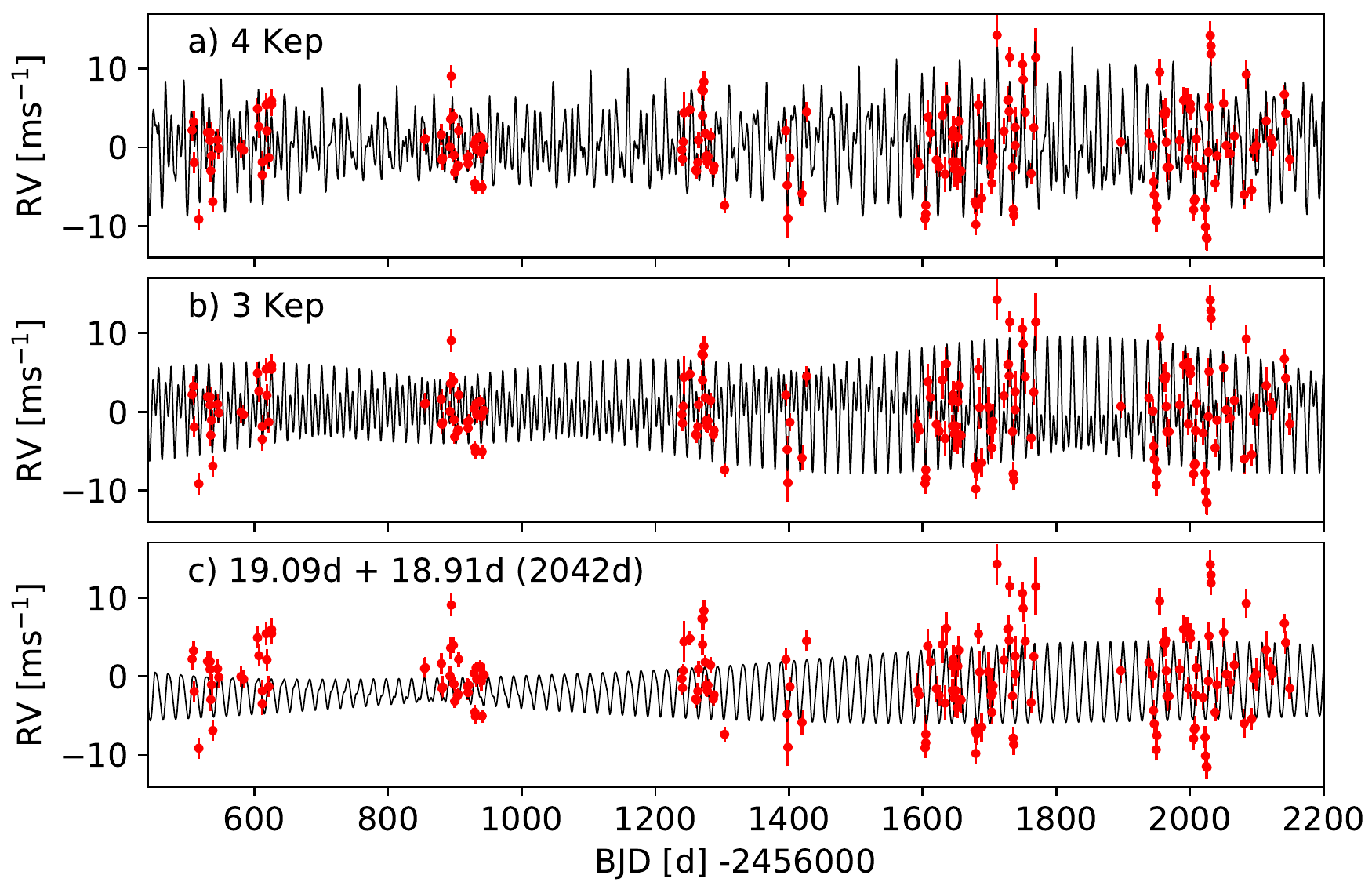}}
	\caption{Four Keplerian best-fit models including all signals (periods of 9.4, 13.9, 18.9, and 19.1\,d; top panel), the three signals connected to the stellar rotation (9.4, 18.9, and 19.1\,d; middle panel), and the two signals around 19\,d (bottom panel), resulting in the amplitude variation following the beat frequency of 2042\,d in the observed time span. The red dots are the HARPS-N S2 to S6 and CARMENES data points.}
	\centering
	\label{F3x}
\end{figure}

As a result, the four-Keplerian model (see top panel of Fig.\,\ref{F3x}) is able to clean the periodogram significantly, and it describes the possible stellar contribution with three Keplerian curves (middle panel) with periods of 9.37, 19.09, and 18.91\,d. For the sakes of this exercise, the best fit and the largest $\ln{L}$, this procedure should be as valid as the usual fit of simple- or double-sinusoidal models for the complex and noisy stellar contribution. The remaining periodicity at 13.85\,d is hidden behind those three strong signals but identified significantly as the third Keplerian. In the best MCMC solution, this periodicity shows a RV amplitude of 2.71\,m\,s$^{-1}$, and an eccentricity of 0.36. Actually, the results are very similar to the outcome of the prewhitening process of the individual datasets using sinusoids in Sect.\,\ref{sig:fa}. We see strong yearly aliases on all peaks in the periodograms showing the response of the time-sampling as seen in the window function. 

Whereas we can explain the signal at 9.4\,d as the first harmonic of the rotational period, we show the combination of the two periods around 19\,d in the bottom panel of Fig.\,\ref{F3x}. The two periodicities are effectively responsible for an amplitude change of the 19\,d-signal, which occurs with the beat frequency at ((19.09)$^{-1}$+(18.91)$^{-1}{\rm)}^{-1}$=2042\,d. We identify this periodicity to be the same as the long-term variation of $>$1500\,d already described, especially in the HARPS-N datasets (see Figs.\,\ref{F1}, \ref{A1}). 

\subsection{Models including correlated RV noise} \index{sig:dm} \label{sig:dm}

A more physical approach to model the stellar contribution is the application of a GP noise term. Here, as already explained, we model the stellar contribution as correlated noise through a covariance matrix representing a damped or stochastically driven harmonic oscillator. In Table\,\ref{T5} we can see that the models are significant including also one Keplerian ($\ln{L}_{\rm 1Kep+GP} - \ln{L}_{\rm GP} >$16.11) and fitting thereby 13 and 12 parameters for {\tt george} and {\tt celerite} codes, respectively. Including a second Keplerian is statistically not important with $\ln{L}_{\rm 2Kep+GP} - \ln{L}_{\rm 1Kep+GP} <$14.24. With the wide prior boundaries as shown in Table\,\ref{T10}, the best model captures both the rotational and the 13.85\,d-signal as expected, indicating the difference in stability of those two periodicities. The more flexible quasi-periodic kernel delivers a larger best $\ln{L}$ value in the MCMC analysis than the simple harmonic oscillator kernel, and we reach up to $-$491.8, resulting in a high significance of$\Delta \ln{L}$=59.1 for our proposed system and planet. If we compare the results with the model of the four Keplerians of Sect.\,\ref{dat:rv}, the amplitude of the 13.85\,d-signal decreases slightly to 2.54\,m\,s$^{-1}$. Such a suppression of the RV amplitude is suspected to be a disadvantage of the GP regression \citep{2016MNRAS.461.2440F, 2018Natur.563..365R}. 

We carry out some additional tests with the selected best model of one Keplerian plus a GP noise term using the quasi-periodic kernel (see Table\,\ref{T5}):

\begin{itemize}
	\item We assume seasonal RV shifts for each instrument and include in our models RV offsets for every season and instrument. The period ranges that we are interested in should not be affected. As we increase the number of parameters from 13 to 20, we achieve practically no change in $\ln{L}$. This is certainly partially due to the absorption of the effect by the GP.
	\item If we apply our model with a circular rather than an eccentric Keplerian orbit, we reach $\ln{L}$=$-$496.9. Comparing this circular model to the eccentric model, we reach a difference of $\Delta \ln{L}$=5.1, which is close to the noise level of $\Delta \ln{L}$=5.
	\item If we include HARPS-N S1 data, the RV amplitude drops and the eccentricity slightly rises. We reach $\ln{L}$=$-$570.3 for 217 data points, which translates to $\ln{L}$=$-$499.3 for 190 data points, indicating a worsening of the fit.	
	\item If we include both HARPS-N S1 and HIRES data, the trends for RV amplitude and eccentricity continue. We reach a translated $\ln{L}$ value of$-$502.3.
\end{itemize}

The results of those exercises reflect the stability of both HARPS-N and CARMENES instruments, the unusual behavior of HARPS-N S1 data, and the poor sampling of the HIRES data.

\subsection{Final model} \index{dat:fm} \label{data:fm}

\begin{figure}
	\resizebox{\hsize}{!}{\includegraphics[width=\textwidth]{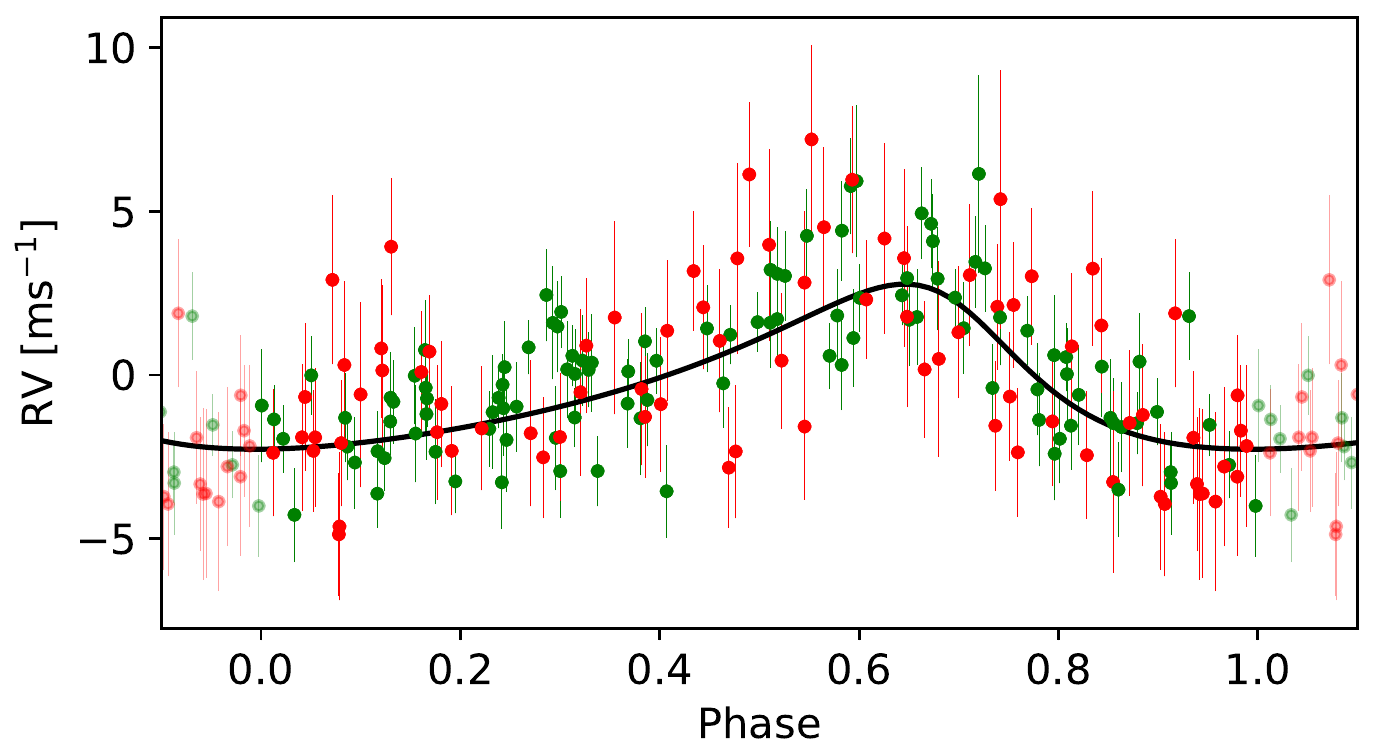}}
	\caption{RV data from HARPS-N (green) and CARMENES (red) of Gl\,49 of seasons S2 to S6 without the stellar contribution as calculated by the GP term of the final model. The data are folded in phase to 13.85\,d to illustrate the impact of the Keplerian orbit as shown by the black line.}
	\centering
	\label{F5x}
\end{figure}

\begin{figure*}
	\centering
	\resizebox{0.7\hsize}{!}{\includegraphics[width=\textwidth]{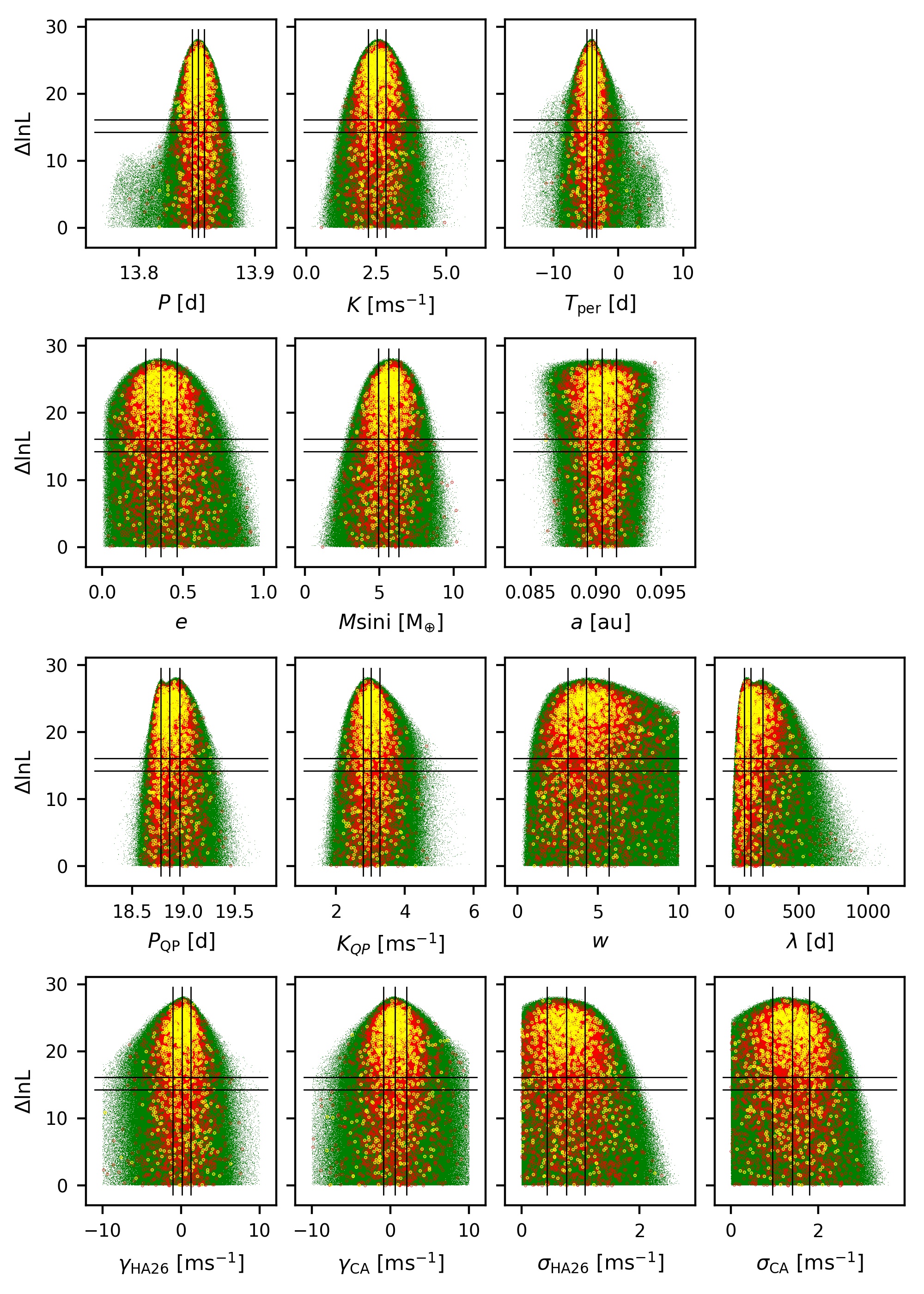}}
	\caption{$\Delta \ln{L}$ distribution of the different parameters of interest following the MCMC procedure. The zero level is set at the best-fit $\ln{L}$ of the model including only the GP$_{\rm QP}$ term (see Table\,\ref{T5}). We show the 0.1 and 1\,\% FAP level calculated by bootstrap randomization using a single Keplerian model (Sect.\,\ref{sig:me}) by the black horizontal lines. The black vertical lines in each image indicate the median value and the rms of each parameter considering the solutions below the 1\,\% FAP level. Whereas the green dots indicate all solutions, we plot in red every 1\,000, and in yellow every 10\,000 solutions, in order to get an insight on the density distribution. We note, that for $M \sin{\rm i}$ and $a$, the uncertainty of the mass of Gl\,49 is considered.}
	\centering
	\label{F7}
\end{figure*}

\begin{figure*}
	\resizebox{\hsize}{!}{\includegraphics[width=\textwidth]{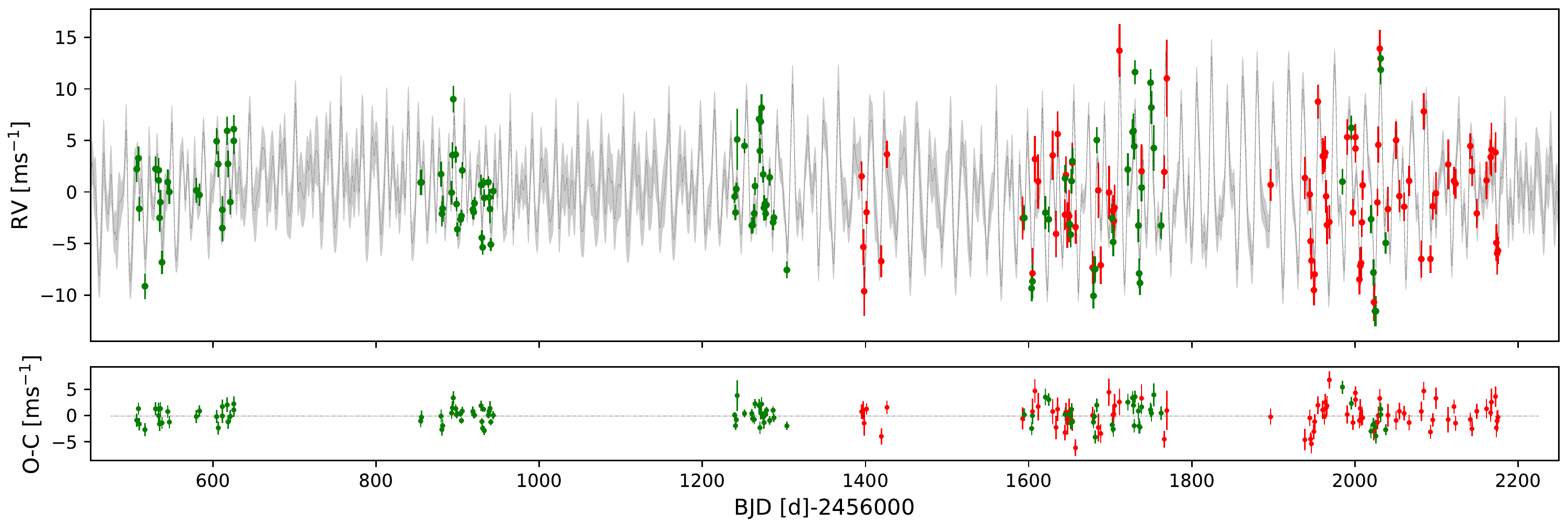}}
	\caption{RV data of Gl\,49 of season S2 to S6 of HARPS-N (green) and CARMENES (red), and the final model (black line) and its 1$\sigma$ uncertainties (gray lines) including a GP term and a quasi-periodic kernel for the stellar contribution and a Keplerian curve. In the bottom panel, the residuals are shown.}
	\centering
	\label{F4x}
\end{figure*}

To find the most probable values and uncertainties for the 13 free parameters of our selected best model including one Keplerian and a GP noise term, we explore 250$\times$50\,000 solutions of our MCMC analysis. We use $\ln{L}$=$-$519.9 of the GP$_{QP}$-term best-fit (see Table\,\ref{T5}) as a reference value. We have 10\,076\,912 solutions above that limit. Including a Keplerian, we consider every solution as tentative which adds to this number at least the 1\% FAP of $\Delta \ln{L}$, which is 14.24. The application of such a limit of $\ln{L} > -$505.7 leaves us with 8\,156\,650 solutions, which we use for our statistical approach. Our most probable model parameters and their uncertainties are then the median and the rms of the selected solutions, respectively. We show the distribution of all tentative solutions for the fitted parameters in the corner plot in Fig.\,\ref{F10} and give the final values in Table\,\ref{T10}. Additionally, we show the $\ln{L}$ distribution per parameter of interest in Fig.\,\ref{F7}. The time series of the final model is shown in Fig.\,\ref{F4x} and the phase-folded curve excluding the stellar GP contribution in Fig.\,\ref{F5x}.

\begin{table}[] 
	\caption{Fitted and derived planetary and stellar parameters.}  \label{T10}
	\small
	\centering
	\begin{tabular}{lcr}
		\hline \hline
		\noalign{\smallskip}
		Parameter & Priors & Results \\
		\noalign{\smallskip}
		\hline
		\noalign{\smallskip}
		\multicolumn{3}{c}{Fitted Keplerian parameters} \\
		\noalign{\smallskip}
		\hline
		\noalign{\smallskip}
		$P$ [d]   & 0-2$\cdot \Delta$T & 13.8508$^{+0.0053}_{-0.0051}$ \\
		\noalign{\smallskip}
		$K$ [m\,s$^{-1}$] & 0-10$\cdot$rms  & 2.52$^{+0.31}_{-0.30}$ \\
		\noalign{\smallskip}
		$e \sin \omega$  & $\pm$1 & 0.23$^{+0.11}_{-0.12}$\\
		\noalign{\smallskip}
		$e \cos \omega$  & $\pm$1 &  0.25$^{+0.11}_{-0.12}$ \\
		\noalign{\smallskip}
		$T_{\rm peri}$  [d] & $\pm \Delta$T  & 2455995.88$^{+0.72}_{-0.78}$ \\
		\noalign{\smallskip}
		\hline
		\noalign{\smallskip}
		\multicolumn{3}{c}{Derived planetary parameters} \\
		\noalign{\smallskip}
		\hline
		\noalign{\smallskip}
		$e$ & &  0.363$^{+0.099}_{-0.096}$ \\   
		\noalign{\smallskip}
		$\omega$ [deg] & &  43$\pm$21 \\    
		\noalign{\smallskip}
		$M \sin{\rm i}$ [M$_{\oplus}$] & &  5.63$^{+0.67}_{-0.68}$ \\ 
		\noalign{\smallskip}
		$a$ [au] & &  0.0905$\pm$0.0011  \\    
		\noalign{\smallskip}
		\hline
		\noalign{\smallskip}
		\multicolumn{3}{c}{Fitted hyper-parameters of quasi-periodic kernel} \\
		\noalign{\smallskip}
		\hline
		\noalign{\smallskip}
		$P_{\rm QP}$  [d] & 0-2$\cdot \Delta$T & 18.864$^{+0.103}_{-0.085}$\\
		\noalign{\smallskip}
		$K_{\rm QP}$ [m\,s$^{-1}$] & 0-10$\cdot$rms & 3.02$^{+0.25}_{-0.23}$\\
		\noalign{\smallskip}
		$w$ &  0-10 & 4.3$^{+1.4}_{-1.1}$	 \\
		\noalign{\smallskip}
		$\lambda$ [d] & 0-2$\cdot \Delta$T & 150$^{+90}_{-50}$ \\
		\noalign{\smallskip}
		\hline
		\noalign{\smallskip}
		\multicolumn{3}{c}{Fitted RV offsets and additional jitters} \\
		\noalign{\smallskip}
		\hline
		\noalign{\smallskip}
		$\gamma_{\rm HN26}$ [m\,s$^{-1}$] & $\pm$10$\cdot$rms & 0.1$^{+1.1}_{-1.2}$ \\
		\noalign{\smallskip}
		$\gamma_{\rm CA}$ [m\,s$^{-1}$] & $\pm$10$\cdot$rms & 0.6$\pm$1.5 \\		
		\noalign{\smallskip}
		$\sigma_{\rm HN26}$ [m\,s$^{-1}$] & 0-10$\cdot$rms & 0.76$^{+0.31}_{-0.32}$ \\
		\noalign{\smallskip}
		$\sigma_{\rm CA}$  [m\,s$^{-1}$] & 0-10$\cdot$rms & 1.41$^{+0.39}_{-0.45}$  \\
		\noalign{\smallskip}
		\hline
	\end{tabular}
	\tablefoot{We show the fitted and derived parameters for the proposed planetary companion, the GP hyper-parameters of the quasi-periodic kernel for the correlated stellar RV contribution, and the offsets and jitters of the two RV datasets (HARPS-N S2 to S6, CARMENES) used for the best model of Gl\,49. The uniform priors are all limited by the largest reasonable boundaries.}
\end{table} 

The periods of both the GP term and the Keplerian signal are constrained very precisely, although we can see still some structure well below any noise level in the $\ln{L}$ distribution of the former (see Fig.\,\ref{F7}). This is also visible in the corner plot (Fig.\,\ref{F10}), where this parameter shows a slight correlation with $\lambda$, which also shows this insignificant bi-modal behavior. The eccentricity of the Keplerian orbit is fitted by $e \sin \omega$ and $e \cos \omega$ and adds up to 0.36. We see evidence at this point for the parameters to be different from 0 within $\Delta \ln{L} <$5 from the peak value. The variable GP-term contribution, which corresponds to the more significant rotational signal, shows a larger impact on the RV data with an amplitude of 3.02\,m\,s$^{-1}$ than the stable contribution of the Keplerian with 2.52\,m\,s$^{-1}$. The important values of $\lambda$ and $w$ are rather poorly constrained. We calculate for the first a value between 100 and 240\,d. The median value of $w$ is 4.3. Although beyond the normal limits used for this parameter, it illustrates the importance of the exponential decay, that is the variation in amplitude, of the signal in comparison to its strong periodic behavior. The additional jitters of our best-fit model show a consistent picture and correlate with the uncertainties of the measurements already in place.

\section{Discussion} \index{dis} \label{dis}

\subsection{The planet Gl\,49b}

We find evidence of a 13.85\,d-signal in our RV data. We have shown that the signal and its yearly aliases are not present in any activity index, and that it is not introduced by the time-sampling or the data treatment using a variety of datasets and models. In comparison to the 18.86\,d-signal also present in the data, it shows great stability in period, phase, and amplitude over the six years of observations used for the final fit. We therefore assume this periodicity to be caused by the orbital motion of a planetary companion Gl\,49b. Its impact on the RV data of its host star is deeply hidden inside the complex stellar activity-induced variations. But with our best noise model, by which we accurately correct for the stellar contribution, the planetary signal is strong, with a significance of $\Delta \ln{L}$=59.1, and the parameters are very well constrained. We find the companion to have a minimum mass of 5.6$\pm$0.7\,M$_{\oplus}$. The super-earth orbits its star in 13.851$\pm$0.005\,d in an eccentric orbit ($e$=0.36$\pm$0.10) with a semi-major axis of 0.090$\pm$0.001\,au. The amplitude it introduces in our RV time-series data is 2.5$\pm$0.3\,m\,s$^{-1}$. 

Although not significantly distinguishable from the noise in our GP approach (see Table\,\ref{T5}), we consider the eccentricity of the orbit of Gl\,49b in our final fit. This is because the distributions of the fitted parameters $e \sin{\omega}$ and $e \cos{\omega}$ (see Fig.\,\ref{F10}), and of the derived eccentricity (see Fig.\,\ref{F7}) are different from zero at the $\Delta \ln{L}$=5 level. Another argument is that in the prewhitening process of the individual datasets (see Table\,\ref{T2}) we detect the yearly aliases at 13.34, and 14.40\,d rather than the orbital period of Gl\,49b at 13.85\,d. Additionally, the eccentricity of the Keplerian signal is very persistent in the variety of models fitted to the whole dataset and to all subsets.

\subsection{Stellar rotation}

We confirm the rotational period of Gl\,49 to be 18.86$^{+0.10}_{-0.09}$\,d and refine thereby the previously reported values by \citet{2018A&A...612A..89S} and \citet{2019A&A...621A.126D}. The signal induced into the RV data is very stable in phase and period in all the different data subsets, but not in amplitude. The idea is that those variations are caused by surface phenomena, which co-rotate independently of their latitude with the rotational period but appear always at specific longitudes in order to explain the stable phase of the signal \citep[see e.g.,][]{2009A&ARv..17..251S}. The strong presence of the first harmonic at 9.4\,d and its persistence in the prewhitening process shows the non-Keplerian nature of this contribution. The effect can be best explained if we assume the stellar surface to be dark-spot dominated and that the bright hot faculae with their effective convective blueshift have only little influence on the measured RVs. This is expected for low-mass stars and should be also in part responsible for the CCF indices not showing great consistency in this study. In the S1 season, on the other hand, where the activity contribution varies, and where the rotational influence dominates, the impact of faculae is supposed to be stronger.

The periodicity is also detected strongly in the photometric data, where the difference in impact on the different filters shows its non-planetary nature. In general, the photometric data are shifted in phase with respect to the RV data by 40 to 80\,deg. We detect the rotational signal also in the prewhitening of the CARMENES activity indices, and in four out of seven HARPS-N indices. A closer look with our GP noise model then reveals the presence of the periodicity in all the chromospheric indices. Those datasets show consistent phase shifts of 75--80\,deg, which is in an expected range following \citet{2017MNRAS.468.4772S}, and slightly smaller than the 120\,deg that we measured for the less active GJ\,3942 \citep{2017A&A...608A..63P}. The CRX index shows a phase shift of 230 to 240\,deg.

\subsection{Evolutionary timescale of dark spots}

Signals of a few rotational periods to some hundred days are detected in many of the datasets, which should be at least in part due to the evolutionary time scales of activity phenomena on the surface of Gl\,49. Most consistent values of the prewhitening process are delivered by the chromospheric activity indices with periodicities of 400 to 550\,d. Such signals are not found in the RVs. To find better constraints on those timescales we fitted our GP model including $P_{\rm life}$ and $\lambda$, but we still lack a definitive translation of those hyper-parameters to the physical world. For the first, we find values from 40 to 80\,d in photometry and chromospheric indices, which match the value of 56.4\,d that we found for the amplitude variation of the 19\,d-signal in S5 and S6. Whereas the EXORAP photometry finds similar values for $\lambda$, the line features set the parameter in a range from 150 to 250\,d. Since the covariance matrix represents a periodic model with changing apparent amplitude, which is exactly what we see in the RV data, we can most certainly trust the $\lambda$=150\,d found by the analysis of those datasets, but scale it down by the Euler number $e$=2.71828..., in order to match $P_{\rm life}$. Those values then are also in agreement with the decay times found for M dwarfs in \citet{2017MNRAS.472.1618G}. The stellar origin of those periodicities can also be seen by the impact shown on the SNO and EXORAP data on this time scale. The contrast of those spots against the photosphere seems to decrease with increasing wavelengths. 

\subsection{Evolution of stellar activity}

The impact of stellar activity on our datasets goes even further, since the RV data shows a varying scatter over the different seasons, which we find to be connected to the amplitude of the stellar signal changing on time scales of at least 1500\,d. We can trace this pattern in all activity indices showing that actual evolution of stellar activity is in progress, and the long-term variation is not found in the frequentist analysis of our data. Instead, this signal can be mathematically described by two Keplerian curves of slightly different periods, resulting in a signal with an amplitude that changes on a 2000\,d-time scale as a consequence of the beat frequency. As already mentioned, this is exactly what the covariance matrices of the GP noise terms represent and how we explain the high value of $w$ in our GP approach with the quasi-periodic kernel. As a consequence of this study, we assume Gl\,49 in S1 to be at a moderate, yet strongly varying activity level, perhaps influenced by bright faculae. The activity level increased until the peak level in S2, which was manifested by an increase in spot number or in the motion of the spots to higher latitudes. Then, until March 2015 (BJD=2\,457\,100\,d, S3-S4), the activity level dropped. A small increase until S5-S6 was followed by a decreasing level in 2018. This also explains that the activity and planetary signals in all CARMENES data are more stable than in the observations before S4.

\section{Conclusions and final remarks} \index{con} \label{con}

In this study, we analyzed 21.5 years of time-series data of the M1.5\,V-type star Gl\,49. Those include 137 epochs of spectroscopic observations with the HARPS-N instrument within the HADES framework, 80 epochs within the M-dwarf program of the CARMENES spectrograph, and additional 21 epochs from archived spectra observed with the HIRES instrument. Only the combination of HARPS-N and CARMENES data and their in-depth analysis made it possible to disentangle the contribution of the exoplanet Gl\,49b from the complex activity contributions of its host star. We furthermore studied simultaneous photometric multiband observations made within the EXORAP program, with the T90 telescope at SNO, and using archival MEarth and ASAS photometry.

We discover the super-Earth Gl\,49b, which orbits its host star with a period of 13.9\,d in an eccentric orbit ($e$=0.36) and with a semi-major axis of 0.090\,au. In all the different datasets that we have analyzed in this study, this periodicity is only present in the RV time series. The rotational velocity that we are able to calculate from the radius and rotational period of the star is in agreement with the upper limit of 2\,km\,s$^{-1}$ given for $v \sin{i}$ in Table\,\ref{T1}, pointing to an orbital inclination close to 90\,deg for an orbit close to the equatorial plane of Gl\,49. We therefore calculate the geometric transit probability with 2.0\,\% as a lower limit. Such an event would take approximately 2.5\,h and show a depth of 1 to 2.5\,mmag for a rocky/gaseous nature of Gl\,49b with $R_{\rm p}$=1.8 and 3\,$R_{\oplus}$. This is challenging to detect with ground-based telescopes, but could be a good target for {\it TESS} \citep[Transiting Exoplanet Survey Satellite;][]{2015JATIS...1a4003R}, or the upcoming {\it CHEOPS} mission \citep[CHaracterising ExOPlanet Satellite;][]{2014CoSka..43..498B}. We furthermore calculate an equilibrium temperature of approximately 350\,K for the planet assuming zero albedo and using the radiative equilibrium temperature $T_{\rm p, eq.}$=$T_{\rm star ,eff} (0.5 \cdot {\rm R_{\rm star} / a})^{0.5}$. Gl\,49b orbits well interior to the 'recent Venus' habitable zone at approximately 0.18\,au (see Table\,\ref{T1}). In the search for exoplanets around M dwarfs, Gl\,49b has characteristics similar to other recently discovered planets with its minimum mass of 5.6\,M$_{\oplus}$ and orbital period of 13.9\,d \citep[see][]{2019A&A...622A.193A}. 

The radial-velocity signal of the planet is hidden among strong signals of various time scales induced by the rotating host star and its evolving surface elements. Recently, such cases have been discovered quite often in the search for exoplanets orbiting early M-type stars \citep[e.g.,][]{2016A&A...593A.117A, 2017A&A...608A..63P}, which shows the importance of the understanding of the stellar contribution to all different kinds of observed time-series data. Whereas there are no signs of differential rotation on the stellar surface of Gl\,49, the activity level varies significantly over the observed timespan. This is reflected in a long-term variation, which we suspect to be connected to the motion and reduction of the number of dark spots. We do not have sufficient data at hand, but if this behavior were periodic as in the magnetic cycle on the Sun, we would estimate its period to be $>$1500\,d, which is consistent with recent findings for early M dwarfs \citep{2018A&A...612A..89S}. We also detect a periodicity of some 40 to 80\,d, which we explain with the timescale for the evolution of spots and faculae. Interestingly, those signals are visible in all our different datasets no matter if they are rather measurements of phenomena produced by actual motions (CCF), or by stellar activity (lines, photometry), which are rather connected to changes in temperature or pressure. From 2012 to 2018, the star went from an unusual season at the beginning up to a peak of activity in 2013--2014. After passing a local minimum in 2015, the level of activity rose to a less-active local maximum in 2016--2017 and is since then again slowly losing strength. In the framework of the CARMENES M-dwarf program, Gl\,49 will be observed in the up-coming semesters in order to monitor this interesting behavior in all the different time-series data.

\begin{acknowledgements}
	M.P., I.R., J.C.M, D.B., E.H, and M.L., acknowledge support from the Spanish Ministry of Economy and Competitiveness (MINECO) and the Fondo Europeo de Desarrollo Regional (FEDER) through grant ESP2016-80435-C2-1-R, as well as the support of the Generalitat de Catalunya/CERCA program.
	G.S. acknowledges financial support from “Accordo ASI–INAF” No. 2013-016-R.0 July 9, 2013 and July 9, 2015.
	The Italian Telescopio Nazionale Galileo (TNG) is operated on the island of La Palma by the INAF - Fundaci\'on Galileo Galilei at the Roque de los Muchachos Observatory of the Instituto de Astrof\'isica de Canarias (IAC).
	The HARPS-N Project is a collaboration between the Astronomical Observatory of the Geneva University (lead), the CfA in Cambridge, the Universities of St. Andrews and Edinburgh, the Queen's University of Belfast, and the TNG-INAF Observatory.
	CARMENES is an instrument for the Centro Astron\'omico Hispano-Alem\'an de Calar Alto (CAHA, Almer\'ia, Spain). CARMENES is funded by the German Max-Planck-Gesellschaft (MPG), the Spanish Consejo Superior de Investigaciones Cient\'ificas (CSIC), the European Union through FEDER/ERF FICTS-2011-02 funds, and the members of the CARMENES Consortium (Max-Planck-Institut für Astronomie, Instituto de Astrof\'isica de Andaluc\'ia, Landessternwarte K\"onigstuhl, Institut de Ci\`encies de l'Espai, Insitut für Astrophysik G\"ottingen, Universidad Complutense de Madrid, Th\"uringer Landessternwarte Tautenburg, Instituto de Astrof\'isica de Canarias, Hamburger Sternwarte, Centro de Astrobiolog\'ia and Centro Astron\'omico Hispano-Alem\'an), with additional contributions by the Spanish Ministry of Science through projects RYC2013-14875, AYA2015-69350-C3-2-P, AYA2016-79425-C3-1/2/3-P, ESP2016-80435-C2-1-R, ESP2017-87143-R, ESP2017-87676-C05-1/2/5-R, and AYA2017-86389-P, the German Science Foundation through the Major Research Instrumentation Program and DFG Research Unit FOR2544 “Blue Planets around Red Stars”, the Klaus Tschira Stiftung, the states of Baden-W\"urttemberg and Niedersachsen, and by the Junta de Andaluc\'ia. Additional support was provided by the European Union FP7/2007-2013 program under grant agreement No. 313014 (ETAEARTH), the Progetto Premiale INAF 2015 FRONTIER, and the "Center of Excellence Severo Ochoa" award for the Instituto de Astrof\'isica de Andaluc\'ia (SEV-2017-0709).
	This work has made use of data from the European Space Agency (ESA) mission {\it Gaia} (\url{https://www.cosmos.esa.int/gaia}), processed by the {\it Gaia} Data Processing and Analysis Consortium (DPAC, \url{https://www.cosmos.esa.int/web/gaia/dpac/consortium}). Funding for the DPAC has been provided by national institutions, in particular the institutions participating in the {\it Gaia} Multilateral Agreement.
	{\bf Gl\,49b wurde vom Entdecker mit dem Namen "Sibel" getauft.}
\end{acknowledgements}

\bibliography{bibtex}{}
\bibliographystyle{aa}

\begin{appendix}
	
	\section{Observational log, data distributions, periodograms, and MCMC solutions}

\tiny
\onecolumn

\begin{landscape}
	\begin{longtable}{lccccccccl}
		\caption{\label{TT1} Radial velocity and activity indicator time-series data of Gl\,49 from HIRES, HARPS-N, and CARMENES instruments.}\\
		
		\hline \hline
		\noalign{\smallskip}
		BJD & RV & CaHK/CaIRT & H$\alpha$ & \ion{Na}{I} & FWHM & CON & BIS & CRX & instrument \\
		$[$d]-2450000 & [m\,s$^{-1}$] &  & &  & [km\,s$^{-1}$] & [\%] & [m\,s$^{-1}$] & [m\,s$^{-1}$\,N$_{\rm p}^{-1}$] & and season \\ 
		\hline
		\noalign{\smallskip}
		\endfirsthead
		\caption{Radial velocity and activity indicator time-series data of Gl\,49 from HIRES, HARPS-N, and CARMENES instruments (cont.).}\\
		
		\hline \hline
		\noalign{\smallskip}
		BJD & RV & CaHK/CaIRT & H$\alpha$ & \ion{Na}{I} & FWHM & CON & BIS & CRX & instrument \\
		$[$d]-2450000 & [m\,s$^{-1}$] &  & & & [km\,s$^{-1}$] & [\%] & [m\,s$^{-1}$] & [m\,s$^{-1}$\,N$_{\rm p}^{-1}$] & and season \\ 
		\hline
		\noalign{\smallskip}
		\endhead
		\hline
		\endfoot		
		
		667.0532 & 0.8110$\pm$1.3300 & … & … & … & … & … & … & … & HIRES S0 \\
		716.0104 & -0.8290$\pm$1.4500 & … & … & … & … & … & … & … & HIRES S0 \\
		1044.0814 & 3.3310$\pm$1.1500 & … & … & … & … & … & … & … & HIRES S0 \\
		1071.0386 & -3.6890$\pm$1.1900 & … & … & … & … & … & … & … & HIRES S0 \\
		1173.8468 & -8.1490$\pm$1.2900 & … & … & … & … & … & … & … & HIRES S0 \\
		1368.0584 & 3.5710$\pm$1.5800 & … & … & … & … & … & … & … & HIRES S0 \\
		1543.7872 & 8.0210$\pm$1.2900 & … & … & … & … & … & … & … & HIRES S0 \\
		1757.0935 & -5.8890$\pm$1.3100 & … & … & … & … & … & … & … & HIRES S0 \\
		1899.8454 & -0.3390$\pm$1.5300 & … & … & … & … & … & … & … & HIRES S0 \\
		2133.0897 & -1.7790$\pm$1.2400 & … & … & … & … & … & … & … & HIRES S0 \\
		2133.9996 & 2.5910$\pm$1.3800 & … & … & … & … & … & … & … & HIRES S0 \\
		2242.8851 & -5.7690$\pm$1.4000 & … & … & … & … & … & … & … & HIRES S0 \\
		2515.0599 & -3.2090$\pm$1.3900 & … & … & … & … & … & … & … & HIRES S0 \\
		2535.9824 & -4.5290$\pm$1.2800 & … & … & … & … & … & … & … & HIRES S0 \\
		2834.0661 & 1.8310$\pm$1.3300 & … & … & … & … & … & … & … & HIRES S0 \\
		3197.1130 & -0.1590$\pm$1.6200 & … & … & … & … & … & … & … & HIRES S0 \\
		5427.0842 & 6.2310$\pm$1.0200 & … & … & … & … & … & … & … & HIRES S0 \\
		5516.0198 & 8.0710$\pm$1.1900 & … & … & … & … & … & … & … & HIRES S0 \\
		5517.9432 & 9.5310$\pm$1.0900 & … & … & … & … & … & … & … & HIRES S0 \\
		5522.9838 & -4.4690$\pm$1.2800 & … & … & … & … & … & … & … & HIRES S0 \\
		5851.9753 & -5.1790$\pm$1.2400 & … & … & … & … & … & … & … & HIRES S0 \\
		6173.7359 & -1.7750$\pm$0.8992 & 4.1398$\pm$0.0089 & 0.8946$\pm$0.0025 & 0.5641$\pm$0.0035 & 3.229$\pm$0.014 & 20.114$\pm$0.081 & -8.60$\pm$0.43 & … & HARPS-N S1 \\
		6176.6680 & -0.3391$\pm$1.4275 & 3.4771$\pm$0.0110 & 0.8555$\pm$0.0027 & 0.5660$\pm$0.0035 & 3.355$\pm$0.013 & 21.718$\pm$0.082 & -2.07$\pm$0.65 & … & HARPS-N S1 \\
		6177.7017 & -1.5611$\pm$1.2475 & 3.6967$\pm$0.0088 & 0.8647$\pm$0.0027 & 0.5653$\pm$0.0035 & 3.286$\pm$0.013 & 20.639$\pm$0.078 & 0.94$\pm$0.48 & … & HARPS-N S1 \\
		6178.6635 & 3.3061$\pm$1.2674 & 3.4641$\pm$0.0087 & 0.8507$\pm$0.0028 & 0.5657$\pm$0.0035 & 3.277$\pm$0.013 & 20.562$\pm$0.080 & -1.22$\pm$0.49 & … & HARPS-N S1 \\
		6178.7327 & 4.0106$\pm$1.3056 & 3.6024$\pm$0.0091 & 0.8599$\pm$0.0027 & 0.5675$\pm$0.0035 & 3.257$\pm$0.013 & 20.318$\pm$0.077 & -2.87$\pm$0.48 & … & HARPS-N S1 \\
		6181.5932 & 17.3224$\pm$1.2894 & 4.0782$\pm$0.0096 & 0.8854$\pm$0.0026 & 0.5693$\pm$0.0035 & 3.229$\pm$0.014 & 19.938$\pm$0.079 & -9.36$\pm$0.45 & … & HARPS-N S1 \\
		6244.5529 & -10.0990$\pm$1.5207 & 4.4908$\pm$0.0171 & 0.9792$\pm$0.0021 & 0.5767$\pm$0.0034 & 3.276$\pm$0.013 & 20.167$\pm$0.078 & -6.09$\pm$0.74 & … & HARPS-N S1 \\
		6255.4344 & 11.6070$\pm$1.3527 & 3.8605$\pm$0.0145 & 0.9420$\pm$0.0023 & 0.5849$\pm$0.0033 & 3.317$\pm$0.013 & 20.838$\pm$0.081 & -6.70$\pm$0.77 & … & HARPS-N S1 \\
		6255.5223 & 11.9556$\pm$1.2206 & 3.8501$\pm$0.0099 & 0.9306$\pm$0.0023 & 0.5700$\pm$0.0035 & 3.231$\pm$0.014 & 19.857$\pm$0.078 & -5.96$\pm$0.44 & … & HARPS-N S1 \\
		6260.4830 & -1.0291$\pm$2.3740 & 4.1800$\pm$0.0303 & 0.9880$\pm$0.0020 & 0.5851$\pm$0.0033 & 3.236$\pm$0.014 & 19.938$\pm$0.080 & -2.18$\pm$2.18 & … & HARPS-N S1 \\
		6260.5346 & 1.3631$\pm$1.9723 & 4.3138$\pm$0.0325 & 0.9919$\pm$0.0020 & 0.5911$\pm$0.0032 & 3.242$\pm$0.014 & 19.915$\pm$0.086 & 1.88$\pm$2.27 & … & HARPS-N S1 \\
		6295.3266 & 17.4268$\pm$1.3872 & 4.6115$\pm$0.0123 & 1.0166$\pm$0.0019 & 0.5803$\pm$0.0033 & 3.286$\pm$0.013 & 20.303$\pm$0.076 & -8.20$\pm$0.56 & … & HARPS-N S1 \\
		6295.4240 & 16.6720$\pm$1.2901 & 5.2617$\pm$0.0131 & 1.0734$\pm$0.0017 & 0.5854$\pm$0.0032 & 3.282$\pm$0.013 & 20.208$\pm$0.079 & -9.41$\pm$0.53 & … & HARPS-N S1 \\
		6296.4111 & 13.1867$\pm$1.1628 & 4.8476$\pm$0.0107 & 1.0567$\pm$0.0018 & 0.5811$\pm$0.0033 & 3.302$\pm$0.013 & 20.287$\pm$0.080 & -6.94$\pm$0.45 & … & HARPS-N S1 \\
		6296.4472 & 12.8119$\pm$1.2740 & 4.6409$\pm$0.0114 & 1.0369$\pm$0.0019 & 0.5804$\pm$0.0033 & 3.319$\pm$0.013 & 20.579$\pm$0.080 & -6.68$\pm$0.51 & … & HARPS-N S1 \\
		6297.3572 & 6.0887$\pm$1.1700 & 4.4540$\pm$0.0087 & 1.0004$\pm$0.0020 & 0.5746$\pm$0.0034 & 3.331$\pm$0.013 & 20.558$\pm$0.080 & -6.38$\pm$0.42 & … & HARPS-N S1 \\
		6297.4095 & 3.9509$\pm$1.2259 & 4.7919$\pm$0.0127 & 1.0269$\pm$0.0019 & 0.5795$\pm$0.0033 & 3.321$\pm$0.013 & 20.593$\pm$0.080 & -9.76$\pm$0.56 & … & HARPS-N S1 \\
		6298.3581 & -5.0397$\pm$1.2660 & 5.0060$\pm$0.0104 & 1.0456$\pm$0.0018 & 0.5822$\pm$0.0033 & 3.348$\pm$0.013 & 20.827$\pm$0.077 & -1.70$\pm$0.48 & … & HARPS-N S1 \\
		6298.4080 & -6.8156$\pm$1.1938 & 4.6789$\pm$0.0105 & 1.0159$\pm$0.0019 & 0.5764$\pm$0.0034 & 3.318$\pm$0.013 & 20.449$\pm$0.076 & -5.68$\pm$0.48 & … & HARPS-N S1 \\
		6299.3826 & -10.7292$\pm$1.1756 & 4.6175$\pm$0.0089 & 1.0194$\pm$0.0019 & 0.5762$\pm$0.0034 & 3.323$\pm$0.013 & 20.434$\pm$0.077 & -3.09$\pm$0.42 & … & HARPS-N S1 \\
		6305.3720 & -5.1877$\pm$1.2125 & 3.5679$\pm$0.0086 & 0.9154$\pm$0.0024 & 0.5672$\pm$0.0035 & 3.326$\pm$0.013 & 20.884$\pm$0.079 & -5.87$\pm$0.46 & … & HARPS-N S1 \\
		6322.3888 & -12.9577$\pm$1.2158 & 4.2915$\pm$0.0095 & 0.9772$\pm$0.0021 & 0.5753$\pm$0.0034 & 3.292$\pm$0.013 & 20.256$\pm$0.078 & -3.15$\pm$0.44 & … & HARPS-N S1 \\
		6322.4528 & -15.3583$\pm$1.1857 & 4.1317$\pm$0.0116 & 0.9633$\pm$0.0022 & 0.5741$\pm$0.0034 & 3.296$\pm$0.013 & 20.369$\pm$0.078 & -2.62$\pm$0.51 & … & HARPS-N S1 \\
		6323.3329 & -11.8810$\pm$1.1963 & 4.0113$\pm$0.0127 & 0.9517$\pm$0.0022 & 0.5724$\pm$0.0034 & 3.299$\pm$0.013 & 20.554$\pm$0.077 & -3.01$\pm$0.61 & … & HARPS-N S1 \\
		6323.4300 & -10.8862$\pm$1.2430 & 4.0261$\pm$0.0144 & 0.9599$\pm$0.0022 & 0.5702$\pm$0.0034 & 3.278$\pm$0.013 & 20.331$\pm$0.078 & -2.44$\pm$0.64 & … & HARPS-N S1 \\
		6324.3587 & -9.4700$\pm$1.8958 & 3.8649$\pm$0.0268 & 0.9380$\pm$0.0023 & 0.5771$\pm$0.0034 & 3.278$\pm$0.013 & 20.485$\pm$0.074 & 1.06$\pm$1.68 & … & HARPS-N S1 \\
		6324.4360 & -6.6637$\pm$1.4805 & 3.8025$\pm$0.0221 & 0.9484$\pm$0.0022 & 0.5768$\pm$0.0034 & 3.285$\pm$0.013 & 20.594$\pm$0.086 & -5.42$\pm$1.21 & … & HARPS-N S1 \\
		6506.6358 & 2.3203$\pm$1.2618 & 4.5983$\pm$0.0149 & 1.0055$\pm$0.0020 & 0.5827$\pm$0.0033 & 3.327$\pm$0.013 & 20.290$\pm$0.079 & -5.89$\pm$0.65 & -12.1$\pm$6.6 & HARPS-N S2 \\
		6508.6885 & 3.3709$\pm$1.1425 & 4.7076$\pm$0.0093 & 1.0147$\pm$0.0019 & 0.5806$\pm$0.0033 & 3.336$\pm$0.013 & 20.332$\pm$0.076 & -4.90$\pm$0.41 & -0.8$\pm$6.3 & HARPS-N S2 \\
		6509.7290 & -1.5308$\pm$1.1858 & 4.7682$\pm$0.0133 & 1.0151$\pm$0.0019 & 0.5797$\pm$0.0034 & 3.294$\pm$0.013 & 20.044$\pm$0.076 & -3.25$\pm$0.54 & -7.6$\pm$6.7 & HARPS-N S2 \\
		6516.6377 & -9.0398$\pm$1.2467 & 4.3511$\pm$0.0129 & 0.9829$\pm$0.0021 & 0.5743$\pm$0.0034 & 3.310$\pm$0.013 & 20.362$\pm$0.077 & -0.66$\pm$0.56 & 3.3$\pm$6.6 & HARPS-N S2 \\
		6529.7158 & 2.3432$\pm$1.2357 & 4.7058$\pm$0.0109 & 1.0116$\pm$0.0020 & 0.5783$\pm$0.0034 & 3.263$\pm$0.013 & 19.608$\pm$0.080 & -6.72$\pm$0.44 & -35.1$\pm$6.7 & HARPS-N S2 \\
		6533.5034 & 1.2240$\pm$1.1833 & 4.7104$\pm$0.0101 & 1.0221$\pm$0.0019 & 0.5768$\pm$0.0034 & 3.362$\pm$0.013 & 20.701$\pm$0.075 & -6.10$\pm$0.45 & -9.5$\pm$7.6 & HARPS-N S2 \\
		6533.6116 & 2.2137$\pm$1.2001 & 4.7227$\pm$0.0097 & 1.0241$\pm$0.0019 & 0.5765$\pm$0.0034 & 3.354$\pm$0.013 & 20.678$\pm$0.078 & -4.55$\pm$0.44 & -5.1$\pm$6.8 & HARPS-N S2 \\
		6534.5612 & -2.4093$\pm$1.3222 & 4.5451$\pm$0.0158 & 1.0102$\pm$0.0020 & 0.5729$\pm$0.0034 & 3.312$\pm$0.013 & 20.325$\pm$0.079 & -3.15$\pm$0.66 & -20.6$\pm$8.0 & HARPS-N S2 \\
		6535.6417 & -0.8885$\pm$1.1881 & 4.7448$\pm$0.0090 & 1.0310$\pm$0.0019 & 0.5780$\pm$0.0034 & 3.356$\pm$0.013 & 20.760$\pm$0.078 & -1.74$\pm$0.41 & -9.5$\pm$6.1 & HARPS-N S2 \\
		6537.5847 & -6.7190$\pm$1.1599 & 4.3228$\pm$0.0103 & 0.9839$\pm$0.0021 & 0.5761$\pm$0.0034 & 3.293$\pm$0.013 & 20.282$\pm$0.074 & -5.89$\pm$0.46 & -13.9$\pm$5.3 & HARPS-N S2 \\
		6544.6457 & 1.0641$\pm$1.2110 & 4.8527$\pm$0.0107 & 1.0238$\pm$0.0019 & 0.5793$\pm$0.0033 & 3.344$\pm$0.013 & 20.428$\pm$0.073 & -5.33$\pm$0.46 & -6.7$\pm$7.0 & HARPS-N S2 \\
		6546.6039 & 0.1094$\pm$1.1692 & 4.6964$\pm$0.0109 & 1.0115$\pm$0.0020 & 0.5779$\pm$0.0034 & 3.345$\pm$0.013 & 20.435$\pm$0.076 & -3.24$\pm$0.48 & -0.7$\pm$6.0 & HARPS-N S2 \\
		6579.7292 & 0.2502$\pm$1.2059 & 4.4994$\pm$0.0106 & 0.9955$\pm$0.0020 & 0.5745$\pm$0.0034 & 3.318$\pm$0.013 & 20.238$\pm$0.079 & -5.09$\pm$0.43 & -14.0$\pm$7.5 & HARPS-N S2 \\
		6583.5733 & -0.1905$\pm$1.0901 & 4.6355$\pm$0.0135 & 1.0195$\pm$0.0019 & 0.5851$\pm$0.0033 & 3.319$\pm$0.013 & 20.146$\pm$0.076 & -2.82$\pm$0.58 & -10.2$\pm$6.2 & HARPS-N S2 \\
		6604.5628 & 5.0178$\pm$1.2712 & 4.7568$\pm$0.0167 & 1.0327$\pm$0.0019 & 0.5813$\pm$0.0033 & 3.311$\pm$0.013 & 20.111$\pm$0.080 & -7.38$\pm$0.67 & 0.5$\pm$8.5 & HARPS-N S2 \\
		6606.6129 & 2.7993$\pm$1.2565 & 4.3281$\pm$0.0157 & 0.9854$\pm$0.0021 & 0.5785$\pm$0.0033 & 3.356$\pm$0.013 & 20.594$\pm$0.075 & 0.62$\pm$0.69 & -6.5$\pm$8.0 & HARPS-N S2 \\
		6611.5716 & -1.6277$\pm$1.3244 & 4.3661$\pm$0.0164 & 0.9921$\pm$0.0020 & 0.5760$\pm$0.0034 & 3.321$\pm$0.013 & 20.251$\pm$0.074 & -4.63$\pm$2.89 & -11.7$\pm$7.5 & HARPS-N S2 \\
		6611.5841 & -3.4033$\pm$1.3301 & 4.4899$\pm$0.0153 & 0.9847$\pm$0.0021 & 0.5756$\pm$0.0034 & 3.310$\pm$0.013 & 20.157$\pm$0.080 & -2.02$\pm$2.64 & -9.3$\pm$8.4 & HARPS-N S2 \\
		6617.5071 & 6.0244$\pm$1.3725 & 4.7050$\pm$0.0161 & 1.0038$\pm$0.0020 & 0.5755$\pm$0.0034 & 3.319$\pm$0.013 & 20.256$\pm$0.079 & -2.96$\pm$0.67 & -15.5$\pm$9.0 & HARPS-N S2 \\
		6618.5505 & 2.8244$\pm$1.2907 & 4.6289$\pm$0.0104 & 0.9942$\pm$0.0020 & 0.5741$\pm$0.0034 & 3.291$\pm$0.013 & 19.872$\pm$0.079 & -5.32$\pm$0.43 & -26.5$\pm$6.0 & HARPS-N S2 \\
		6621.3900 & -0.8679$\pm$1.1977 & 4.5393$\pm$0.0109 & 0.9875$\pm$0.0021 & 0.5739$\pm$0.0034 & 3.295$\pm$0.013 & 19.931$\pm$0.078 & -8.12$\pm$0.46 & -10.6$\pm$6.7 & HARPS-N S2 \\
		6625.5642 & 6.2012$\pm$1.3841 & 5.3185$\pm$0.0210 & 1.0799$\pm$0.0017 & 0.5893$\pm$0.0032 & 3.323$\pm$0.013 & 20.257$\pm$0.082 & -4.22$\pm$3.26 & -14.2$\pm$9.7 & HARPS-N S2 \\
		6625.5748 & 5.0547$\pm$1.3092 & 5.3753$\pm$0.0198 & 1.0685$\pm$0.0018 & 0.5869$\pm$0.0032 & 3.327$\pm$0.013 & 20.243$\pm$0.078 & -5.24$\pm$3.04 & -2.2$\pm$8.2 & HARPS-N S2 \\
		6854.6643 & 1.0187$\pm$1.2246 & 4.1574$\pm$0.0114 & 0.9685$\pm$0.0021 & 0.5714$\pm$0.0034 & 3.277$\pm$0.013 & 20.864$\pm$0.079 & -2.76$\pm$0.47 & 0.9$\pm$7.7 & HARPS-N S3 \\
		6855.6874 & 1.0357$\pm$1.2324 & 4.1842$\pm$0.0137 & 0.9728$\pm$0.0021 & 0.5735$\pm$0.0034 & 3.267$\pm$0.013 & 20.870$\pm$0.083 & -5.76$\pm$0.55 & -0.9$\pm$7.7 & HARPS-N S3 \\
		6879.6776 & 1.8299$\pm$1.2149 & 4.4878$\pm$0.0119 & 1.0058$\pm$0.0020 & 0.5785$\pm$0.0034 & 3.300$\pm$0.013 & 21.260$\pm$0.082 & -3.04$\pm$0.48 & -7.9$\pm$6.9 & HARPS-N S3 \\
		6880.6694 & -2.0162$\pm$1.2035 & 4.1942$\pm$0.0146 & 0.9813$\pm$0.0021 & 0.5753$\pm$0.0034 & 3.287$\pm$0.013 & 21.201$\pm$0.082 & -1.26$\pm$0.58 & 7.0$\pm$6.3 & HARPS-N S3 \\
		6881.6113 & -1.5174$\pm$1.2451 & 3.9777$\pm$0.0138 & 0.9587$\pm$0.0022 & 0.5706$\pm$0.0035 & 3.331$\pm$0.013 & 21.797$\pm$0.084 & -2.25$\pm$0.61 & 20.3$\pm$6.8 & HARPS-N S3 \\
		6892.6492 & 0.0351$\pm$1.1524 & 4.0618$\pm$0.0102 & 0.9650$\pm$0.0021 & 0.5765$\pm$0.0034 & 3.287$\pm$0.013 & 21.087$\pm$0.079 & -5.45$\pm$0.44 & -2.1$\pm$5.6 & HARPS-N S3 \\
		6893.6273 & 3.6747$\pm$1.2640 & 4.2764$\pm$0.0145 & 0.9901$\pm$0.0020 & 0.5782$\pm$0.0034 & 3.297$\pm$0.013 & 21.355$\pm$0.082 & -4.28$\pm$0.60 & 3.3$\pm$7.3 & HARPS-N S3 \\
		6894.6469 & 9.1049$\pm$1.3112 & 4.2697$\pm$0.0174 & 0.9957$\pm$0.0020 & 0.5781$\pm$0.0033 & 3.271$\pm$0.013 & 21.034$\pm$0.084 & -5.59$\pm$0.71 & 0.5$\pm$8.7 & HARPS-N S3 \\
		6897.6410 & 3.7368$\pm$0.8017 & 4.1894$\pm$0.0120 & 0.9732$\pm$0.0021 & 0.5781$\pm$0.0034 & 3.333$\pm$0.013 & 21.763$\pm$0.080 & -2.19$\pm$0.53 & 19.7$\pm$6.5 & HARPS-N S3 \\
		6898.6249 & -1.0731$\pm$0.6758 & 4.2248$\pm$0.0107 & 0.9725$\pm$0.0021 & 0.5742$\pm$0.0034 & 3.312$\pm$0.013 & 21.453$\pm$0.080 & -3.05$\pm$0.47 & -4.1$\pm$6.7 & HARPS-N S3 \\
		6899.6327 & -3.5197$\pm$0.6746 & 4.1557$\pm$0.0119 & 0.9643$\pm$0.0022 & 0.5751$\pm$0.0034 & 3.298$\pm$0.013 & 21.317$\pm$0.084 & -2.87$\pm$0.51 & 9.5$\pm$7.3 & HARPS-N S3 \\
		6903.6598 & -2.5603$\pm$0.6348 & 4.1328$\pm$0.0087 & 0.9531$\pm$0.0022 & 0.5706$\pm$0.0035 & 3.297$\pm$0.013 & 21.423$\pm$0.082 & -3.74$\pm$0.38 & 11.1$\pm$6.9 & HARPS-N S3 \\
		6904.6622 & -2.2273$\pm$0.6020 & 4.1667$\pm$0.0114 & 0.9486$\pm$0.0022 & 0.5707$\pm$0.0035 & 3.229$\pm$0.014 & 20.627$\pm$0.082 & -6.66$\pm$0.45 & -8.5$\pm$5.7 & HARPS-N S3 \\
		6905.6381 & 2.1950$\pm$0.7970 & 4.1096$\pm$0.0117 & 0.9510$\pm$0.0022 & 0.5701$\pm$0.0035 & 3.296$\pm$0.013 & 21.494$\pm$0.082 & -3.73$\pm$0.50 & 21.2$\pm$8.8 & HARPS-N S3 \\
		6918.7069 & -1.6008$\pm$0.9263 & 4.4555$\pm$0.0162 & 0.9760$\pm$0.0021 & 0.5733$\pm$0.0034 & 3.265$\pm$0.013 & 21.032$\pm$0.083 & -3.26$\pm$0.62 & 0.1$\pm$8.1 & HARPS-N S3 \\
		6919.6480 & -1.8544$\pm$0.7472 & 4.1628$\pm$0.0126 & 0.9571$\pm$0.0022 & 0.5726$\pm$0.0034 & 3.298$\pm$0.013 & 21.432$\pm$0.080 & -6.63$\pm$0.54 & -1.7$\pm$6.7 & HARPS-N S3 \\
		6920.6746 & -1.0033$\pm$0.5990 & 4.1949$\pm$0.0139 & 0.9666$\pm$0.0021 & 0.5761$\pm$0.0034 & 3.262$\pm$0.013 & 20.987$\pm$0.079 & -6.64$\pm$0.57 & 1.6$\pm$6.4 & HARPS-N S3 \\
		6928.7029 & 0.7874$\pm$1.0055 & 4.2716$\pm$0.0165 & 0.9658$\pm$0.0021 & 0.5748$\pm$0.0034 & 3.172$\pm$0.014 & 20.086$\pm$0.083 & -5.32$\pm$0.57 & -22.1$\pm$8.2 & HARPS-N S3 \\
		6929.7220 & -4.3450$\pm$0.7519 & 4.3170$\pm$0.0133 & 0.9757$\pm$0.0021 & 0.5746$\pm$0.0034 & 3.212$\pm$0.014 & 20.336$\pm$0.081 & -1.69$\pm$0.47 & -4.0$\pm$6.3 & HARPS-N S3 \\
		6930.7049 & -5.2705$\pm$0.7044 & 4.3434$\pm$0.0114 & 0.9907$\pm$0.0020 & 0.5775$\pm$0.0034 & 3.277$\pm$0.013 & 21.074$\pm$0.082 & -4.52$\pm$0.46 & 5.2$\pm$6.6 & HARPS-N S3 \\
		6931.7233 & 0.9824$\pm$0.4821 & 4.3169$\pm$0.0095 & 0.9830$\pm$0.0021 & 0.5749$\pm$0.0034 & 3.286$\pm$0.013 & 21.170$\pm$0.082 & -2.78$\pm$0.40 & 8.2$\pm$4.7 & HARPS-N S3 \\
		6932.6824 & -0.4654$\pm$0.8408 & 4.1917$\pm$0.0139 & 0.9641$\pm$0.0022 & 0.5715$\pm$0.0034 & 3.287$\pm$0.013 & 21.265$\pm$0.082 & -2.38$\pm$0.56 & 23.0$\pm$7.5 & HARPS-N S3 \\
		6937.6482 & 1.0489$\pm$0.5631 & 5.1195$\pm$0.0104 & 1.0510$\pm$0.0018 & 0.5825$\pm$0.0033 & 3.268$\pm$0.013 & 20.872$\pm$0.080 & -6.18$\pm$0.38 & 8.5$\pm$5.7 & HARPS-N S3 \\
		6938.6521 & -0.4214$\pm$0.8151 & 4.3081$\pm$0.0123 & 0.9806$\pm$0.0021 & 0.5736$\pm$0.0034 & 3.237$\pm$0.014 & 20.598$\pm$0.085 & -2.64$\pm$0.48 & 10.2$\pm$7.4 & HARPS-N S3 \\
		6939.6873 & -1.5479$\pm$1.3899 & 3.9546$\pm$0.0240 & 0.9644$\pm$0.0022 & 0.6005$\pm$0.0031 & 3.320$\pm$0.013 & 21.432$\pm$0.083 & -0.54$\pm$1.41 & 1.8$\pm$12.5 & HARPS-N S3 \\
		6940.6449 & -4.9940$\pm$0.6380 & 4.2667$\pm$0.0115 & 0.9755$\pm$0.0021 & 0.5756$\pm$0.0034 & 3.294$\pm$0.013 & 21.339$\pm$0.085 & -3.96$\pm$0.47 & 2.6$\pm$5.5 & HARPS-N S3 \\
		6943.6512 & 0.1842$\pm$0.7668 & 4.0988$\pm$0.0128 & 0.9528$\pm$0.0022 & 0.5702$\pm$0.0034 & 3.255$\pm$0.013 & 20.962$\pm$0.085 & -3.51$\pm$0.52 & 0.4$\pm$7.2 & HARPS-N S3 \\
		7239.6224 & -0.3466$\pm$0.6098 & 4.1558$\pm$0.0112 & 0.9505$\pm$0.0022 & 0.5707$\pm$0.0035 & 3.258$\pm$0.013 & 20.975$\pm$0.077 & -5.46$\pm$0.45 & 0.1$\pm$6.3 & HARPS-N S4 \\
		7240.6229 & -1.8935$\pm$0.7604 & 4.1907$\pm$0.0130 & 0.9552$\pm$0.0022 & 0.5740$\pm$0.0034 & 3.278$\pm$0.013 & 21.238$\pm$0.081 & -4.00$\pm$0.52 & 19.5$\pm$7.3 & HARPS-N S4 \\
		7241.6285 & 0.3910$\pm$0.6258 & 4.0443$\pm$0.0137 & 0.9480$\pm$0.0022 & 0.5729$\pm$0.0034 & 3.272$\pm$0.013 & 21.240$\pm$0.081 & -3.80$\pm$0.56 & 8.4$\pm$7.3 & HARPS-N S4 \\
		7242.6981 & 5.1989$\pm$2.9559 & 5.6241$\pm$0.0587 & 0.9627$\pm$0.0022 & 0.5999$\pm$0.0031 & 3.257$\pm$0.013 & 21.005$\pm$0.071 & -6.75$\pm$3.06 & 22.7$\pm$19.2 & HARPS-N S4 \\
		7251.6764 & 4.5748$\pm$0.7146 & 4.0754$\pm$0.0118 & 0.9451$\pm$0.0022 & 0.5719$\pm$0.0034 & 3.267$\pm$0.013 & 20.976$\pm$0.082 & -3.51$\pm$0.49 & 5.4$\pm$7.2 & HARPS-N S4 \\
		7260.6041 & -3.1498$\pm$0.8176 & 4.2801$\pm$0.0128 & 0.9699$\pm$0.0021 & 0.5736$\pm$0.0034 & 3.285$\pm$0.013 & 21.383$\pm$0.084 & -1.72$\pm$0.53 & 9.1$\pm$7.2 & HARPS-N S4 \\
		7261.6234 & -3.1311$\pm$0.7851 & 4.2175$\pm$0.0143 & 0.9667$\pm$0.0021 & 0.5724$\pm$0.0034 & 3.237$\pm$0.014 & 20.861$\pm$0.081 & -2.69$\pm$0.57 & 1.9$\pm$8.2 & HARPS-N S4 \\
		7263.5969 & -1.9881$\pm$0.9299 & 4.2493$\pm$0.0144 & 0.9805$\pm$0.0021 & 0.5799$\pm$0.0034 & 3.309$\pm$0.013 & 21.661$\pm$0.080 & -6.21$\pm$0.62 & 5.6$\pm$8.6 & HARPS-N S4 \\
		7264.5952 & 0.6718$\pm$0.8729 & 3.9212$\pm$0.0135 & 0.9423$\pm$0.0023 & 0.5753$\pm$0.0034 & 3.294$\pm$0.013 & 21.479$\pm$0.081 & -4.96$\pm$0.59 & 18.9$\pm$8.4 & HARPS-N S4 \\
		7269.6047 & 7.1802$\pm$1.2289 & 4.4411$\pm$0.0126 & 0.9848$\pm$0.0021 & 0.5799$\pm$0.0033 & 3.254$\pm$0.013 & 20.737$\pm$0.083 & -3.24$\pm$0.49 & 9.7$\pm$7.9 & HARPS-N S4 \\
		7270.5880 & 4.0779$\pm$1.1772 & 4.1420$\pm$0.0098 & 0.9603$\pm$0.0022 & 0.5746$\pm$0.0034 & 3.274$\pm$0.013 & 20.970$\pm$0.083 & -5.45$\pm$0.41 & 10.8$\pm$6.4 & HARPS-N S4 \\
		7271.6229 & 6.9219$\pm$1.2472 & 4.0244$\pm$0.0107 & 0.9519$\pm$0.0022 & 0.5746$\pm$0.0034 & 3.241$\pm$0.014 & 20.633$\pm$0.085 & -5.73$\pm$0.43 & 4.7$\pm$6.5 & HARPS-N S4 \\
		7272.6319 & 8.2715$\pm$1.3163 & 4.1495$\pm$0.0129 & 0.9558$\pm$0.0022 & 0.5754$\pm$0.0034 & 3.240$\pm$0.014 & 20.692$\pm$0.080 & -7.72$\pm$0.52 & -4.4$\pm$8.2 & HARPS-N S4 \\
		7274.5833 & 1.8002$\pm$0.7870 & 4.2728$\pm$0.0142 & 0.9640$\pm$0.0022 & 0.5749$\pm$0.0034 & 3.256$\pm$0.013 & 20.960$\pm$0.081 & -4.69$\pm$0.56 & 11.2$\pm$7.2 & HARPS-N S4 \\
		7275.5797 & -1.4385$\pm$1.2342 & 4.2273$\pm$0.0134 & 0.9585$\pm$0.0022 & 0.5745$\pm$0.0034 & 3.277$\pm$0.013 & 21.194$\pm$0.081 & -4.38$\pm$0.56 & 5.6$\pm$7.7 & HARPS-N S4 \\
		7276.5819 & -1.0573$\pm$0.5538 & 4.5012$\pm$0.0115 & 0.9880$\pm$0.0020 & 0.5807$\pm$0.0033 & 3.278$\pm$0.013 & 21.037$\pm$0.083 & -2.14$\pm$0.48 & -17.7$\pm$6.5 & HARPS-N S4 \\
		7277.5762 & -1.9999$\pm$0.6468 & 4.3553$\pm$0.0126 & 0.9762$\pm$0.0021 & 0.5766$\pm$0.0034 & 3.245$\pm$0.014 & 20.691$\pm$0.079 & -1.68$\pm$0.50 & 1.7$\pm$6.3 & HARPS-N S4 \\
		7278.6016 & -1.1620$\pm$0.6489 & 4.1069$\pm$0.0122 & 0.9603$\pm$0.0022 & 0.5740$\pm$0.0034 & 3.269$\pm$0.013 & 21.079$\pm$0.084 & -5.41$\pm$0.51 & 1.1$\pm$6.2 & HARPS-N S4 \\
		7282.5963 & 1.5189$\pm$0.7917 & 4.0596$\pm$0.0092 & 0.9542$\pm$0.0022 & 0.5739$\pm$0.0034 & 3.295$\pm$0.013 & 21.364$\pm$0.082 & -2.08$\pm$0.41 & -14.4$\pm$20.9 & HARPS-N S4 \\
		7286.7258 & -2.8379$\pm$0.7695 & 3.7668$\pm$0.0120 & 0.9332$\pm$0.0023 & 0.5723$\pm$0.0034 & 3.275$\pm$0.013 & 21.137$\pm$0.086 & -5.39$\pm$0.53 & 4.9$\pm$8.0 & HARPS-N S4 \\
		7287.7279 & -2.3853$\pm$0.7732 & 3.7496$\pm$0.0151 & 0.9257$\pm$0.0023 & 0.5747$\pm$0.0034 & 3.270$\pm$0.013 & 21.142$\pm$0.081 & -3.30$\pm$0.66 & 15.6$\pm$7.2 & HARPS-N S4 \\
		7303.5930 & -7.4601$\pm$0.7978 & 4.0589$\pm$0.0116 & 0.9624$\pm$0.0022 & 0.5731$\pm$0.0034 & 3.312$\pm$0.013 & 21.503$\pm$0.084 & -3.63$\pm$0.50 & -5.1$\pm$8.1 & HARPS-N S4 \\
		7594.6447 & -2.4083$\pm$1.2137 & 3.7958$\pm$0.0123 & 0.9499$\pm$0.0022 & 0.5751$\pm$0.0034 & 3.242$\pm$0.014 & 20.901$\pm$0.084 & -6.56$\pm$0.53 & 7.3$\pm$5.9 & HARPS-N S5 \\
		7603.6552 & -9.2349$\pm$1.2514 & 4.2816$\pm$0.0117 & 0.9881$\pm$0.0020 & 0.5732$\pm$0.0034 & 3.221$\pm$0.014 & 20.713$\pm$0.082 & -6.33$\pm$0.48 & 4.3$\pm$7.3 & HARPS-N S5 \\
		7604.6475 & -8.5705$\pm$1.6629 & 3.6779$\pm$0.0259 & 0.9565$\pm$0.0022 & 0.5771$\pm$0.0034 & 3.252$\pm$0.014 & 21.208$\pm$0.094 & -10.97$\pm$1.56 & -6.7$\pm$13.0 & HARPS-N S5 \\
		7620.5566 & -1.9061$\pm$1.5985 & 4.2421$\pm$0.0196 & 0.9975$\pm$0.0020 & 0.5788$\pm$0.0034 & 3.253$\pm$0.014 & 21.183$\pm$0.081 & -0.73$\pm$0.84 & 5.1$\pm$11.7 & HARPS-N S5 \\
		7624.6652 & -2.5524$\pm$1.2270 & 3.7167$\pm$0.0104 & 0.9433$\pm$0.0022 & 0.5697$\pm$0.0035 & 3.240$\pm$0.014 & 21.028$\pm$0.082 & -2.46$\pm$0.48 & -1.1$\pm$6.8 & HARPS-N S5 \\
		7644.6659 & 1.4244$\pm$1.3018 & 3.8637$\pm$0.0115 & 0.9598$\pm$0.0022 & 0.5706$\pm$0.0035 & 3.210$\pm$0.014 & 20.760$\pm$0.085 & -4.23$\pm$0.49 & -14.6$\pm$8.4 & HARPS-N S5 \\
		7650.5464 & -3.0408$\pm$1.2297 & 3.9801$\pm$0.0125 & 0.9684$\pm$0.0021 & 0.5722$\pm$0.0034 & 3.260$\pm$0.013 & 21.330$\pm$0.086 & -5.54$\pm$0.55 & 13.8$\pm$6.5 & HARPS-N S5 \\
		7651.5936 & -4.0341$\pm$1.2429 & 3.7489$\pm$0.0136 & 0.9549$\pm$0.0022 & 0.5702$\pm$0.0034 & 3.225$\pm$0.014 & 20.925$\pm$0.081 & -3.02$\pm$0.60 & 11.6$\pm$6.7 & HARPS-N S5 \\
		7652.6097 & 1.1585$\pm$1.1979 & 3.7852$\pm$0.0118 & 0.9569$\pm$0.0022 & 0.5697$\pm$0.0034 & 3.199$\pm$0.014 & 20.569$\pm$0.087 & -6.23$\pm$0.51 & 0.1$\pm$9.4 & HARPS-N S5 \\
		7653.6167 & 3.0905$\pm$1.2506 & 3.7946$\pm$0.0122 & 0.9635$\pm$0.0022 & 0.5701$\pm$0.0035 & 3.273$\pm$0.013 & 21.458$\pm$0.080 & -5.48$\pm$0.56 & 4.7$\pm$6.6 & HARPS-N S5 \\
		7679.5004 & -9.9809$\pm$1.2040 & 4.2530$\pm$0.0116 & 0.9842$\pm$0.0021 & 0.5764$\pm$0.0034 & 3.236$\pm$0.014 & 20.845$\pm$0.085 & -3.62$\pm$0.49 & -8.3$\pm$7.3 & HARPS-N S5 \\
		7680.5461 & -7.4510$\pm$1.3384 & 4.1485$\pm$0.0116 & 0.9768$\pm$0.0021 & 0.5734$\pm$0.0034 & 3.302$\pm$0.013 & 21.774$\pm$0.087 & -2.91$\pm$0.53 & 17.4$\pm$7.9 & HARPS-N S5 \\
		7681.5852 & -7.3649$\pm$1.2513 & 4.2132$\pm$0.0095 & 0.9918$\pm$0.0020 & 0.5740$\pm$0.0034 & 3.307$\pm$0.013 & 21.738$\pm$0.083 & -3.32$\pm$0.44 & 4.4$\pm$9.0 & HARPS-N S5 \\
		7683.5344 & 5.1337$\pm$1.2800 & 3.9168$\pm$0.0125 & 0.9552$\pm$0.0022 & 0.5714$\pm$0.0034 & 3.271$\pm$0.013 & 21.334$\pm$0.080 & -7.17$\pm$0.58 & 13.1$\pm$7.4 & HARPS-N S5 \\
		7702.4450 & -2.4638$\pm$1.4217 & 4.0400$\pm$0.0199 & 0.9798$\pm$0.0021 & 0.5716$\pm$0.0034 & 3.261$\pm$0.013 & 21.246$\pm$0.083 & -7.27$\pm$0.94 & -2.7$\pm$9.8 & HARPS-N S5 \\
		7703.6205 & -4.7537$\pm$1.3979 & 4.0622$\pm$0.0162 & 0.9679$\pm$0.0021 & 0.5709$\pm$0.0034 & 3.253$\pm$0.013 & 21.131$\pm$0.082 & -3.33$\pm$0.68 & 7.5$\pm$8.1 & HARPS-N S5 \\
		7721.5566 & 2.2866$\pm$1.5925 & 4.3654$\pm$0.0243 & 0.9876$\pm$0.0021 & 0.5808$\pm$0.0033 & 3.190$\pm$0.014 & 20.571$\pm$0.090 & -8.56$\pm$1.12 & -11.9$\pm$10.8 & HARPS-N S5 \\
		7727.4281 & 5.9223$\pm$1.2366 & 4.2273$\pm$0.0116 & 0.9834$\pm$0.0021 & 0.5759$\pm$0.0034 & 3.276$\pm$0.013 & 21.325$\pm$0.080 & -6.57$\pm$0.50 & 0.3$\pm$7.0 & HARPS-N S5 \\
		7728.5223 & 6.0193$\pm$1.7022 & 4.0941$\pm$0.0195 & 0.9784$\pm$0.0021 & 0.5750$\pm$0.0034 & 3.244$\pm$0.014 & 21.025$\pm$0.075 & -6.09$\pm$0.84 & 3.9$\pm$10.8 & HARPS-N S5 \\
		7729.4161 & 4.5281$\pm$1.2825 & 4.2733$\pm$0.0157 & 0.9985$\pm$0.0020 & 0.5745$\pm$0.0034 & 3.245$\pm$0.014 & 20.994$\pm$0.080 & -5.31$\pm$0.65 & 5.1$\pm$8.5 & HARPS-N S5 \\
		7730.3968 & 11.7289$\pm$1.1568 & 4.1357$\pm$0.0100 & 0.9674$\pm$0.0021 & 0.5714$\pm$0.0034 & 3.190$\pm$0.014 & 20.399$\pm$0.086 & -8.54$\pm$0.39 & -11.3$\pm$6.5 & HARPS-N S5 \\
		7734.5708 & -3.1547$\pm$1.5998 & 4.0872$\pm$0.0152 & 0.9882$\pm$0.0020 & 0.5823$\pm$0.0033 & 3.230$\pm$0.014 & 20.656$\pm$0.083 & -2.86$\pm$0.64 & 14.3$\pm$10.2 & HARPS-N S5 \\
		7735.4488 & -7.8042$\pm$1.4418 & 4.5482$\pm$0.0177 & 1.0293$\pm$0.0019 & 0.5774$\pm$0.0033 & 3.234$\pm$0.014 & 20.867$\pm$0.087 & -5.66$\pm$0.71 & -7.4$\pm$10.9 & HARPS-N S5 \\
		7736.4463 & -8.7253$\pm$1.1742 & 4.3167$\pm$0.0112 & 0.9975$\pm$0.0020 & 0.5730$\pm$0.0034 & 3.236$\pm$0.014 & 20.857$\pm$0.085 & -4.61$\pm$0.47 & 8.3$\pm$6.7 & HARPS-N S5 \\
		7738.4334 & 0.5254$\pm$1.3399 & 4.2030$\pm$0.0138 & 0.9895$\pm$0.0020 & 0.5720$\pm$0.0034 & 3.251$\pm$0.014 & 21.098$\pm$0.081 & -4.59$\pm$0.58 & 7.2$\pm$7.4 & HARPS-N S5 \\
		7749.4581 & 10.7016$\pm$1.3292 & 4.0114$\pm$0.0151 & 0.9830$\pm$0.0021 & 0.5748$\pm$0.0034 & 3.270$\pm$0.013 & 21.288$\pm$0.084 & -4.18$\pm$0.66 & 1.5$\pm$8.8 & HARPS-N S5 \\
		7750.4766 & 8.2896$\pm$1.5970 & 4.0536$\pm$0.0224 & 0.9862$\pm$0.0021 & 0.5738$\pm$0.0034 & 3.251$\pm$0.014 & 21.179$\pm$0.087 & -9.61$\pm$1.06 & -3.4$\pm$10.8 & HARPS-N S5 \\
		7753.4789 & 4.3611$\pm$2.2210 & 4.3139$\pm$0.0331 & 0.9898$\pm$0.0020 & 0.5880$\pm$0.0032 & 3.208$\pm$0.014 & 20.805$\pm$0.093 & -2.18$\pm$1.88 & 1.0$\pm$14.0 & HARPS-N S5 \\
		7762.4218 & -3.1596$\pm$1.2899 & 4.4177$\pm$0.0161 & 1.0063$\pm$0.0020 & 0.5756$\pm$0.0034 & 3.247$\pm$0.014 & 21.043$\pm$0.080 & -2.76$\pm$0.66 & -2.1$\pm$7.8 & HARPS-N S5 \\
		7984.6323 & 1.0962$\pm$1.2496 & 3.8280$\pm$0.0105 & 0.9394$\pm$0.0023 & 0.5710$\pm$0.0035 & 3.263$\pm$0.013 & 21.070$\pm$0.084 & -3.78$\pm$0.49 & -5.1$\pm$6.6 & HARPS-N S6 \\
		7995.6865 & 6.3189$\pm$1.1839 & 4.2605$\pm$0.0122 & 0.9831$\pm$0.0021 & 0.5738$\pm$0.0034 & 3.197$\pm$0.014 & 20.406$\pm$0.080 & -4.55$\pm$0.49 & -3.3$\pm$5.7 & HARPS-N S6 \\
		8019.7325 & -2.5380$\pm$1.3792 & 3.9744$\pm$0.0164 & 0.9538$\pm$0.0022 & 0.5744$\pm$0.0034 & 3.249$\pm$0.014 & 21.004$\pm$0.077 & -4.68$\pm$0.72 & -3.9$\pm$8.7 & HARPS-N S6 \\
		8022.6843 & -7.7117$\pm$1.2749 & 3.8366$\pm$0.0124 & 0.9528$\pm$0.0022 & 0.5738$\pm$0.0034 & 3.227$\pm$0.014 & 20.741$\pm$0.082 & -5.24$\pm$0.55 & 1.7$\pm$6.7 & HARPS-N S6 \\
		8024.6455 & -11.4591$\pm$1.4515 & 3.6429$\pm$0.0114 & 0.9241$\pm$0.0023 & 0.5695$\pm$0.0035 & 3.223$\pm$0.014 & 20.722$\pm$0.083 & -1.98$\pm$0.52 & -1.6$\pm$8.8 & HARPS-N S6 \\
		8025.6317 & -11.4648$\pm$1.4488 & 3.6327$\pm$0.0110 & 0.9122$\pm$0.0024 & 0.5670$\pm$0.0035 & 3.189$\pm$0.014 & 20.408$\pm$0.083 & -4.17$\pm$0.49 & 1.3$\pm$9.8 & HARPS-N S6 \\
		8031.5560 & 13.0554$\pm$1.2623 & 3.7735$\pm$0.0115 & 0.9377$\pm$0.0023 & 0.5697$\pm$0.0035 & 3.249$\pm$0.014 & 21.006$\pm$0.083 & -7.90$\pm$0.53 & -0.6$\pm$8.1 & HARPS-N S6 \\
		8031.6216 & 11.9601$\pm$1.4224 & 3.7346$\pm$0.0140 & 0.9387$\pm$0.0023 & 0.5704$\pm$0.0034 & 3.255$\pm$0.013 & 21.122$\pm$0.081 & -8.02$\pm$0.66 & -0.1$\pm$10.9 & HARPS-N S6 \\
		8037.6901 & -4.8464$\pm$1.0332 & 4.1820$\pm$0.0175 & 0.9778$\pm$0.0021 & 0.5796$\pm$0.0033 & 3.255$\pm$0.013 & 21.120$\pm$0.080 & -3.62$\pm$0.78 & 16.0$\pm$8.9 & HARPS-N S6 \\
		7395.3132 & 2.1218$\pm$1.4092 & 1.7454$\pm$0.0025 & 0.9357$\pm$0.0015 & 0.6425$\pm$0.0038 & 4.875$\pm$0.022 & 20.633$\pm$0.065 & -2.52$\pm$0.92 & -22.7$\pm$12.7 & CARMENES S4 \\
		7397.3347 & -4.7407$\pm$1.7413 & 1.7998$\pm$0.0030 & 0.9741$\pm$0.0018 & 0.6701$\pm$0.0049 & 4.905$\pm$0.019 & 20.608$\pm$0.057 & 6.95$\pm$1.08 & 8.6$\pm$11.9 & CARMENES S4 \\
		7398.3453 & -9.0309$\pm$2.3998 & 1.7047$\pm$0.0034 & 0.9045$\pm$0.0021 & … & 4.900$\pm$0.022 & 20.673$\pm$0.065 & 4.64$\pm$1.31 & 21.6$\pm$20.5 & CARMENES S4 \\
		7401.2735 & -1.3684$\pm$1.1132 & 1.6720$\pm$0.0014 & 0.8741$\pm$0.0008 & 0.6139$\pm$0.0021 & 4.893$\pm$0.021 & 20.581$\pm$0.062 & 5.01$\pm$0.53 & 14.6$\pm$8.4 & CARMENES S4 \\
		7419.3799 & -6.1178$\pm$1.5674 & 1.6630$\pm$0.0022 & 0.8734$\pm$0.0011 & 0.5899$\pm$0.0031 & 4.884$\pm$0.021 & 20.761$\pm$0.063 & 5.04$\pm$0.69 & 15.0$\pm$15.3 & CARMENES S4 \\
		7426.3531 & 4.2565$\pm$1.2943 & 1.6725$\pm$0.0018 & 0.8821$\pm$0.0010 & 0.6134$\pm$0.0031 & 4.871$\pm$0.021 & 20.781$\pm$0.064 & -2.83$\pm$0.63 & 27.9$\pm$10.8 & CARMENES S4 \\
		7592.6888 & -1.9402$\pm$2.0841 & 1.7091$\pm$0.0017 & 0.9154$\pm$0.0010 & 0.6451$\pm$0.0024 & 4.898$\pm$0.021 & 20.263$\pm$0.061 & 2.59$\pm$0.63 & 17.4$\pm$11.9 & CARMENES S5 \\
		7604.6843 & -7.2872$\pm$2.4385 & 1.6454$\pm$0.0023 & 0.8784$\pm$0.0013 & 0.5340$\pm$0.0030 & 4.888$\pm$0.022 & 20.763$\pm$0.064 & 1.05$\pm$0.82 & 30.6$\pm$13.3 & CARMENES S5 \\
		7607.6875 & 3.7822$\pm$2.2250 & 1.6488$\pm$0.0022 & 0.8797$\pm$0.0012 & 0.5920$\pm$0.0029 & 4.889$\pm$0.022 & 20.628$\pm$0.064 & -4.88$\pm$0.76 & -0.4$\pm$16.6 & CARMENES S5 \\
		7611.6100 & 1.6209$\pm$2.6504 & 1.6228$\pm$0.0041 & 0.8715$\pm$0.0025 & 0.5876$\pm$0.0077 & 4.862$\pm$0.023 & 20.640$\pm$0.068 & 8.58$\pm$1.59 & 75.8$\pm$21.8 & CARMENES S5 \\
		7629.4806 & 4.1486$\pm$2.3723 & 1.6599$\pm$0.0040 & 0.8903$\pm$0.0025 & 0.5834$\pm$0.0074 & 4.868$\pm$0.024 & 20.676$\pm$0.069 & 0.05$\pm$1.54 & -2.9$\pm$22.2 & CARMENES S5 \\
		7633.5910 & -3.4593$\pm$2.2615 & 1.6559$\pm$0.0041 & 0.8866$\pm$0.0024 & 0.5877$\pm$0.0066 & 4.899$\pm$0.018 & 20.671$\pm$0.052 & 0.26$\pm$1.52 & 3.3$\pm$22.2 & CARMENES S5 \\
		7635.5553 & 6.2193$\pm$2.1956 & 1.6756$\pm$0.0041 & 0.9065$\pm$0.0024 & 0.6010$\pm$0.0067 & 4.868$\pm$0.021 & 20.643$\pm$0.061 & 6.60$\pm$1.52 & -36.4$\pm$21.1 & CARMENES S5 \\
		7644.6005 & -1.6044$\pm$1.4913 & 1.6824$\pm$0.0023 & 0.9030$\pm$0.0014 & 0.5316$\pm$0.0033 & 4.874$\pm$0.020 & 20.789$\pm$0.059 & -6.56$\pm$0.87 & 2.5$\pm$13.7 & CARMENES S5 \\
		7645.6629 & 2.2504$\pm$1.8186 & 1.6488$\pm$0.0027 & 0.8785$\pm$0.0016 & 0.5879$\pm$0.0043 & 4.869$\pm$0.021 & 20.703$\pm$0.063 & -2.27$\pm$1.01 & 11.4$\pm$14.7 & CARMENES S5 \\
		7647.4826 & -2.6226$\pm$2.2216 & 1.6700$\pm$0.0036 & 0.8997$\pm$0.0022 & 0.5884$\pm$0.0062 & 4.867$\pm$0.022 & 20.797$\pm$0.064 & 1.87$\pm$1.38 & 2.5$\pm$21.6 & CARMENES S5 \\
		7649.6315 & -1.7279$\pm$2.5159 & 1.6661$\pm$0.0042 & 0.8962$\pm$0.0024 & 0.5306$\pm$0.0065 & 4.898$\pm$0.019 & 20.961$\pm$0.055 & 1.16$\pm$1.48 & -26.8$\pm$22.9 & CARMENES S5 \\
		7653.5346 & 3.4239$\pm$1.9264 & 1.6683$\pm$0.0040 & 0.9105$\pm$0.0023 & 0.5994$\pm$0.0061 & 4.855$\pm$0.018 & 20.722$\pm$0.054 & -1.13$\pm$1.45 & -8.3$\pm$18.0 & CARMENES S5 \\
		7657.4684 & -2.7971$\pm$1.6143 & 1.6691$\pm$0.0028 & 0.8905$\pm$0.0017 & 0.5907$\pm$0.0044 & 4.893$\pm$0.021 & 20.771$\pm$0.063 & -7.78$\pm$1.05 & -4.2$\pm$14.4 & CARMENES S5 \\
		7678.3952 & -6.7376$\pm$1.5953 & 1.7491$\pm$0.0026 & 0.9395$\pm$0.0016 & 0.6308$\pm$0.0040 & 4.891$\pm$0.019 & 20.556$\pm$0.056 & 8.16$\pm$0.96 & 0.6$\pm$15.3 & CARMENES S5 \\
		7685.3671 & 0.7547$\pm$2.6824 & 1.7304$\pm$0.0046 & 0.9193$\pm$0.0029 & 0.6345$\pm$0.0088 & 4.854$\pm$0.023 & 20.553$\pm$0.066 & -7.22$\pm$1.77 & -24.0$\pm$25.3 & CARMENES S5 \\
		7688.4499 & -6.5007$\pm$1.8055 & 1.6320$\pm$0.0029 & 0.8650$\pm$0.0017 & 0.5835$\pm$0.0045 & 4.900$\pm$0.019 & 20.658$\pm$0.056 & 3.99$\pm$1.10 & -4.5$\pm$16.5 & CARMENES S5 \\
		7698.4636 & 0.5480$\pm$2.5962 & 1.7500$\pm$0.0047 & 0.9549$\pm$0.0029 & 0.6692$\pm$0.0088 & 4.876$\pm$0.018 & 20.787$\pm$0.054 & 9.11$\pm$1.78 & 12.1$\pm$25.9 & CARMENES S5 \\
		7703.4153 & -1.2194$\pm$1.5290 & 1.6781$\pm$0.0027 & 0.8865$\pm$0.0016 & 0.5484$\pm$0.0040 & 4.876$\pm$0.022 & 20.774$\pm$0.065 & 7.30$\pm$1.00 & 27.6$\pm$13.2 & CARMENES S5 \\
		7704.4039 & -2.2163$\pm$1.6587 & 1.6849$\pm$0.0028 & 0.8890$\pm$0.0016 & 0.5513$\pm$0.0041 & 4.891$\pm$0.020 & 20.813$\pm$0.059 & 2.15$\pm$1.03 & 31.1$\pm$15.5 & CARMENES S5 \\
		7705.3366 & -0.9255$\pm$2.2560 & 1.7074$\pm$0.0045 & 0.8984$\pm$0.0027 & 0.6381$\pm$0.0083 & 4.850$\pm$0.020 & 20.464$\pm$0.059 & 8.94$\pm$1.74 & 34.4$\pm$21.6 & CARMENES S5 \\
		7711.3008 & 14.3086$\pm$2.5642 & 1.6976$\pm$0.0051 & 0.9071$\pm$0.0030 & 0.6619$\pm$0.0097 & 4.867$\pm$0.022 & 20.657$\pm$0.064 & -2.66$\pm$1.90 & 1.3$\pm$25.5 & CARMENES S5 \\
		7738.4123 & 2.6132$\pm$2.6322 & 1.7044$\pm$0.0042 & 0.9096$\pm$0.0026 & 0.6233$\pm$0.0080 & 4.853$\pm$0.022 & 20.524$\pm$0.065 & 14.10$\pm$1.66 & -45.6$\pm$19.8 & CARMENES S5 \\
		7766.2878 & 2.5401$\pm$1.6133 & 1.6645$\pm$0.0026 & 0.8855$\pm$0.0015 & 0.5390$\pm$0.0036 & 4.906$\pm$0.020 & 20.891$\pm$0.059 & -3.45$\pm$0.93 & 13.9$\pm$14.0 & CARMENES S5 \\
		7769.3293 & 11.6179$\pm$3.7318 & 1.6781$\pm$0.0062 & 0.8983$\pm$0.0038 & 0.6474$\pm$0.0132 & 4.865$\pm$0.018 & 20.713$\pm$0.054 & 19.80$\pm$2.47 & -64.2$\pm$26.9 & CARMENES S5 \\
		7896.6670 & 1.2891$\pm$1.5408 & 1.6949$\pm$0.0016 & 0.8967$\pm$0.0010 & 0.6712$\pm$0.0026 & 4.871$\pm$0.021 & 20.439$\pm$0.060 & -7.29$\pm$0.61 & 10.4$\pm$10.8 & CARMENES S6 \\
		7938.6472 & 1.9646$\pm$2.0459 & 1.7010$\pm$0.0018 & 0.8936$\pm$0.0011 & 0.6142$\pm$0.0027 & 4.929$\pm$0.017 & 20.622$\pm$0.051 & -1.81$\pm$0.70 & -11.5$\pm$10.9 & CARMENES S6 \\
		7944.6704 & 0.3636$\pm$1.5841 & 1.7074$\pm$0.0018 & 0.9068$\pm$0.0011 & 0.6176$\pm$0.0026 & 4.912$\pm$0.021 & 20.512$\pm$0.062 & 8.72$\pm$0.68 & -17.9$\pm$11.7 & CARMENES S6 \\
		7945.6153 & -4.1719$\pm$1.2965 & 1.7137$\pm$0.0017 & 0.9142$\pm$0.0011 & 0.6294$\pm$0.0025 & 4.919$\pm$0.019 & 20.495$\pm$0.054 & 0.86$\pm$0.65 & -4.1$\pm$8.9 & CARMENES S6 \\
		7946.6684 & -6.0562$\pm$1.8075 & 1.7194$\pm$0.0026 & 0.9115$\pm$0.0016 & 0.6390$\pm$0.0039 & 4.891$\pm$0.021 & 20.465$\pm$0.061 & 12.53$\pm$0.97 & -14.6$\pm$15.8 & CARMENES S6 \\
		7949.6309 & -8.9183$\pm$1.4644 & 1.6200$\pm$0.0018 & 0.8410$\pm$0.0011 & 0.5794$\pm$0.0025 & 4.880$\pm$0.020 & 20.590$\pm$0.059 & -0.02$\pm$0.68 & 21.5$\pm$10.2 & CARMENES S6 \\
		7950.5906 & -7.3860$\pm$1.4416 & 1.6190$\pm$0.0017 & 0.8427$\pm$0.0010 & 0.5774$\pm$0.0025 & 4.885$\pm$0.020 & 20.595$\pm$0.060 & -0.83$\pm$0.66 & -9.2$\pm$10.9 & CARMENES S6 \\
		7954.6366 & 9.3590$\pm$1.6650 & 1.6387$\pm$0.0017 & 0.8747$\pm$0.0010 & 0.5927$\pm$0.0025 & 4.920$\pm$0.018 & 20.639$\pm$0.053 & -2.54$\pm$0.65 & 9.4$\pm$12.3 & CARMENES S6 \\
		7960.5181 & 4.0781$\pm$1.7588 & 1.7452$\pm$0.0019 & 0.9416$\pm$0.0012 & 0.6298$\pm$0.0029 & 4.903$\pm$0.022 & 20.642$\pm$0.065 & 9.14$\pm$0.71 & 37.8$\pm$15.0 & CARMENES S6 \\
		7962.6565 & 4.0516$\pm$1.5272 & 1.7730$\pm$0.0018 & 0.9631$\pm$0.0011 & 0.6543$\pm$0.0027 & 4.901$\pm$0.021 & 20.520$\pm$0.061 & 5.10$\pm$0.67 & 12.8$\pm$13.6 & CARMENES S6 \\
		7963.6746 & 4.4156$\pm$1.6314 & 1.6999$\pm$0.0017 & 0.9042$\pm$0.0010 & 0.6178$\pm$0.0024 & 4.922$\pm$0.022 & 20.526$\pm$0.063 & 5.67$\pm$0.64 & 16.5$\pm$13.8 & CARMENES S6 \\
		7964.6424 & 0.1782$\pm$1.5924 & 1.7259$\pm$0.0018 & 0.9166$\pm$0.0011 & 0.6377$\pm$0.0027 & 4.897$\pm$0.022 & 20.495$\pm$0.065 & 5.28$\pm$0.68 & -15.9$\pm$13.9 & CARMENES S6 \\
		7965.6673 & -2.6009$\pm$1.8440 & 1.6853$\pm$0.0018 & 0.8901$\pm$0.0011 & 0.6166$\pm$0.0026 & 4.883$\pm$0.022 & 20.514$\pm$0.065 & 9.31$\pm$0.68 & 30.8$\pm$15.0 & CARMENES S6 \\
		7968.6252 & -2.3128$\pm$1.6405 & 1.6294$\pm$0.0017 & 0.8526$\pm$0.0010 & 0.5803$\pm$0.0024 & 4.903$\pm$0.019 & 20.671$\pm$0.057 & -0.29$\pm$0.64 & 31.3$\pm$14.0 & CARMENES S6 \\
		7990.5161 & 5.9262$\pm$1.7334 & 1.6427$\pm$0.0017 & 0.8702$\pm$0.0010 & 0.5911$\pm$0.0025 & 4.900$\pm$0.022 & 20.559$\pm$0.064 & 0.91$\pm$0.65 & 11.1$\pm$14.9 & CARMENES S6 \\
		7997.5845 & -1.4014$\pm$1.3662 & 1.6962$\pm$0.0017 & 0.9030$\pm$0.0010 & 0.6098$\pm$0.0024 & 4.906$\pm$0.020 & 20.599$\pm$0.060 & -1.86$\pm$0.64 & 2.7$\pm$9.9 & CARMENES S6 \\
		8000.5295 & 5.9164$\pm$1.2737 & 1.6825$\pm$0.0018 & 0.8946$\pm$0.0011 & 0.6061$\pm$0.0025 & 4.906$\pm$0.020 & 20.631$\pm$0.060 & -2.32$\pm$0.66 & -31.1$\pm$9.0 & CARMENES S6 \\
		8000.6631 & 4.8073$\pm$1.3662 & 1.6715$\pm$0.0021 & 0.8842$\pm$0.0013 & 0.6014$\pm$0.0033 & 4.908$\pm$0.021 & 20.537$\pm$0.061 & -0.51$\pm$0.82 & -28.9$\pm$10.4 & CARMENES S6 \\
		8005.5147 & -7.8766$\pm$1.4763 & 1.6410$\pm$0.0018 & 0.8635$\pm$0.0011 & 0.5966$\pm$0.0028 & 4.893$\pm$0.021 & 20.602$\pm$0.060 & 0.86$\pm$0.73 & -0.2$\pm$13.0 & CARMENES S6 \\
		8006.5883 & -6.5700$\pm$1.7950 & 1.6366$\pm$0.0030 & 0.8577$\pm$0.0018 & 0.5852$\pm$0.0049 & 4.852$\pm$0.019 & 20.560$\pm$0.057 & -7.97$\pm$1.16 & -13.3$\pm$15.5 & CARMENES S6 \\
		8007.5236 & -6.2947$\pm$1.5762 & 1.6204$\pm$0.0022 & 0.8483$\pm$0.0013 & 0.5774$\pm$0.0031 & 4.900$\pm$0.019 & 20.746$\pm$0.055 & -8.01$\pm$0.83 & 3.5$\pm$14.0 & CARMENES S6 \\
		8008.5384 & -2.3464$\pm$1.4351 & 1.6346$\pm$0.0017 & 0.8649$\pm$0.0010 & 0.5910$\pm$0.0024 & 4.889$\pm$0.020 & 20.518$\pm$0.057 & -10.92$\pm$0.65 & -12.5$\pm$10.1 & CARMENES S6 \\
		8009.4829 & 1.2484$\pm$1.4247 & 1.6204$\pm$0.0018 & 0.8453$\pm$0.0011 & 0.5821$\pm$0.0026 & 4.893$\pm$0.020 & 20.558$\pm$0.057 & -4.95$\pm$0.69 & -28.6$\pm$11.4 & CARMENES S6 \\
		8023.3099 & -10.0960$\pm$1.8404 & 1.6410$\pm$0.0021 & 0.8680$\pm$0.0012 & 0.5854$\pm$0.0032 & 4.877$\pm$0.022 & 20.694$\pm$0.064 & -0.79$\pm$0.77 & -1.4$\pm$16.1 & CARMENES S6 \\
		8027.5570 & -0.4132$\pm$1.3167 & 1.6033$\pm$0.0016 & 0.8400$\pm$0.0009 & 0.5700$\pm$0.0021 & 4.900$\pm$0.019 & 20.674$\pm$0.056 & -12.06$\pm$0.59 & 0.6$\pm$10.5 & CARMENES S6 \\
		8028.5945 & 5.1740$\pm$1.7496 & 1.6025$\pm$0.0016 & 0.8439$\pm$0.0010 & 0.5651$\pm$0.0022 & 4.903$\pm$0.020 & 20.722$\pm$0.059 & -15.13$\pm$0.60 & -21.7$\pm$11.9 & CARMENES S6 \\
		8030.4346 & 14.5024$\pm$1.8300 & 1.6886$\pm$0.0017 & 0.9029$\pm$0.0010 & 0.6255$\pm$0.0024 & 4.907$\pm$0.019 & 20.682$\pm$0.056 & -13.35$\pm$0.63 & 0.4$\pm$13.0 & CARMENES S6 \\
		8040.5104 & -1.0588$\pm$2.1842 & 1.6694$\pm$0.0017 & 0.8819$\pm$0.0010 & 0.5854$\pm$0.0022 & 4.890$\pm$0.020 & 20.494$\pm$0.057 & -4.25$\pm$0.61 & 28.0$\pm$11.8 & CARMENES S6 \\
		8050.4911 & 5.6164$\pm$1.8266 & 1.6460$\pm$0.0017 & 0.8664$\pm$0.0010 & 0.5960$\pm$0.0024 & 4.907$\pm$0.019 & 20.621$\pm$0.057 & -1.74$\pm$0.63 & 5.8$\pm$13.0 & CARMENES S6 \\
		8054.4417 & 0.1965$\pm$1.5933 & 1.6647$\pm$0.0019 & 0.8786$\pm$0.0011 & 0.6044$\pm$0.0025 & 4.884$\pm$0.022 & 20.582$\pm$0.063 & -0.76$\pm$0.67 & 5.7$\pm$11.3 & CARMENES S6 \\
		8060.3782 & -0.8298$\pm$1.4019 & 1.6498$\pm$0.0022 & 0.8630$\pm$0.0013 & 0.5992$\pm$0.0031 & 4.881$\pm$0.023 & 20.515$\pm$0.066 & 6.36$\pm$0.80 & -10.9$\pm$11.9 & CARMENES S6 \\
		8066.4227 & 1.6766$\pm$1.4739 & 1.5857$\pm$0.0022 & 0.8225$\pm$0.0014 & 0.5870$\pm$0.0035 & 4.886$\pm$0.021 & 20.731$\pm$0.061 & 4.68$\pm$0.87 & -13.2$\pm$11.4 & CARMENES S6 \\
		8081.3706 & -5.9032$\pm$1.8106 & 1.6116$\pm$0.0025 & 0.8407$\pm$0.0014 & 0.5831$\pm$0.0035 & 4.869$\pm$0.019 & 20.705$\pm$0.055 & 1.08$\pm$0.94 & 13.5$\pm$15.0 & CARMENES S6 \\
		8084.4080 & 8.4098$\pm$1.7798 & 1.5845$\pm$0.0018 & 0.8243$\pm$0.0010 & 0.5714$\pm$0.0024 & … & … & … & -23.5$\pm$11.0 & CARMENES S6 \\
		8092.5559 & -5.8962$\pm$1.3455 & 1.6467$\pm$0.0015 & 0.8788$\pm$0.0008 & 0.6046$\pm$0.0020 & 4.891$\pm$0.021 & 20.744$\pm$0.063 & 5.32$\pm$0.52 & 5.2$\pm$8.6 & CARMENES S6 \\
		8095.3943 & -0.7760$\pm$1.3138 & 1.6389$\pm$0.0016 & 0.8645$\pm$0.0009 & 0.5479$\pm$0.0021 & 4.913$\pm$0.019 & 20.823$\pm$0.058 & 1.65$\pm$0.59 & -22.2$\pm$9.6 & CARMENES S6 \\
		8099.2958 & 0.4674$\pm$2.0542 & 1.6184$\pm$0.0039 & 0.8497$\pm$0.0024 & 0.5055$\pm$0.0072 & 4.874$\pm$0.020 & 20.839$\pm$0.059 & 4.25$\pm$1.53 & 16.3$\pm$18.4 & CARMENES S6 \\
		8114.3012 & 3.2727$\pm$2.4292 & 1.6598$\pm$0.0036 & 0.8842$\pm$0.0022 & 0.5076$\pm$0.0062 & 4.892$\pm$0.020 & 20.947$\pm$0.059 & -2.19$\pm$1.37 & 1.0$\pm$19.2 & CARMENES S6 \\
		8121.4035 & 1.6909$\pm$1.3377 & 1.5951$\pm$0.0016 & 0.8329$\pm$0.0009 & 0.5265$\pm$0.0021 & 4.892$\pm$0.020 & 20.755$\pm$0.059 & -5.02$\pm$0.59 & -4.7$\pm$9.7 & CARMENES S6 \\
		8123.3299 & 1.3996$\pm$1.4461 & 1.6447$\pm$0.0018 & 0.8572$\pm$0.0010 & 0.5508$\pm$0.0024 & 4.899$\pm$0.019 & 20.740$\pm$0.056 & 0.54$\pm$0.65 & 24.0$\pm$10.1 & CARMENES S6 \\
		8141.4369 & 5.0553$\pm$1.2530 & 1.6591$\pm$0.0016 & 0.8683$\pm$0.0009 & 0.5579$\pm$0.0022 & 4.881$\pm$0.020 & 20.821$\pm$0.059 & 3.65$\pm$0.56 & -10.5$\pm$8.9 & CARMENES S6 \\
		8143.4314 & 2.6230$\pm$1.4587 & 1.6530$\pm$0.0016 & 0.8606$\pm$0.0009 & 0.5476$\pm$0.0024 & 4.897$\pm$0.020 & 20.791$\pm$0.060 & 3.51$\pm$0.57 & 3.7$\pm$9.2 & CARMENES S6 \\
		8149.3838 & -1.4678$\pm$1.4172 & 1.6744$\pm$0.0020 & 0.8939$\pm$0.0012 & 0.5463$\pm$0.0028 & 4.881$\pm$0.022 & 20.719$\pm$0.064 & 0.77$\pm$0.71 & 13.7$\pm$11.4 & CARMENES S6 \\
		8161.3404 & 1.7213$\pm$1.8407 & 1.6459$\pm$0.0021 & 0.8631$\pm$0.0013 & 0.5284$\pm$0.0032 & 4.898$\pm$0.019 & 20.761$\pm$0.057 & -0.87$\pm$0.79 & -4.0$\pm$11.9 & CARMENES S6 \\
		8166.3726 & 3.9816$\pm$1.7122 & 1.6575$\pm$0.0024 & 0.8741$\pm$0.0014 & 0.5293$\pm$0.0040 & 4.884$\pm$0.022 & 20.825$\pm$0.065 & 5.21$\pm$0.90 & 11.6$\pm$14.7 & CARMENES S6 \\
		8167.3469 & 4.6990$\pm$2.5945 & 1.6835$\pm$0.0045 & 0.8940$\pm$0.0027 & 0.5320$\pm$0.0088 & 4.819$\pm$0.023 & 20.702$\pm$0.068 & 13.88$\pm$1.68 & -11.1$\pm$21.9 & CARMENES S6 \\
		8172.2861 & 4.4350$\pm$1.9599 & 1.6413$\pm$0.0034 & 0.8740$\pm$0.0021 & 0.5157$\pm$0.0058 & 4.866$\pm$0.026 & 20.723$\pm$0.077 & -4.98$\pm$1.29 & 5.0$\pm$14.6 & CARMENES S6 \\
		8173.2864 & -4.3313$\pm$1.7845 & 1.6852$\pm$0.0018 & 0.8987$\pm$0.0011 & 0.6105$\pm$0.0026 & 4.884$\pm$0.020 & 20.737$\pm$0.058 & 3.85$\pm$0.67 & 5.4$\pm$12.4 & CARMENES S6 \\
		8174.2934 & -5.3571$\pm$2.0300 & 1.6451$\pm$0.0025 & 0.8703$\pm$0.0015 & 0.6049$\pm$0.0041 & 4.898$\pm$0.019 & 20.789$\pm$0.055 & 6.17$\pm$0.96 & -3.6$\pm$11.3 & CARMENES S6 \\
		8175.3066 & -5.0708$\pm$1.3183 & 1.6411$\pm$0.0020 & 0.8697$\pm$0.0011 & 0.5949$\pm$0.0028 & 4.878$\pm$0.019 & 20.817$\pm$0.056 & 5.11$\pm$0.73 & -8.6$\pm$9.1 & CARMENES S6 \\
		\noalign{\smallskip}
		\hline
	\end{longtable}
\end{landscape}

	\begin{figure*}
		\resizebox{\hsize}{!}{\includegraphics[width=\textwidth]{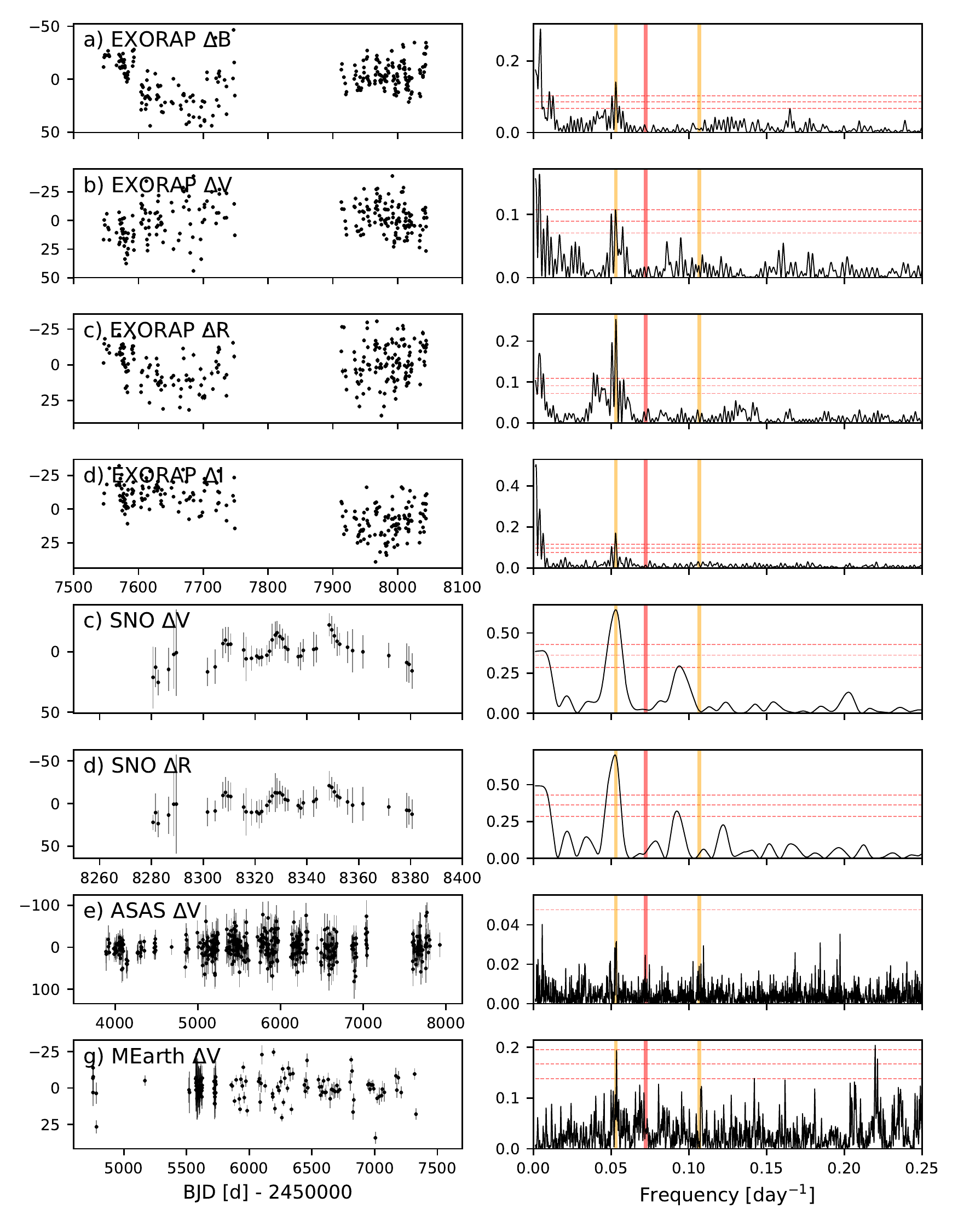}}
		\caption{Differential photometric data of Gl\,49 in mmag (left panels; we note the inverted $y$ axis) and GLS periodograms (right panels) including EXORAP, SNO, ASAS, and MEarth. The vertical orange lines indicate periods of 9.37 and 18.86\,d, the red vertical line a period of 13.85\,d, and the horizontal dashed red lines the 0.1, 1, and 10\,\% analytical FAP.}
		\centering
		\label{A3}
	\end{figure*}
	
	\begin{figure*}
		\resizebox{\hsize}{!}{\includegraphics[width=\textwidth]{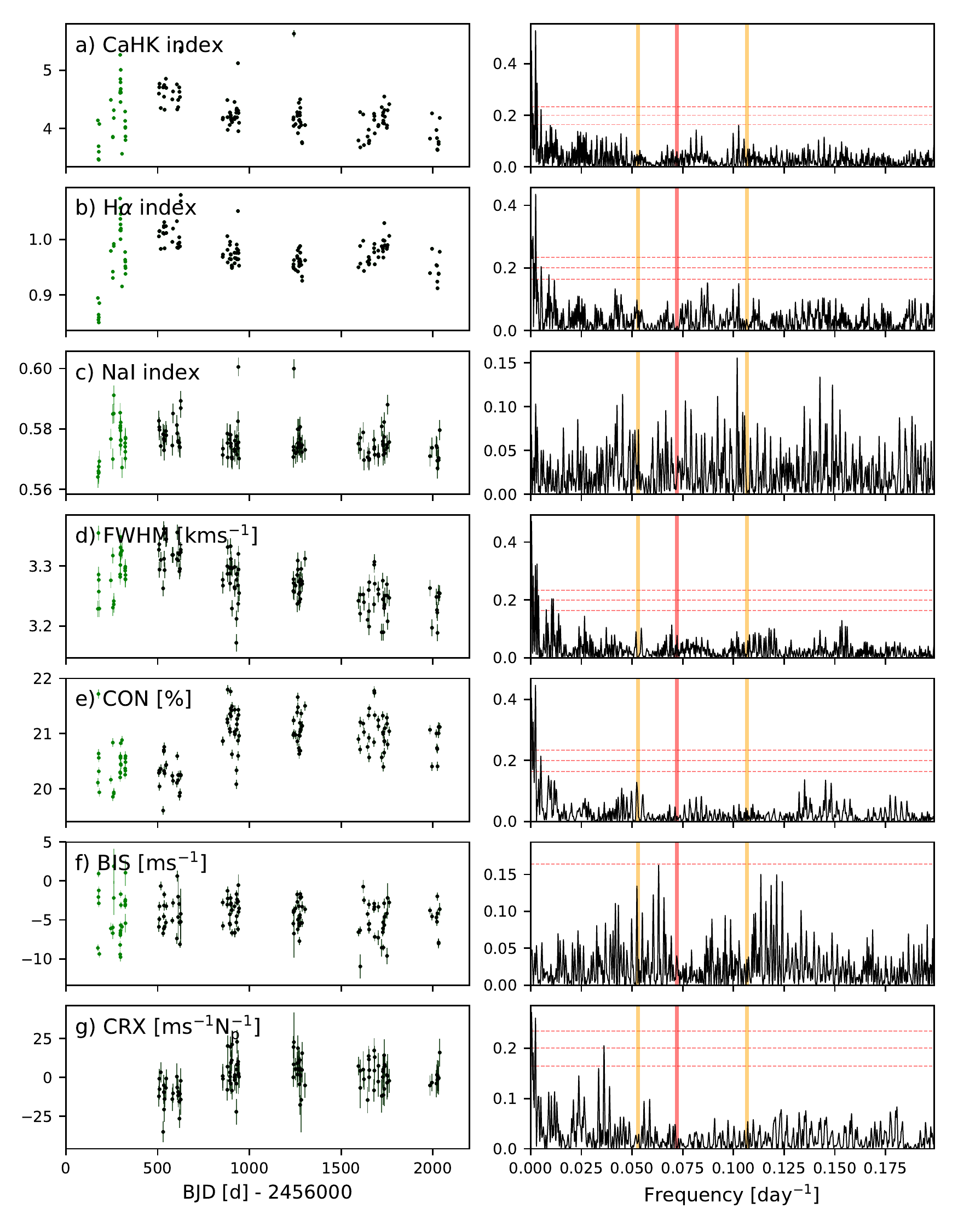}}
		\caption{Seven activity indices calculated from HARPS-N spectra. On the left we show the time series of (from top to bottom) the CaHK, H$\alpha$, NaI, FWHM, CON, BIS, and CRX indices. On the right we show the periodograms of those datasets excluding HARPS-N S1 (marked in green on the left). For more details see Fig\,\ref{A3}.}
		\centering
		\label{A1}
	\end{figure*}
	
	\begin{figure*}
		\resizebox{\hsize}{!}{\includegraphics[width=\textwidth]{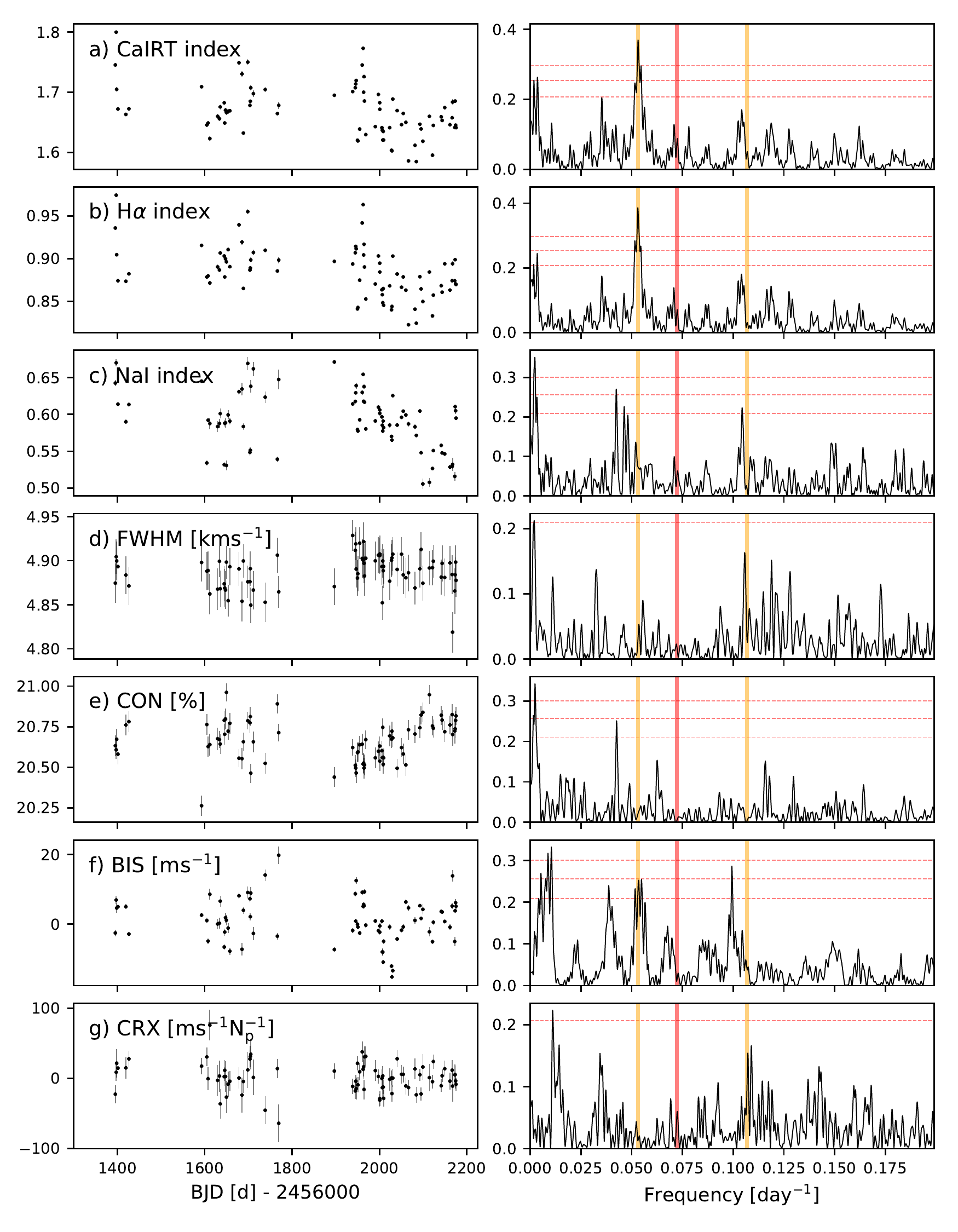}}
		\caption{Same as Fig.\,\ref{A1} but for the CARMENES data.}
		\centering
		\label{A2}
	\end{figure*}
	
	\begin{figure*}
		\resizebox{\hsize}{!}{\includegraphics[width=\textwidth]{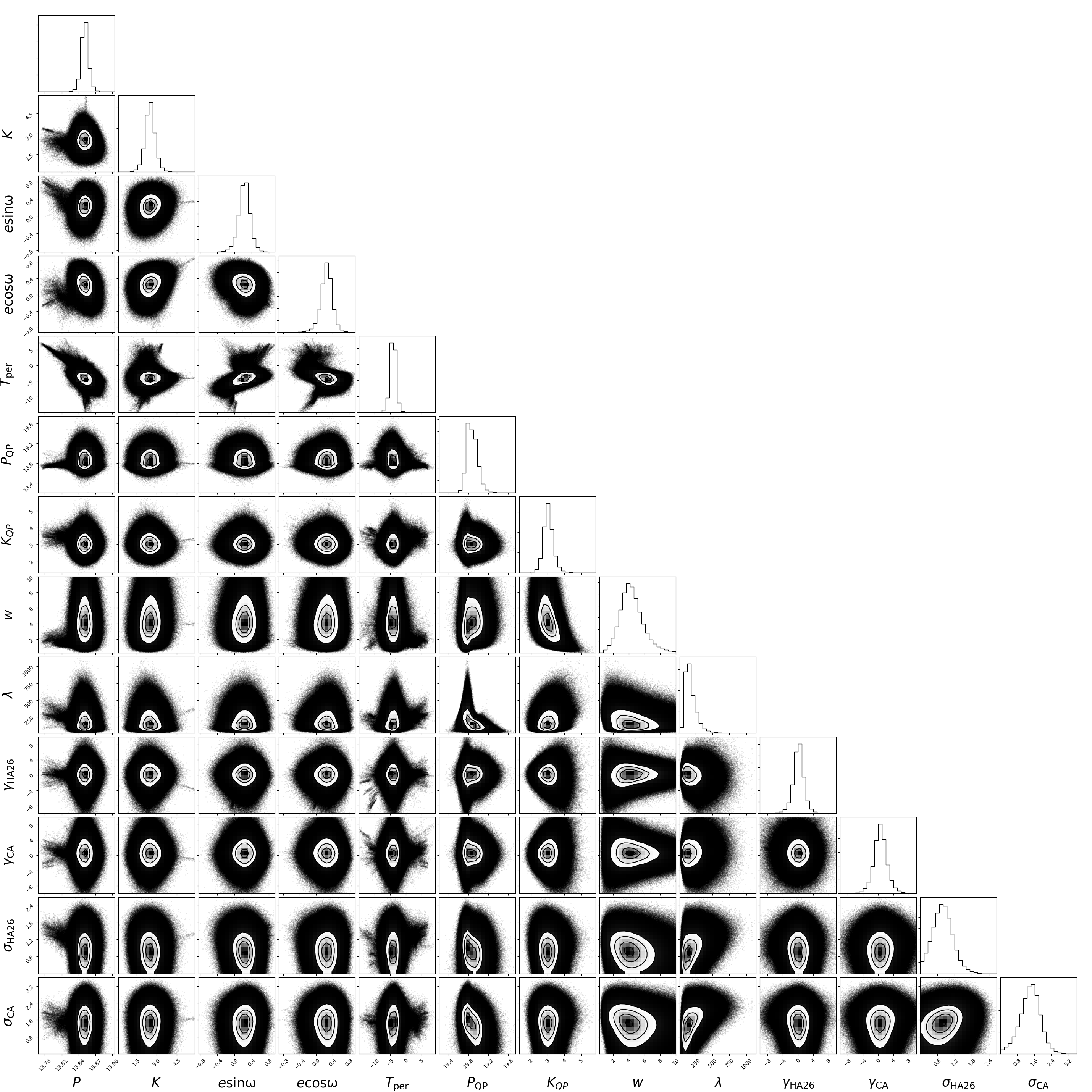}}
		\caption{Corner plot of the fitted parameters of the MCMC solutions using a Keplerian orbit (parameters $P$, $K$, $e \sin{\omega}$, $e \cos{\omega}$, $T_{\rm per}$) and a GP noise term (hyper-parameters $P_{\rm QP}$, $K_{\rm QP}$, $w$, $\lambda$) to the RVs of HARPS-N S2 to S6 and CARMENES of Gl\,49. The solutions included in this plot exceed the $\ln{L}$ value of a fit with a GP term only (see Table\,\ref{T5}) plus the 1\,\% FAP value that we calculate empirically in Sect.\,\ref{sig:me}.}
		\centering
		\label{F10}
	\end{figure*}
		
\end{appendix}

\end{document}